\documentclass[twoside,8pt]{article}

\usepackage{blindtext}

\usepackage[preprint]{jmlr2e}
\usepackage{amsmath}
\usepackage{pgf,tikz}  %画DAG专用包
\usetikzlibrary{arrows,shapes.arrows,shapes.geometric,
shapes.multipart,decorations.pathmorphing,positioning,
swigs}
\newcommand{\indep}{\rotatebox[origin=c]{90}{$\models$}}

\newcommand{\argmin}{\mathop{\mathrm{arg\,min}}}

\usepackage[ruled,vlined]{algorithm2e} %alogrithm
\usepackage{algpseudocode} %使得算法里面有数字标头

\usepackage{graphicx} % 支持 \resizebox

\usepackage{booktabs}
\usepackage{float}
\usepackage{multirow}

\usepackage{enumitem}  % 这个包允许更灵活的列表设置

\usepackage{csquotes}

\usepackage{hyperref}

\newcommand{\mysection}[1]{%
  \section*{#1}%
  \addcontentsline{toc}{section}{#1}%
}

\usepackage{lastpage}
\jmlrheading{00}{2025}{1-\pageref{LastPage}}{0/00; Revised 0/00}{0/00}{21-0000}{Tian, Patel, and Burgess}

\ShortHeadings{Stratification IV for Nonlinear Effects}{Tian, Patel, and Burgess}
\firstpageno{1}

\begin{document}

\title{\LARGE Stratification-based Instrumental Variable Analysis Framework for Nonlinear Effect Analysis}

\author{\normalfont{ \centering \large Haodong Tian\textsuperscript{1,2,*}, Ashish Patel\textsuperscript{3}, Stephen Burgess\textsuperscript{3,4} \vspace{2mm} \\
\normalsize \textsuperscript{1} Center for Genomic Medicine, Massachusetts General Hospital\vspace{0mm}\\
\normalsize \textsuperscript{2} Broad Institute of MIT and Harvard\vspace{0mm}\\
\normalsize \textsuperscript{3} MRC Biostatistics Unit, University of Cambridge \vspace{0mm} \\
\normalsize \textsuperscript{4} BHF Cardiovascular Epidemiology Unit, University of Cambridge \vspace{0.5mm}\\
\normalsize \textsuperscript{*}\href{email:htian2@mgh.harvard.edu}{htian2@mgh.harvard.edu} \vspace{0mm}\\
}
}
\date{ }

\maketitle

\begin{abstract}%   <- trailing '%' for backward compatibility of .sty file
Nonlinear causal effects are prevalent in many research scenarios involving continuous exposures, and instrumental variables (IVs) can be employed to investigate such effects, particularly in the presence of unmeasured confounders. However, common IV methods for nonlinear effect analysis, such as IV regression or the control-function method, have inherent limitations, leading to either low statistical power or potentially misleading conclusions. 
In this work, we propose an alternative IV framework for nonlinear effect analysis, which has recently emerged in genetic epidemiology and addresses many of the drawbacks of existing IV methods. The proposed IV framework consists of up to three key `S' elements: (i) the \underline{S}tratification approach, which constructs multiple strata that are sub-samples of the population in which the IV core assumptions remain valid, (ii) the \underline{S}calar-on-function model and \underline{S}calar-on-scalar model, which connect local stratum-specific information to global effect estimation, and (iii) the \underline{S}um-of-single-effects method for effect estimation. This framework enables study of the effect function while avoiding unnecessary model assumptions. In particular, it facilitates the identification of change points or threshold values in causal effects.  
Through a wide variety of simulations, we demonstrate that our framework outperforms other representative nonlinear IV methods in predicting the effect shape when the instrument is weak and can accurately estimate the effect function as well as identify the change point and predict its value under various structural model and effect shape scenarios.
We further apply our framework to assess the nonlinear effect of alcohol consumption on systolic blood pressure using a genetic instrument (i.e. Mendelian randomization) with UK Biobank data. Our analysis detects a threshold beyond which alcohol intake exhibits a clear causal effect on the outcome. Our results are consistent with published medical guidelines.
\end{abstract}


\section{Introduction}

Investigating causal effects has become a central topic in many studies.  
In many scenarios, the exposure or treatment is continuous-valued, and the corresponding causal effect on the outcome can generally be defined as the counterfactual contrast between two treatment or exposure levels, $x_1$ and $x_2$, using the potential outcome framework \citep{rubin1974estimating,holland1986statistics}:  
$
\tau(x_1,x_2) := \mathbb{E}( Y(x_2) - Y(x_1) )
$,
where $Y(x)$ denotes the counterfactual outcome under exposure level $x$.  
When the outcome structural model is linear with respect to the exposure level, with an average effect intensity $\beta$, the causal effect $\tau(x_1,x_2)$ simplifies to a function of a single argument:  
$
\tau( \Delta ) := \beta \Delta
$
where $\Delta := x_2 - x_1$ represents only the intervention size and is independent of the current exposure level. In this case, the average causal effect $\beta$, which is a scalar, can be defined without ambiguity.  
However, when the assumption of a linear outcome model is inappropriate—a common occurrence in many real-world scenarios—the causal effect is better defined as a functional quantity that depends on the exposure level, as in \cite{florens2008identification}:  
\begin{equation}\label{effect_density}
    \beta(x) := \lim_{ \Delta \to 0 } \frac{  \mathbb{E}( Y(x+\Delta)  - Y(x) )    }{  \Delta }
\end{equation}
which represents the derivative of the average counterfactual outcome $\mathbb{E}( Y(x) )$. Given that $\mathbb{E}( Y(x) )$ is continuous in $x$, the causal effect can be expressed as  
$
\tau(x_1,x_2) = \int_{x_1}^{x_2}  \beta(x) \,dx
$.

When unmeasured confounding factors exist in the exposure-outcome association, instrumental variables (IVs) serve as a valuable tool for studying causal effects. For a valid instrument $Z$, three core assumptions must be satisfied:  
(i) the instrument is associated with the exposure;  
(ii) the instrument is not associated with the outcome via a confounding pathway; and  
(iii) the instrument affects the outcome only indirectly through the exposure, with no direct effect \citep{greenland2000, martens2006}.  
These core assumptions are sufficient to test the sharp causal null hypothesis—that no causal effect exists for any individual—by simply testing the instrument-outcome association. Further inference on the causal effect of interest requires additional model assumptions.  
When the exposure is a naturally binary variable, an additional monotonicity assumption is typically introduced to enable IV analysis to identify the local average treatment effect \citep{imbens1994identification}. For continuous exposures, one can impose a structural outcome model assumption, such as  
$
Y = h(X) + U,
$  
under the stable unit treatment value assumption (SUTVA) \citep{rubin1980randomization}, where $h(\cdot)$ represents the effect function (also referred to as the dose-response curve in some literature), and $U$ is a mean-zero error term independent of $Z$. Under some regularity conditions, the effect function $h(\cdot)$ can be identified via IV regression:  
\begin{equation}\label{IV_regression}
    \mathbb{E}(  Y \mid Z \in \mathcal{Z}  ) =  \mathbb{E}(  h(X) \mid Z \in \mathcal{Z}  ) = \int_{\mathcal{X}} h(x) \, \text{d} F(x \mid Z \in \mathcal{Z} ),  
\end{equation}
where $\mathcal{Z}$ is a sub-domain of the original $Z$ domain.  
To address the ill-posed inverse problem—where the functional objective $h(\cdot)$ remains ill-identified even when consistent estimators of other functional covariates are available—additional basis or sieve function assumptions on $h(\cdot)$ are often further introduced for inference \citep{horowitz2014ill}.   

IV regression has been a longstanding and widely used method for nonlinear effect analysis within the IV framework \citep{newey2003}. Its core idea has been extended to various nonparametric methods and integrated into machine learning approaches \citep{hartford2017deep,singh2019kernel,he2023delivr}. However, despite its conciseness and flexibility, IV regression for nonlinear effect analysis struggles to gain traction in scenarios where the number of distinct instrument values and the number of IVs are low, or where the instrumental strength is weak (i.e., the proportion of exposure variation explained by the instrument is low).  
A representative example is Mendelian randomization (MR), a popular epidemiological method that leverages genetic variants as IVs for health-related causal effect analysis \citep{davey2003mendelian,lawlor2008mendelian}. Since genetic variants in MR are typically single-nucleotide polymorphisms (SNPs) that take values of $0,1,2$—representing the number of effect alleles in a diploid genome—the number of distinct instrument values is inherently limited. Moreover, the genetic strength is often weak; in many applications, the $R^2$ from regressing exposure on the genetic variant(s) is below $0.1$. Compared to a simple model of homogeneous causal effects, identifying heterogeneous causal effects is more demanding on instrument strength \citep{brinch2017beyond}, since weak instruments would map to only a limited support of exposure values, making it difficult to identify nonlinear causal effects.

As an alternative approach, the control function or residual inclusion method \citep{heckman1985alternative} offers a different strategy for studying the effect shape. In a simple setting, instrumental variables can be used to extract the exposure’s endogenous variation, yielding a residual term, and this residual is then added as a covariate in the outcome regression on exposure to eliminate endogeneity bias in the estimated exposure–outcome causal effect. The general form of the control function method involves fitting the regression model  
\begin{equation}
    Y \sim f_1(X) + \cdots + f_K(X) +  r_1 + \cdots + r_L,
\end{equation}
where $\{ f_k( \cdot ) \}$ represent basis functions assumed for the effect shape $h(\cdot)$, and $\{r_l\}$ are residual terms. Each residual term can either be a transformation of the residual from regressing the exposure on the instruments or the residual from regressing a transformed exposure on the instruments.  
For instance, in the case where $L = K$ and each $r_l$ is the residual from regressing $f_l(X)$ on the instruments, the control function estimation becomes equivalent to two-stage least squares (2SLS) estimation, treating $\{ f_k(X) \}$ as $K$ separate exposures. In this case, identification typically requires at least $K$ valid, distinct instruments. Alternatively, one can define $\{ r_l \}$ as polynomial terms of the conventional residual (i.e., the residual from regressing $X$ on the instrument). This latter approach aligns with the core intuition of the control function method: including the residual acts as an adjustment for unmeasured exposure-outcome confounders, thereby mitigating the endogeneity issue that arises from directly regressing $Y$ on $\{ f_k(X) \}$ \citep{terza2008two}. Here, higher-order polynomial terms of the conventional residual approximate the effect function of the exposure-outcome confounder via a Taylor series expansion.  
A key advantage of this polynomial-based strategy is that it requires only a single valid instrument to identify all parameters. Partially due to this reason, such a control function method has gained traction in the MR field \citep{sulc2022polynomial}. However, this approach relies on strong parametric assumptions—both for the exposure model and for the functional form of the confounder's effect on the outcome, implying that a single scalar residual (confounder) can adequately approximate the total confounding effect, which is often implausible in IV contexts where confounders are unmeasured, and their structural relationships with other variables are unknown and can be complex.  
Moreover, in scenarios where the control function estimator is equivalent to a 2SLS estimator with an augmented set of instrumental variables \citep{guo2016control}, the additional instrumental variables may be invalid without strong modeling assumptions. These limitations pose significant challenges for applying control function methods in nonlinear causal effect analysis.  

Facing the key challenges discussed above—namely, low dimensionality of the IV, weak instrument strength, and the high burden of structural assumptions on confounding—a different approach, originally developed in the context of MR, has recently emerged for nonlinear causal effect analysis. This approach introduces the concept of stratification, where the sample is divided into multiple strata, and stratum-specific estimates are obtained. Each stratum-specific statistic captures localized information over the exposure domain, which is then aggregated to infer the global effect shape. We refer to methods based on this idea as \textit{stratification-based methods}.
Importantly, the stratification-based method shifts multiple assumptions, originally required by other common nonlinear IV methods, into the stratification process (i.e., the reliability of the final result is primarily determined by the stratification assumption, rather than by other common conditions, including IV type, IV strength, confounding pattern, and underlying effect shape). As a result, the plausibility of the assumptions in the stratification-based method can be more easily controlled and evaluated in practice than in other methods.
Due to their strong ability to address many of the key challenges, their effectiveness in estimating effect shapes, and their simplicity and extensibility, stratification-based methods have rapidly gained popularity in MR. These methods have been applied to study effect shapes for various phenotypes in many recent studies \citep{sofianopoulou2024estimating,yang2024dose,kassaw2024alcohol}. This paper aims to formally introduce this class of methods for nonlinear effect analysis and extend the original stratification concept by integrating additional techniques to a comprehensive framework, making them applicable to a broader range of IV scenarios beyond genetic epidemiology.  

\begin{figure}[tb]  % 'h' 表示图片位置，h 表示尽可能放在当前位置
    \centering  % 图片居中
    \includegraphics[width=1.0\textwidth]{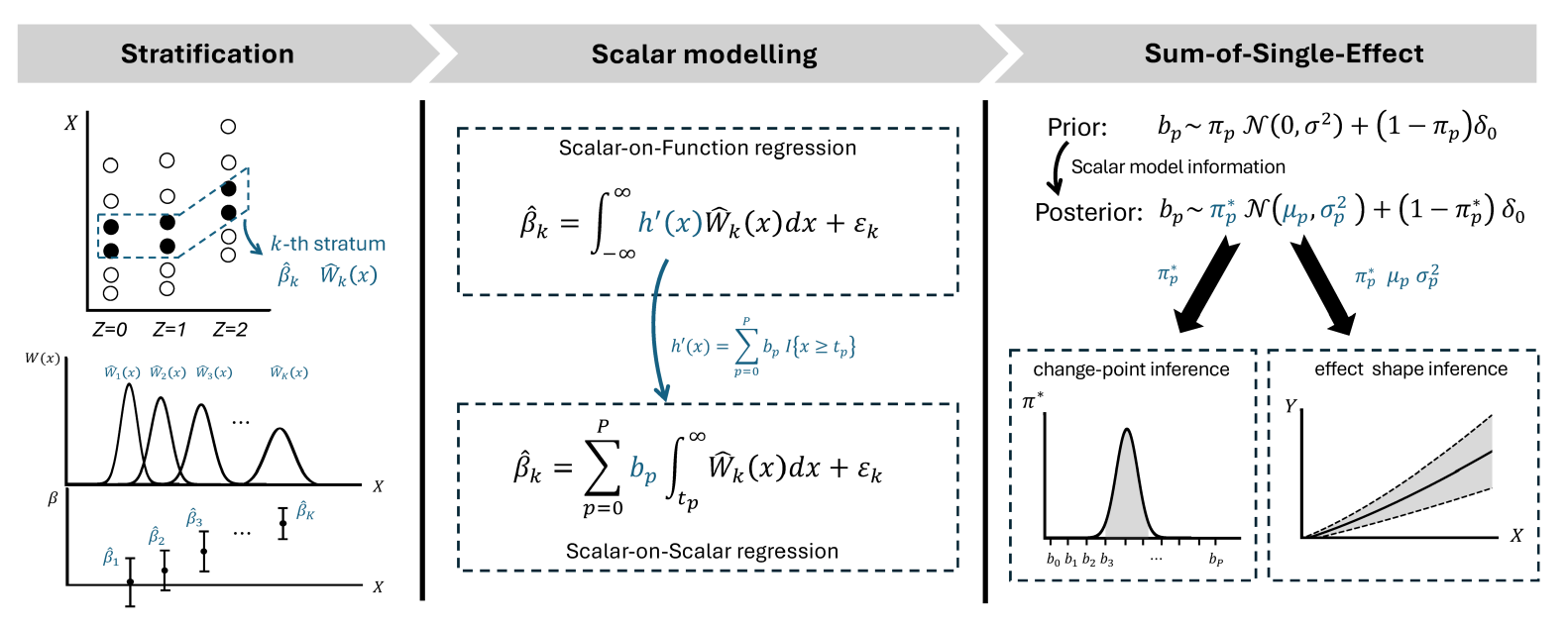}  % 图片路径和大小
    \vspace{-1em}
    \caption{\normalfont The graphical overview of the “SSS” framework for complete nonlinear effect analysis. From left to right: (Stratification) Multiple strata are constructed using a stratification approach that ensures the IV assumption holds within each stratum. The stratum-specific IV estimators $\hat{\beta}_k$, and the weight functions, $\hat{W}_k(x)$, are recorded. (Scalar modelling) A scalar-on-function regression is specified and then transformed into a scalar-on-scalar regression via a nonparametric representation of the underlying effect function $h'(x)$. (Sum-of-Single-Effect) A Bayesian approach with spike-and-slab priors is applied to infer change-points from the posterior inclusion probabilities $\pi^\ast$, and to estimate the effect shape function based on all posterior parameters.}
     \vspace{-0em}
    % 图片标题
    \label{SSS_overview}  % 图片标签，方便引用
\end{figure}

In this article, we propose a complete analysis pipeline for comprehensive nonlinear effect analysis with multiple objectives, including testing effect linearity, predicting counterfactual outcomes, estimating effect shapes, and identifying effect change-points. Our nonlinear effect analysis is structured as a three-layer framework, where each layer is based on an `S'-named method, hence referred to as the three-`S' or SSS framework. Specifically, given the data containing the instrument, exposure, and outcome, we conduct the following sequential analyses:
\begin{itemize}
    \item We first implement \underline{S}tratification approaches to divide the sample into multiple subgroups (strata with different average levels of the exposure), ensuring within-stratum exchangeability—such that the instrument remains independent of confounders within each stratum. We introduce stratification methods that are robust to collider/selection bias, ensuring that selection based on exposure does not induce selection bias.

    \item We then construct a \underline{S}calar-on-function or \underline{S}calar-on-scalar regression model, which connects stratum-specific estimates (scalar information) with the effect shape function (functional information). This model can be applied in general IV settings, including those with invalid instruments. We further introduce strategies for conducting reliable nonlinear analysis based on the scalar-on-function model in both valid and multiple invalid instrument scenarios.

    \item Finally, we incorporate multiple effect function forms for inference, including conventional functional data analysis approaches and, in particular, the change-point effect model. By transforming the scalar-on-function model into a \underline{S}um-of-single-effect (SuSiE) models, we frame the nonlinear analysis as a Bayesian variable selection problem, facilitating both effect shape estimation and change-point detection.
\end{itemize}
A graphical overview of the framework is provided in Figure \ref{SSS_overview}. We will demonstrate that our framework can yield meaningful conclusions in nonlinear effect analysis, even in the presence of challenges that are difficult to address using IV-regression and control-function-based methods. We apply our framework to investigate the effect of alcohol intake on systolic blood pressure (SBP), with a particular focus on the change-point problem—a question of significant interest in epidemiology. Our analysis can be reliably implemented using our code, available at \url{https://github.com/HDTian/SSS}.

The remainder of the article is structured as follows. Section \ref{Stratification} introduces the stratification method in IV analysis. Section \ref{Scalar-on-function regression} presents how to connect stratum-specific information to global effect shape information via scalar-on-function and scalar-on-scalar regression for multiple objectives. We discuss key properties of the scalar-on-function model, introduce the inference strategies, and extend it to complex scenarios involving invalid instruments. We specifically consider methods for estimating and identifying change-points using the sum of single effects model.
Section \ref{sec_susie} introduces SuSiE, a Bayesian nonparametric method, for inferring the change-point and conduct general effect function estimation.
The complete algorithm and workflow for conducting our nonlinear analysis are provided in the end of Section \ref{sec_susie}. We evaluate our methods through various simulations in Section \ref{simulation} and apply them to real data from the UK Biobank in Section \ref{application}. Finally, we conclude with a discussion in Section \ref{discussion}.

\section{The first `S': Stratification}\label{Stratification}
\begin{figure}[tbp]  %[tbp] or [htp]: 一般排法   [H]: 强排
	\centering
     \includegraphics[width=1.0\textwidth]{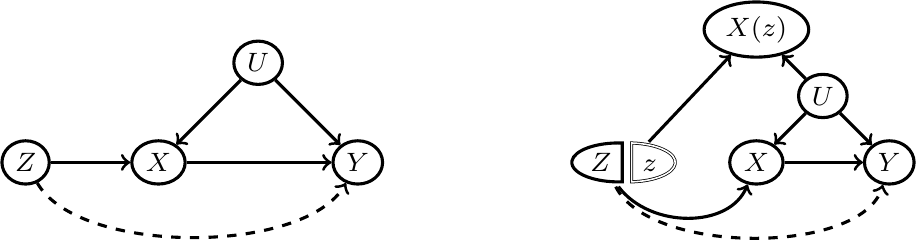} 
	\caption{\normalfont Left: Causal Directed Acyclic Graph \citep{pearl2009causality} for model setup in this article. $Z$, $X$, $Y$, $U$ represent the instrument, the exposure, the outcome, and the confounders, respectively. Right: The corresponding Single World Intervention Graph \citep{richardson2013single} for the stratification idea. $X(z)$ represents the counterfactual exposure give the instrument level $z$. $X(z)$ is not a collider of $Z$ and $U$, and $X(z) \indep Z$, indicated by the graph structure.}
	\label{f1}
\end{figure}

We propose that the model is represented by a directed acyclic graph (DAG), as shown in the left panel of Figure \ref{f1}, with the instrument $Z$, the continuous exposure $X$, the continuous outcome $Y$, and the confounder $U$. This DAG assumes exchangeability, $ Z \indep U $ \citep{didelez2007mendelian}, where we adopt the independence notation '$\indep$' proposed by \cite{dawid1979conditional}. However, we allow for the possible violation of the exclusion restriction assumption, meaning that the instrument may have a direct effect on the outcome (denoted by the dashed arrow). The model represented by the DAG is plausible in most experiments and many quasi-experiments.  
In a general case, we assume the structural outcome model:  
\begin{equation}\label{simple_model}
    %Y   = h(X,Z,\epsilon) + g( U,Z, \epsilon ) 
    Y = h(X) + U 
\end{equation}
where $\epsilon$ represents an exogenous error term, and $g(\cdot)$ is an arbitrary unknown function. The functional objective of interest, $h(\cdot)$, is referred to as the effect function. We call the partial derivative of $h(X)$ with respect to $X$ evaluated at $X=x$, $h'(x)$, the effect intensity function. We aim to test, estimate, or infer either $h(x)$ or $h'(x)$. In this section, we first focus on stratification, which lays the foundation for downstream analyses.

\subsection{Definitions and assumptions}
{\bf Definition (Stratification)} {\it In the context of nonlinear effect analysis for the exposure $X$ using the instrumental variable $Z$ that is independent of the confounder $U$, we define the stratification as the process of dividing the population into multiple subgroups, each of which is called a stratum, with two key features, in an asymptotical sense, 
\begin{itemize}[noitemsep]
    \item[1.] The instrument $Z$ is independent of the confounder $U$ within each stratum
    \item[2.] The exposure $X$ is distributed differently across strata.
\end{itemize}}

{Feature 1 is the stratum-specific exchangeability assumption and ensures that we can conduct an IV analysis for the local data in the stratum as valid as the IV analysis for the population. Feature 2 allows each stratum to provide uniquely useful local information. Trivially, one can completely randomly select arbitrary subgroups from the population, where the IV is still independent of the confounder, but 
such stratification would not yield distinct local information of the exposure distribution and hence be of no use to non-linear studies.}

A straightforward way to achieve stratification is by selecting a measurable covariate that is associated with the exposure and is not a common effect of the instrument and confounders, or by selecting the instrument directly. In both cases, Features 1 and 2 should be satisfied. However, we require the distribution of exposure in each stratum, determined by stratifying the measurable covariate or instrument, to be as distinct as possible in order to increase the power of our downstream nonlinear effect analysis. This becomes particularly challenging when the association between the exposure and the measurable covariate or instrument is weak. Additionally, it may be difficult to measure covariates that satisfy the stratification features. In this article, we will focus on the more challenging stratification scenario where only the data information of the instrument, exposure, and outcome are available, and the instrument strength is relatively weak. When appropriate covariates or instruments are available for stratification, our proposed methods can be directly applied, but in this article we will not expect this.

% Remark: stratification of cousrce can be achieved by selecting on other measurable variable; for example, the covariate or/and the instrument. For covariate, like age or gender, it should not be the collider of the IV and confounder; for IV, this idea is similar as the IV-regression. In either case, we should expect the distribution of exposure in each stratum led by the covariate or IV is as different as possible; however, it is difficult when the effect of voariate or IV on the exposure is weak; in such scenarios, selecting on the counterfactual exposure provide the solution to maximize the difference across stratum-specific exposure distribution.

\subsection{Methods for stratification}
Now performing stratification while satisfying Features 1 and 2 is not straightforward. The primary challenge arises from the fact that the exposure is a common effect of both the instrument and the confounder, as illustrated in Figure \ref{f1}, where the exposure acts as a collider. 
A naive approach would be to stratify directly based on the exposure; however, this induces \textit{collider bias} \citep{hernan2004structural,cole2010}. Specifically, after stratification on the exposure, the instrument may become correlated with the confounder, even if they were independent in the population before stratification. This collider bias leads to nonzero covariance between the instrument and the confounder, which may also vary across strata. Such bias can severely distort nonlinear effect analysis, making it unreliable.

To address this key challenge, we propose a stratification approach based on the counterfactual exposure, denoted as $X(z)$, where $z$ is a reference instrument level. As illustrated by the Single World Intervention Graph (SWIG) in the right panel of Figure \ref{f1}, SWIG represents an augmented DAG that simultaneously considers both observed and counterfactual variables. Notably, $X(z)$ is independent of $Z$ and does not serve as a collider between $Z$ and $U$. Consequently, stratifying based on $X(z)$ should not introduce collider bias, while still preserving the conditional exchangeability property:  
\begin{equation}
    Z \indep U \mid X(z) \in \mathcal{S} \qquad \text{for any } z,
\end{equation}
where $\mathcal{S}$ represents a set defined based on the distribution of $X(z)$ for selection. 
Furthermore, since $X(z)$ is associated with the observed exposure $X$, stratification based on $X(z)$ ensures that different strata exhibit distinct distributions of $X$, thereby satisfying Feature 2. Thus, the core idea of this stratification approach is to achieve selection based on the counterfactual exposure $X(z)$. In the following, we introduce two types of stratification methods.

\medskip 

\noindent
{\bf Prediction Method.} The prediction method for selecting $X(z)$ aims to estimate the counterfactual value of $X(z)$ for each individual based on observed data. This approach typically relies on assumptions about the structural exposure model:  
\begin{equation}\label{prediction_model}
  X \equiv X(Z) = t\big( f(Z) + g_X( U, \epsilon_X ) \big),
\end{equation}
where $t[\cdot]$ is a known monotonic bijective transformation function (often simply $t[x] \equiv x$), and $g_X(\cdot)$ is an arbitrary function that depends on the confounder $U$ and the exogenous error term $\epsilon_X$. Since $Z \indep g_X(U, \epsilon_X)$, the instrument-exposure curve $f(\cdot)$ is identifiable under certain parametric smoothing conditions. 
Thus, the predicted counterfactual exposure for an individual is given by  
\begin{equation}
    \hat{X}(z) := t\big( \hat{f}(z) + t^{-1}(X) - \hat{f}(Z) \big)
    \overset{p}{\to} X(z) \quad \text{as } \hat{f}(\cdot) \text{ consistently estimates } f(\cdot).
\end{equation}
A simple and widely used example assumes a linear exposure model, $X = \alpha Z + g_X(U, \epsilon_X)$. In this case, the residual from regressing $X$ on $Z$ serves as a counterfactual exposure prediction. This approach, known as \textit{residual stratification} in MR \citep{burgess2014instrumental}, has been the predominant method for nonlinear MR studies. The corresponding algorithm is provided in Appendix A of the Supplementary Materials.

\medskip

\noindent
{\bf Matching Method.} The matching method leverages the fact that we only need to achieve selection on the counterfactual exposure $X(z)$ without necessarily knowing its exact value or using it for downstream nonlinear effect analysis. Instead, we seek a measurable surrogate variable that can be matched with $X(z)$, allowing us to indirectly stratify on $X(z)$ by selecting based on this surrogate.
One strategy relies on the \textit{rank-preserving assumption} (RPA), a concept well-known in instrumental quantile regression \citep{chernozhukov2006instrumental}. In our context, the assumption states that an individual's counterfactual exposure maintains the same rank ordering for different instrument values. Formally, for each individual and any instrument levels $z$,  
\begin{equation}
   %  X(z_1) \overset{RPA}{\Longleftrightarrow} X(z_2)
   F_{X(z)}(  x_i(z)   ) \,\, \text{is constant for all } z \in Z
\end{equation}
where $F_{X(z)}(\cdot)$ is the cumulative distribution function for $X(z)$ and $ x_i(z)$ is the $i$-th individual's counterfactual exposure level. The RPA implies the stratum-specific RPA within a fixed instrument level $Z=z$, that is,
\begin{equation}
    F_{X(z)}(  x_i(z)   )=   F_{X}(  x_i   ) \mid Z = z, \quad \forall z.
\end{equation}
for the individuals with $Z=z$.
Thus, we can use the rank information of the observed exposure $X$ within each instrument level $z$ to approximate the rank of $X(z)$, ensuring a stratification process that is robust to selection bias.
A practical implementation of this idea is as follows:
\begin{itemize}
    \item First, rank the instrument values and divide them into multiple equally-sized pre-strata, each containing similar instrument levels. Within each pre-stratum, the rank-preserving assumption approximately holds.
    \item Then, rank the exposure values within each pre-stratum. The final strata are constructed such that each stratum contains individuals with the same rank across all pre-strata. This ensures the strata have different average levels of the exposure.
\end{itemize}
This approach, known as \textit{doubly-ranked stratification} \citep{tian2023relaxing}, uses a simple two-step ranking procedure to achieve stratification while satisfying the required features. Since the instrument is independent of $U$ and $X(z)$, each pre-stratum (constructed based on $Z$ alone) has the same sampling distribution for $U$ and $X(z)$. Consequently, within each final stratum, $U$ remains i.i.d. and independent of $Z$ in an asymptotical sense. The full algorithm is provided in Appendix B of the Supplementary Materials.
The matching method, which relies on the rank-preserving assumption, imposes weaker conditions than the prediction approach, which is based on a structural exposure model. Specifically, all models conforming to Equation \eqref{prediction_model} also satisfy the rank-preserving assumption. Due to this, doubly-ranked stratification is generally recommended over residual stratification in most MR applications \citep{burgess2023violation}. If the exposure model assumption can be strongly justified, residual stratification may remain preferable due to its more concise algorithm (only rank once).

We list some potential use cases and discuss the plausibility of the rank-preserving assumption in some real examples (see Appendix B of the Supplementary Materials). We also applied our analysis framework to the real data of \cite{angrist1991does}. They are given in Appendix C of the Supplementary Materials. Note that the rank-preserving assumption is a counterfactual-variable-based assumption, like many others in causal inference, and is therefore essentially unverifiable. Investigators are encouraged to leverage domain knowledge or employ triangulation strategies, such as the negative control design, to assess its plausibility \citep{hamilton2024non}. We will present these steps in the real application in Section \ref{application}.

% IV example

\section{The second `S': Scalar-on-function and scalar-on-scalar model}\label{Scalar-on-function regression}
In this section, we consider two models for linking the stratum-specific estimates obtained from the stratification approach introduced in Section \ref{Stratification} to the effect function of interest: the scalar-on-function regression model (Section \ref{SoF_sec}) and the scalar-on-scalar regression model (Section \ref{SoS_sec}).

A fundamental aspect of our approach is its focus on the functional parameter $h'(x)$ (effect intensity) rather than $h(x)$ (effect shape). However, $h'(x)$ is closely related to $h(x)$ and is sufficient for nearly all objectives in nonlinear studies. We first introduce the key objectives, and then the scalar-on-function regression model, along with the inference strategies.

% \subsection{Objectives in nonlinear effect analysis}\label{objectives}

\medskip 
\noindent
{\bf Objective 1 (linearity test).} {\it The primary objective in many studies considering nonlinear effect as additional rather main analysis is to test whether the causal effect is linear. In some cases, this is the only goal, particularly in studies that rely on linear models rather than detailed nonlinear effect analysis. When the effect is truly linear, simple linear models achieve higher statistical power than flexible nonlinear models with many degrees of freedom. This is equivalent to testing whether the effect intensity $h'(x)$ is constant.}

\medskip 
\noindent
{\bf Objective 2 (counterfactual prediction).} {\it Another key objective is predicting the counterfactual outcome $Y(x)$ or its expectation $\mathbb{E}(Y(x))$. Given an estimate of the effect density $\hat{h}'(x)$, the counterfactual outcome at any exposure level $x$ can be estimated as:
\begin{equation}
   \hat{\mathbb{E}}( Y(x) ) = n^{-1} \sum_{i=1}^n \hat{Y}_i( x ) =  n^{-1} \sum_{i=1}^n \left[ Y_i  +  \hat{\tau}(X_i,x) \right]  = n^{-1} \sum_{i=1}^n\left[ Y_i  +  \int_{X_i}^{x} \hat{h}'(s) \text{d}s \right],
\end{equation}
where $ \tau(x_1,x_2) $ represents the treatment effect of changing the exposure from $x_1$ to $x_2$.}

\medskip 
\noindent 
{\bf Objective 3 (effect shape estimation).} {\it If the goal is to estimate the effect shape, say, $h(x):= \mathbb{E}(  Y(x) ) - \mathbb{E}(  Y(0) )$, it can be obtained via counterfactual contrast or direct integration:
\begin{equation}
    \hat{h}(x) :=  \hat{\mathbb{E}}(  Y(x) ) - \hat{\mathbb{E}}(  Y(0) ) =  
    n^{-1} \sum_{i=1}^n\left[  \int_{X_i}^{x} \hat{h}'(s) \text{d}s  -  \int_{X_i}^{0} \hat{h}'(s) \text{d}s \right] = \int_0^x \hat{h}'(s) \text{d}s. 
\end{equation}}

\medskip 
\noindent
{\bf Objective 4 (change-point inference).} {\it In some cases, it is crucial to identify or infer specific aspects of the effect shape, such as change points where the causal effect differs below or above a certain threshold. Specifically, we assume $h(x) = \sum_p \beta_p (x -  t_p)_+ $ or equivalently $h'(x)=\sum_p b_p I\{x \geq t_p\}$ with multiple change-points $\{t_p\}$. We are interested in the true information regarding $\{t_p\}$, including its existence, number, and precise values.}

\medskip
\noindent
{For Objective 1, we can test whether $h'(x)$ is constant by examining the differences across the stratum-specific IV estimates.  
For Objectives 2, 3, and 4, additional assumptions on the functional form of $h'(x)$ are usually required.}

\medskip

\subsection{Scalar-on-function regression}\label{SoF_sec}
We now illustrate how to leverage stratification for nonlinear effect analysis using instrumental variables via scalar-on-function regression. 
Starting with the simplified model \( Y = h(X) + U \), where we are interested in \( h(\cdot) \) or \( h'(\cdot) \), and the valid instrument \( Z \) such that \( \text{Cov}(Z, X) \neq 0 \) and \( Z \indep U \), denote the instrumental association with the exposure and the outcome by $\hat{\alpha} $ and $ \hat{\theta} $;
the Wald ratio estimator \( \hat{\beta} \) satisfies, under regular conditions, 
\begin{equation}\label{e1}
    \hat{\beta} :=  \frac{\hat{\theta} }{ \hat{\alpha}  } = \frac{ \widehat{\text{Cov}}(Z, Y) }{ \widehat{\text{Cov}}(Z, X) } \overset{p}{\to} \int_{\mathcal{X}} h'(x) W(x) \, \text{d}x \qquad \text{with } W(x) := \frac{ \text{Cov}(Z, I\{X \geq x\}) }{ \text{Cov}(Z, X) }
\end{equation}
where \( I\{\cdot\} \) is the indicator function, and \( W(x) \) is the weight function, which satisfies the following properties (without loss of generality, assuming \( \text{Cov}(Z, X) > 0 \)):
\begin{itemize}
    \item[(1)] \( \int_{\mathcal{X}} W(x) \, \text{d}x = 1 \) (i.e., weights are standardized),
    \item[(2)] \( W(x) \geqslant 0 \) under linearity (or weak monotonicity) and homogeneity assumptions of the instrument-exposure model,
    \item[(3)] \( W(x) \equiv f_X(x) \), the density of \( X \), when \( \{Z, X\} \) are jointly normally distributed.
\end{itemize}

\noindent 
\textbf{Proof.} See Appendix D of the Supplementary Materials.

\noindent 
\textbf{Remark 1.} {\it Property (1) states that the scalar IV estimate $\beta$ is the standard weighted sum of \( h'(x) \), corresponding to the effect density defined in Equation \eqref{effect_density}. Specifically, if \( h'(x) \) is constant and equal to \( \beta \), which corresponds to the average causal effect (ACE) of the conventionally linear exposure-outcome model, the IV estimator is consistent with the ACE \( \beta \).}

\noindent 
\textbf{Remark 2.} {\it Property (2) indicates that when the effect shape \( h(x) \) is monotonic (i.e., \( h'(x) \leq 0 \) or \( h'(x) \geq 0 \) for all \( x \)), and one is interested in the direction of the effect (whether the treatment is beneficial or harmful), the IV estimator \( \hat{\beta} \) can be directly used for inference on the effect direction. Note that the asymptotic sign of \( \hat{\beta} \) may not be consistent with the effect direction if \( W(x) \geq 0 \) does not hold.}

\noindent
\textbf{Remark 3.} {\it Properties (1) and (2) imply that the IV estimator is constrained by the minimal and maximal values of the effect density $h'(x)$ over the exposure domain in partial identification \citep{tamer2010partial}.}

\noindent 
\textbf{Remark 4.} {\it Property (3) supports completely parametric modeling for the weight function based on the exposure distribution. It also implies some useful properties: \( W(\mathbb{E}(X)) \geqslant W(x) \) for any \( x \), and \( W(x) \) is symmetric around \( \mathbb{E}(X) \). This means the weight function assigns the largest weight at \( \mathbb{E}(X) \), and when \( h'(x) \) is linear (i.e., the effect function \( h(x) \) is second-order), the limiting value of the IV estimator $\hat{\beta}$ is \( h'(\mathbb{E}(X)) \). }

\noindent
\textbf{Remark 5.}{\it Property (3) will motivate us to consider simpler models than the scalar-on-function model, such as the scalar-on-scalar model discussed in Section \ref{SoS_sec}. Additionally, this property allows us to use only the mean and variance of the exposure to sufficiently capture the functional information, which provides computational savings in large-scale data analysis (e.g. in modern genetics).}

\medskip

Now, with stratification such that for each stratum \( S \), \( \text{Cov}(Z, X | S) \neq 0 \) and \( Z \indep U | S \) hold and the stratum-specific instrumental associations, with the exposure and outcome $ \{ \hat{\alpha}_k, \hat{\theta}_k \} $, we have a regression based on Equation \eqref{e1}:
\begin{equation}\label{SoF}
    \hat{\beta}_k :=   \frac{ \hat{\theta}_k }{  \hat{\alpha}_k }  = \frac{ \widehat{\text{Cov}}(Z, Y | S = k) }{ \widehat{\text{Cov}}(Z, X | S = k) } = \int_{\mathcal{X}} h'(x) W_k(x) \, \text{d}x + \epsilon_k  \qquad \text{with } \epsilon_k \sim \mathcal{N}( 0, \text{s.e.}(\hat{\beta}_k)^2) 
\end{equation}
for \( k = 1, \ldots, K \), where \( K \) represents the number of strata after stratification; \( \hat{\beta}_k \) represents the \( k \)-th stratum-specific IV estimate, and \( \text{s.e.}(\hat{\beta}_k) \) is the corresponding standard error. The regression \eqref{SoF} involves scalar responses (the naive IV estimate $\hat{\beta}$), a functional objective (the weight function $W(x)$), and the functional parameter of interest (the effect derivative function $h'(x)$). This type of regression is known as scalar-on-function (SoF) regression, a key element of functional data analysis \citep{ramsay2005functional}. Since the stratum-specific IV estimate is the average causal effect conditioning on a subgroup defined by counterfactual exposure, the idea of stratification in our article can also be understood as similar to principal stratification \citep{frangakis2002principal}, except that we stratify on the counterfactual value of the continuous exposure $X(z)$ (which, by the rank-preserving assumption, is the same stratification for any value of $Z=z$) rather than on the set of counterfactuals $[X(0),X(1)]$ as in principal stratification. In this article, our interest is not in the subgroups themselves, but rather in the global effect function \( h(x) \) or \( h'(x) \). Heterogeneity in the effect function could reflect non-linearity or effect modification; we return to this point in the Discussion.

The regression in Equation \eqref{SoF} is highly similar to the regression used in inverse-variance weighting (IVW) estimation \citep{hartung2011statistical}, which is commonly applied in meta-analysis and MR. In most applications, one can simply let 
\begin{equation}
  \text{s.e.}(\hat{\beta}_k) = \frac{\text{s.e.}(\hat{\theta}_k)}{|\hat{\alpha}_k|}   
\end{equation}
or
\begin{equation}
  \text{s.e.}(\hat{\beta}_k) = \sqrt{ \frac{\text{s.e.}(\hat{\theta}_k)^2}{\hat{\alpha}_k^2} + \left( \frac{\hat{\theta}_k}{\hat{\alpha}_k} \right)^2 \frac{\text{s.e.}(\hat{\alpha}_k)^2}{\hat{\alpha}_k^2} }   
\end{equation}
where they correspond to the well-known first-order and second-order regression errors in IVW \citep{bowden2019improving}, based on the approximation precision in the delta method for the ratio term \( \hat{\beta}_k \). 

Note that the normality approximation of the error term \( \epsilon_k \) in this regression assumes the instrument is sufficiently strong. Specifically, the strength of the instrumental association with the exposure should not be close to zero relative to its estimator's variability or uncertainty \citep{nelson1988some, nelson1988distribution, andrews2019weak}. In practice, weak instruments can be mitigated either by using large sample sizes or by ensuring strong instrumental associations via experiment design. In most applications, the normality approximation of \( \epsilon_k \) is sufficiently reliable when applying either first-order or second-order methods. 

%Various well-developed methods exist for inferring $h'(x)$ within the SoF framework. We first introduce the parametric methods and then introduce the nonparametric way for inference.

%\subsection{instrument strength in the SoF framework}
%We will illustrate why the SoF by stratification is appealing to nonlinear effect analysis.

%\subsection{Inference}\label{Effect_shape_inference}

\subsection*{Parametric inference} \label{Parametric_fitting}
In this section, we focus on the basic model, $ Y = h(X) + U $. Inference methods for more complex model scenarios are provided in Appendix F of the Supplementary Materials. We use our main text to focus on the estimation task for the effect intensity $h'(x)$ or/and effect function $h(x)$. The linearity test is instead introduced in Appendix E of the Supplementary Materials.

In the parametric scenario where the effect shape is assumed to be a linear additive combination of given basis functions, with weights to be estimated, the scalar-on-function regression can be transformed into a traditional regression without functional variables. Specifically, suppose that 
\begin{equation}
    h'(x) = \sum_{l=1}^L b_l \phi_l(x),
\end{equation}
where $ \{ \phi_l(x) \}_{l=1,\ldots,L} $ are known basis functions (e.g., B-spline basis). Let the functional inner product of any two functions $f(\cdot)$ and $g(\cdot)$ over $\mathcal{X}$ be denoted as $ \langle f , g \rangle $. From the initial SoF regression \eqref{SoF}, we obtain the conventional regression formulation for the parameters of interest $\{ b_l \}$, which determine the effect shape:
\begin{equation} \label{SoF_para}
    \hat{\beta}_k = \int_{\mathcal{X}} h'(x) W_k(x) \text{d} x  +  \epsilon_k = \sum_{l=1}^L  b_l\,  \langle  \phi_l , W_k  \rangle +\epsilon_k, \quad \text{with } \epsilon_k \sim \mathcal{N}( 0, s.e.(   \hat{\beta}_k )^2 ).
\end{equation}

When employing flexible basis functions, such as B-splines, it may be beneficial to introduce additional regularization or a roughness penalty to ensure smoothness of the estimated effect shape. A common choice is the 2-norm penalty 
\begin{equation}
    \int (\mathcal{D}^m h'(x))^2dx,
\end{equation}
where $\mathcal{D}^m$ denotes the $m$-th derivative operator. This leads to a regularized scalar-on-scalar regression problem, with the parameters $\{ \hat{b}_l \}$ estimated via a weighted least squares approach:
\begin{equation} \label{fitting_estimation}
\begin{split}
      \hat{\boldsymbol{b}} 
      &= \argmin_{ \boldsymbol{b} }  \left[   \sum_{k=1}^K \frac{    (    \hat{\beta}_k -  \sum_{l=1}^L  b_l\,  \langle  \phi_l , W_k  \rangle      )^2   }{  s.e.( \hat{\beta}_k )^2  }   +  \lambda \int \left(\sum_{l=1}^L    b_l \,  \mathcal{D}^m \phi _l(x)  \right)^2dx    \right] \\
      &= \argmin_{ \boldsymbol{b} }  \left[   \sum_{k=1}^K \frac{    (    \hat{\beta}_k -  \sum_{l=1}^L  b_l\,  \langle  \phi_l , W_k  \rangle      )^2   }{  s.e.( \hat{\beta}_k )^2  }   +  \lambda \, \boldsymbol{b}^T   
 \boldsymbol{R} \boldsymbol{b}   \right] \\
      &=  \argmin_{ \boldsymbol{b} }  \left[   (    \hat{\boldsymbol{\beta}}  -   \boldsymbol{X}  \boldsymbol{b}   )^T \boldsymbol{\Sigma}^{-1}  (    \hat{\boldsymbol{\beta}}  -   \boldsymbol{X}  \boldsymbol{b}   )  +  \lambda \, \boldsymbol{b}^T   
 \boldsymbol{R} \boldsymbol{b}   \right].
\end{split}
\end{equation}
where
\begin{itemize}[noitemsep, nosep]
    \item $ \hat{\boldsymbol{\beta}} $ is the $K$-length vector stacking the stratum-specific Wald ratio estimates

    \item $\boldsymbol{X}$ is a $K \times L$ design matrix with entries $\boldsymbol{X}_{k,l} = \langle \phi_l , W_k \rangle$

    \item $\boldsymbol{\Sigma}$ is the variance-covariance matrix of $\hat{\boldsymbol{\beta}}$

    \item $\boldsymbol{R}$ is an $L \times L$ penalty matrix with entries $\boldsymbol{R}_{i,j} = \langle \mathcal{D}^m \phi_i , \mathcal{D}^m \phi_j \rangle$

    \item $\lambda$ is a tuning parameter controlling the degree of regularization
\end{itemize}  

This formulation allows us to estimate the effect shape efficiently while incorporating smoothness constraints through regularization.
The closed-form solution for the estimation of $\boldsymbol{b}$ is straightforward:
\begin{equation}\label{sof_fitting}
    \hat{\boldsymbol{b}} = \left( \boldsymbol{X}^T \boldsymbol{\Sigma}^{-1} \boldsymbol{X}  +  \lambda \boldsymbol{R} \right)^{-1}  \boldsymbol{X}^T  \boldsymbol{\Sigma}^{-1}  \hat{\boldsymbol{\beta}}.
\end{equation}

When the basis functions are determined a priori (e.g., polynomials of a certain order), regularization is unnecessary, so $\lambda = 0$. However, if smoothing is required, the value of $\lambda$ can be selected via cross-validation or, more efficiently, by minimizing the generalized cross-validation (GCV) criterion \citep{craven1978smoothing}:
\begin{equation} \label{GCV_ds}
    GCV(\lambda) = \left(\frac{K}{K - \text{tr}(  \boldsymbol{ H}_ {\lambda }   ) } \right)
    \left(\frac{SSE _{\lambda}  }{  K- \text{tr}(  \boldsymbol{ H}_ {\lambda }    )  }  \right),
\end{equation}
where
\begin{itemize}[noitemsep, nosep]
    \item The hat matrix is given by
    \begin{equation}
        \boldsymbol{ H}_ {\lambda } = \boldsymbol{\Sigma}^{-1/2}\boldsymbol{X} \left( \boldsymbol{X}^T \boldsymbol{\Sigma}^{-1}\boldsymbol{X} + \lambda \boldsymbol{R} \right)^{-1} \boldsymbol{X}^T \boldsymbol{\Sigma}^{-1/2}.
    \end{equation}
    \item The residual sum of squares is 
    \begin{equation}
        SSE_\lambda = \left\|\boldsymbol{\Sigma}^{-1/2} \hat{ \boldsymbol{\beta} }    
        -  \boldsymbol{ H}_ {\lambda }\boldsymbol{\Sigma}^{-1/2} \hat{ \boldsymbol{\beta} } \right\|^2.
    \end{equation}
    \item The notation $\text{tr}(\boldsymbol{ H}_ {\lambda }  )$ represents the trace of the hat matrix.
\end{itemize}

Once $\lambda$ is determined, inference on $\boldsymbol{b}$ and the effect shape can be performed directly. The estimated effect function and its derivative at any exposure level $x$ are given by
\begin{align}
    \hat{h}'(x) &= \boldsymbol{\phi}^T(x) \hat{\boldsymbol{b}} \\
    \hat{h}(x) &= \int_0^x \hat{h}'(s) \text{d}s = \left[\int_0^x  \boldsymbol{\phi}(s) \text{d}s \right]^T \hat{\boldsymbol{b}}.
\end{align}
Under the frequentist inference framework, the distribution of $\hat{\boldsymbol{\beta}}$ 
allows for the uncertainty quantification based on formula \eqref{sof_fitting}.

\subsection{Scalar-on-scalar regression}\label{SoS_sec}
The scalar-on-function regression model contains the functional covariates, which require more storage and computational complexity. Sometimes, we may build scalar-on-scalar regression model that can still be useful in inferring effect function with scalar information. The idea is based on the property of the weight function in Equation \eqref{e1}; namely, the approximation equation with respect to $ h'(x) $ or the basis functions $\{b_l(x)\}$ for each stratum:
\begin{equation} \label{approx}
    \int_{\mathcal{X}}  h'(x) W_k(x)  \text{d}x \approx h'(   \mathbb{E}_k( X    ) )   \approx h'(   \bar{ X}_k    ) 
\end{equation}
where $ \bar{ X}_k$ represents the average value of the exposure in the $k$-th stratum.
This intuition of this equation is that the weighted effect intensity may be approximated by the effect intensity value at the mean exposure for that stratum. For certain scenarios like those in Remark 4, the approximation equation strictly holds. 

The approximation equation greatly simplified our model, so all the formula in the parametric fitting procedure in the previous subsection can be degenetrated into the conventional scalar variables; say, with the basis function assumption $  h'(x)=\sum_{l=1}^L b_l \phi_l(x)  $, the regression \eqref{SoF_para} becomes
\begin{equation} \label{SoF_est}
    \hat{\beta}_k = \sum_{l=1}^L b_l \int_{\mathcal{X}} \phi_l(x) W_k(x) \text{d} x \approx \sum_{l=1}^L b_l \phi_l(  \bar{x}_k )  + \epsilon_s   \qquad \text{with } \epsilon_k \sim \mathcal{N}( 0, s.e.( \hat{\beta}_k )^2 )
\end{equation}
and the estimated effect intensity is easily
$ \hat{h}'(x) = \boldsymbol{\phi}^T(x) \hat{\boldsymbol{b}}  $ with $\hat{\boldsymbol{b}}$ similar to Equation \eqref{SoF_est}
\begin{equation}\label{sos_fitting}
    \hat{\boldsymbol{b}} = \left( \boldsymbol{A}^T \boldsymbol{\Sigma}^{-1} \boldsymbol{A}  +  \lambda \boldsymbol{R} \right)^{-1}  \boldsymbol{A}^T  \boldsymbol{\Sigma}^{-1}  \hat{\boldsymbol{\beta}}.
\end{equation}
with the approximated design matrix $  \boldsymbol{A} \in \mathbb{R}^{ K \times L } $ such that $ \boldsymbol{A}_{k,l} = \phi_l( \bar{x}_k )   $ and the covariance matrix of $ \hat{\boldsymbol{\beta}} $ and the penalty matrix if applicable. Further inference is straightforward.

The convenient fitting procedure based on the approximated scalar-on-scalar regression has been widely implemented in the nonlinear MR practice \citep{staley2017semiparametric}. However, it may suffer from structural bias due to the poor approximation in Equation \eqref{approx}, which usually happens when the shape of $h'(x) $ or $ b_l(x) $ is irregular; for example, $ b_l(x) \propto I\{ x \geq x^\ast \} $ for the threshold value $ x^\ast $ inside the exposure domain. Although our simulation evidence shows that scalar-on-scalar regressions can accurately estimate causal effect curves in simple models, we will show that the basis function with the form $ I\{ x \geq x^\ast \} $ is important in nonparametric fitting and hence the scalar-on-scalar approximation may not always be appropriate. In other general scenarios, both scalar-on-function and scalar-on-scalar regression should work well.

\subsection{Nonparametric change-point model}\label{point_inference}
In practice, particularly for researchers without domain knowledge about the appropriate basis functions for the effect shape, model selection or comparison across different basis functions may be necessary. B-splines are widely used in functional data analysis and are jointly determined by the placement of interior knots (both their number and positions) and the order of the spline. 

Even if we fix the order to 1 (i.e., piecewise constant basis functions), different choices of interior knot positions result in a vast number of possible basis function scenarios. Suppose there are $P$ candidate positions where knots can be placed (typically, $P \gg K$, where $K$ is the number of strata). If the number of knots is unrestricted, the total number of possible basis function configurations is
\begin{equation}
    \sum_{j=0}^{K-1} \binom{P}{j} = \left( \begin{array}{c} P \\ 0 \end{array} \right)  +   \left( \begin{array}{c} P \\ 1 \end{array} \right)  + \cdots +   \left( \begin{array}{c} P \\ K-1 \end{array} \right).
\end{equation}
Exhaustively comparing all possible models would be computationally intensive. While more efficient approaches, such as stepwise model selection or LASSO (e.g., using the solution path of least-angle regression), can reduce the search space, these methods carry the risk of missing the globally optimal model. 

Moreover, basis-function-based methods do not allow direct inference on the knot locations themselves, which can be particularly relevant in epidemiologic research. For instance, while alcohol intake is widely recognized as a risk factor for many health outcomes, some studies suggest the existence of ``safe region'' where alcohol consumption below a certain threshold does not significantly affect health. This implies that the causal effect of alcohol intake on certain outcomes may exhibit nonlinearity near the threshold. In this context, the threshold level, which corresponds to the knot position in our model, becomes a key quantity of interest.

Given the considerations above, we reformulate the inference problem as a nonparametric way with change-point model, which can be expressed using sparse regression. Assuming the effect function $h(x)$ is continuous, we approximate its derivative $h'(x)$ as piecewise constant over minor grid segments defined by $P$ interior knots $\{ t_1, \dots, t_P \} \subset \mathcal{X}$. Specifically, we express $h'(x)$ as:
\begin{equation} \label{change_point_general}
    h'(x) = \sum_{p=0}^P  b_p I\{  x \geq t_p \}, 
\end{equation}
where $I\{\cdot\}$ is the indicator function. The knots $t_p$ are typically chosen as the $ p/(P+1) $ quantiles of $X$. 
Incorporating this form into the SoF model leads to the following change-point model:
\begin{equation}\label{change_point_model}
     \hat{\beta}_k =   \int_{\mathcal{X}} h'(x) W_k(x) \text{d} x  +  \epsilon_k = b_0 + \sum_{p=1}^P  b_p \int_{t_p}^{\infty}  W_k(x) \text{d} x + \epsilon_s, \quad \epsilon_k \sim \mathcal{N}( 0, s.e.(   \hat{\beta}_k )^2 ),  
\end{equation}
for $k = 1, \dots, K$.
The change-point model \eqref{change_point_model} now contains $P+1$ parameters, which is typically larger than the number of strata $K$. However, many of these parameters (or potentially all except $b_0$ if the effect shape is globally linear) are zero, reducing the problem to a classical sparse regression framework.

Notably, our objective is not only to fit a sparse regression model for counterfactual prediction or effect shape estimation, but also to conduct inference, such as obtaining interval estimates for the potential change points. These goals naturally align with Bayesian variable selection methods.
In the remainder of the article, we will focus on the Sum of Single Effects (SuSiE) approach \citep{wang2020simple}, a widely used Bayesian method for change-point detection. SuSiE is particularly known for its applications in fine-mapping in genetics, where the goal is to identify causal genetic variants among a high-dimensional set of candidates. More generally, SuSiE provides estimates for the number of nonzero parameters and constructs credible sets for their locations, both of which align well with our objectives in the change-point model \eqref{change_point_model}. In addition, SuSiE gives the Bayesian nonparametric solution for estimating the effect function with quite low computational burden.

\section{The third `S': Sum of single effect analysis for change-point model}\label{sec_susie}
Recall that we have transformed the scalar-on-function regression, which is based on the stratum-specific scalar variables (IV point estimates) and the functional variables (weight function), into a change-point model. This model contains high-dimensional covariates with parameters corresponding to the underlying effect shape. We aim to adopt a Bayesian nonparametric approach, SuSiE, to fit this sparse model for investigating the change points (both detecting their existence and estimating their precise values) and for estimating the effect shape.

\subsection{Single-effect model}
The SuSiE model can be viewed as a natural extension of the single-effect model, which assumes that only one parameter has a potentially nonzero effect in each regression. Under the change-point model \eqref{change_point_model}, we now express it in matrix notation as:
\begin{equation}\label{change_point_model_matrix_notation}
    \hat{\boldsymbol{\beta}} = \boldsymbol{X} \boldsymbol{b} + \boldsymbol{\epsilon},   \quad \text{where} \quad \boldsymbol{\epsilon} \sim \mathcal{N}( \boldsymbol{0}, \boldsymbol{\Sigma}),
\end{equation}
with the effect parameters defined as $\boldsymbol{b} := (  b_0 , b_1, \ldots, b_P )^T$. 
Given the single-effect assumption, we specify the prior distribution for $\boldsymbol{b}$ as:
\begin{equation}  \label{prior1}
    \boldsymbol{b} =  \boldsymbol{\gamma} \, b,
\end{equation}
\begin{equation}\label{prior2}
    b \sim \mathcal{N}(  0, \sigma_0^2 ),
\end{equation}
\begin{equation}\label{prior3}
    \boldsymbol{\gamma} \in \{0 ,1\}^{P+1}   \sim   \text{Multinomial}( 1, \boldsymbol{\pi} ).
\end{equation}
Here, the vector $\boldsymbol{\gamma}$ represents the selection of a single nonzero effect among the $P+1$ parameters, and $\boldsymbol{\pi} =  (  \pi_0, \pi_1,\ldots, \pi_P )$ denotes the prior inclusion probabilities, where each $\pi_p$ represents the probability that only the corresponding parameter has a nonzero effect. We usually make the uniform prior setting that $\pi_0 = \cdots = \pi_P = (P+1)^{-1}$.

Denote the observed data by $  \boldsymbol{O} := \{  \{ \hat{\beta}_s , s.e.( \hat{\beta} ) \}_s   , \,  \{  
  \int_{t_p}^\infty  W_s(d) \text{d}x  \}_p     \} $.
Given the prior formulations, we can directly express the posterior distribution:
\begin{equation}
\boldsymbol{\gamma}  |  \boldsymbol{O}, \sigma^2_0   \sim \text{Multinomial}( 1, \boldsymbol{\pi}^\ast ),
\end{equation}
where $\boldsymbol{\pi}^\ast$ represents the posterior inclusion probabilities of each parameter. These posterior probabilities can be used to construct a credible set for change points in the single-effect model, which consists of parameter positions (i.e., change-point positions) with cumulative posterior probabilities equal to or exceeding a nominal level.
Furthermore, the posterior distribution of the $p$-th parameter, conditional on its inclusion, is given by
\begin{equation}
b  |  \boldsymbol{O}, \sigma^2_0, \gamma_p =1 \sim \mathcal{N}(  \mu^\ast_{p} , \sigma_{p}^{\ast 2} ).
\end{equation}
The values of $\boldsymbol{\pi}^\ast$ and $\{  \mu^\ast_{p} , \sigma_{p}^{\ast 2} \}_p$ can be easily and efficiently computed.

\subsection{Sum of single-effect model}
Since multiple nonzero effects may exist, SuSiE extends this framework by modeling the overall effect as the sum of multiple single-effect models—hence its name. The prior setting now takes the form:
\begin{equation}
\boldsymbol{b} = \sum_{l=1}^L \boldsymbol{b}_l,  \qquad  \boldsymbol{b}_l \overset{\text{i.i.d.} }{ \sim } \text{Priors}( \sigma^2_0  , \boldsymbol{\pi} ),
\end{equation}
where each $\boldsymbol{b}_l$ follows the same independent priors as defined in \eqref{prior1}-\eqref{prior3}.

The hyperparameter $L$ represents the maximum number of possible nonzero parameters (i.e., change points). Thus, at most $L$ credible sets can be constructed for detected change points. The SuSiE fitting procedure is carried out via the Iterative Bayesian Stepwise Selection (IBSS) algorithm, which is described in detail by \cite{wang2020simple}.
Typically, $L$ is pre-specified before fitting the SuSiE model. If domain knowledge suggests a specific number of change points, one can set $L$ accordingly (e.g., $L=1$ when exactly one change point is assumed, reducing SuSiE to the single-effect model). In cases where this information is unavailable, a relatively large value of $L$ (e.g., $L=10$ in our examples) can be chosen. Importantly, due to Bayesian regularization, an overestimated $L$ does not significantly affect posterior inference, as redundant components beyond the actual number of change points remain non-informative.
The IBSS algorithm also determines the effective number of detected change points, denoted as $L^\ast$ ($L^\ast \leq L$), which represents the number of iterations required to sufficiently fit the model.

\subsection{Inference}
In our nonlinear effect analysis using the change-point model \eqref{change_point_model_matrix_notation}, we require the following results from the SuSiE framework to estimate the change-points, effect function, and obtain their Bayesian uncertainty.
\begin{itemize}
    \item $\boldsymbol{\pi}^\ast \in \mathbb{R}^{ L \times (P+1) }$ or $\mathbb{R}^{ L^\ast \times (P+1) }$: this matrix contains the posterior inclusion probabilities for the change-point position, where each row corresponds to one detected change-point and records its posterior inclusion probabilities across the full set of candidate locations ${t_0, t_1, \ldots, t_P}$.
    
    \item $\boldsymbol{\mu}^\ast \in \mathbb{R}^{ L \times (P+1) }$ or $\mathbb{R}^{ L^\ast \times (P+1) }$: this matrix contains the posterior means, where each row represents the vector of posterior means of the parameters, conditional on their inclusion.
    
    \item $\boldsymbol{\sigma}^\ast \in \mathbb{R}^{ L \times (P+1) }$ or $\mathbb{R}^{ L^\ast \times (P+1) }$: this matrix contains the posterior standard deviations, where each row represents the vector of posterior standard deviations of the parameters, conditional on their inclusion.
\end{itemize}

The credible sets for the $l^\ast$-th detected change point can be easily constructed using the posterior inclusion probabilities $\boldsymbol{\pi}^\ast_{l^\ast}$. For the $l^\ast$-th detected change point, the predicted value can be taken as, say, the posterior mean or the mode.

For the counterfactual prediction of $Y(x^\ast)$ or/and $ \mathbb{E}(  Y( x^\ast ) ) $ with a specified exposure level $x^\ast$ (e.g. $ x^\ast = 0 $), we denote the outcome vector by $ \boldsymbol{Y} = [  Y_i(X_i)  ]_{i=1,\ldots,n}  \in \mathbb{R}^{n \times 1} $. The individual counterfactual outcome at $x^\ast$ is the sum of the present observed outcome $Y_i$ and the $X_i$-to-$x^\ast$ integral part of the effect change-point model \eqref{change_point_general}, 
\begin{equation}
    \int_{X_i}^{x^\ast} h'(x) \text{d}x =  \sum_{p=0}^P b_p \underbrace{   \int_{X_i}^{x^\ast} I\{s \geq t_p\}\text{d}s }_{ f(  x^\ast  ; X_i, t_{p}  )  } 
\end{equation}
where $f(x^\ast; X_i, t_p)$ represents the individual partial counterfactual contrast resulting from an intervention that shifts the exposure from $X_i$ to $x^\ast$, attributable to the change-point at $t_p$ with a unit effect size. Note that $f(  x^\ast  ; X_i, t_{p}  )  = ( x^\ast  - t_p )_+  - (  
X_i  - t_p )_+        $ where the $(\cdot)_+$ is the positive part function. 
We define the partial counterfactual matrix $ \boldsymbol{F}(x^\ast):=  \left[ f(  x^\ast  ; X_i, t_{p}  ) \,    \right]_{i=1,\dots,n; \, p=0,1,\dots,P}  $ and aim to estimate the counterfactual outcome vector $\boldsymbol{Y}(x^\ast)$. 

For each detected change-point $t_p$, we quantify the uncertainty in both its location and its associated effect size, as captured by the posterior outputs of SuSiE, given by $\{\boldsymbol{\pi}^\ast , \boldsymbol{\mu}^\ast, \boldsymbol{\sigma}^\ast\}$ or $\{\boldsymbol{\pi} , \boldsymbol{\mu}, \boldsymbol{\sigma}\}$. By accounting for the uncertainty in $f(x^\ast; X_i, t_p)$, we derive the following estimates of the counterfactual outcome $\boldsymbol{Y}(x^\ast)$ or/and its expectation $\mathbb{E}(Y(x^\ast))$, based on the SuSiE fitting results $\boldsymbol{\pi}^\ast$ and $\boldsymbol{\mu}^\ast$,
\begin{equation}
      \hat{  \boldsymbol{Y} }(  x^\ast ) = \boldsymbol{Y} + 
 \boldsymbol{F}(x^\ast) \underbrace{  (  \boldsymbol{\pi}^\ast    \circ  \boldsymbol{\mu}^\ast       )^T \boldsymbol{1}_{L}}_{  \text{Row sum of $ (\boldsymbol{\pi}^\ast    \circ  \boldsymbol{\mu}^\ast)^T $ }  }
\end{equation}
where $ \boldsymbol{1}_{L}    $ is the $L$-length column vector with all entries equal to $1$, and $\boldsymbol{F}(x^\ast)(\boldsymbol{\pi}^\ast \circ \boldsymbol{\mu}^\ast)^\top \boldsymbol{1}_L$ represents the SuSiE posterior mean of the counterfactual contrast resulting from an intervention that shifts the observed exposure to $x^\ast$ for each individual.
Therefore, $ \hat{  \boldsymbol{Y} }(  x^\ast )$ corresponds to the individual-level posterior mean of the counterfactual outcome $ Y(x^\ast) $. We have the estimated effect shape at any given exposure level $ x^\ast $ using the posterior mean
\begin{equation} \label{poterior_mean}
\begin{split}
    \hat{h}(x^\ast) :=  \hat{\mathbb{E}}(  Y(x^\ast) ) - \hat{\mathbb{E}}(  Y(0) ) = &   
n^{-1} \boldsymbol{1}_n^T ( \hat{\boldsymbol{Y}}(  x^\ast )  - \hat{\boldsymbol{Y}}(  0 )  )  \\
   =& n^{-1} \boldsymbol{1}_n^T ( \boldsymbol{F}(x^\ast) - \boldsymbol{F}(0)  ) (  \boldsymbol{\pi}^\ast    \circ  \boldsymbol{\mu}^\ast       )^T \boldsymbol{1}_{L}  \\
   =& n^{-1} \boldsymbol{1}_n^T [  f( x^\ast ; 0, t_p )   ]_{i=1,\dots,n;p=0,1,\ldots,P} \, (  \boldsymbol{\pi}^\ast    \circ  \boldsymbol{\mu}^\ast       )^T \boldsymbol{1}_{L} \\
   =& [  f( x^\ast ; 0, t_p )   ]_{p=0,1,\ldots,P} \, (  \boldsymbol{\pi}^\ast    \circ  \boldsymbol{\mu}^\ast       )^T \boldsymbol{1}_{L} 
\end{split}
\end{equation}

Since the estimated effect function is expressed as a sum of partial counterfactual contrasts over the detected change-points—each incorporating uncertainty in the candidate location and a Gaussian posterior distribution for the effect size given each candidate location—the posterior distribution of each partial counterfactual contrast follows a mixture of normal distributions. This distribution is completely characterized by $f( x^\ast ; 0, t_p ) $ and the SuSiE results $\{\boldsymbol{\pi}^\ast , \boldsymbol{\mu}^\ast, \boldsymbol{\sigma}^\ast\} $ or $\{\boldsymbol{\pi} , \boldsymbol{\mu}, \boldsymbol{\sigma}\}$. Consequently, we can obtain the Bayesian uncertainty for the effect function value $h(x^\ast)$ via posterior sampling, where the corresponding point-wise $95\%$ credible interval (CI) for $h(x^\ast)$ is given by
\begin{equation} \label{credible_interval1}
   \text{CI}[  h(x^\ast)  ] = \left[  F^{-1}_{  h_p( x^\ast )    } (0.025 ) , \, F^{-1}_{  h_p( x^\ast )    } ( 0.975 ) \right]
\end{equation}
with the posterior of $ h(x^\ast) $, denoted by $  h_p( x^\ast )   $, as
\begin{equation}  \label{credible_interval2}
    h_p( x^\ast ) =  \sum_{l=1}^{L^\ast}  \mathcal{MN}(  \boldsymbol{\pi}^\ast_l , \boldsymbol{f}(x^\ast)  \circ  \boldsymbol{\mu}^\ast_l ,     |\boldsymbol{f}(x^\ast)|  \circ \boldsymbol{\sigma}^\ast_l    )
\end{equation}
with $\boldsymbol{f}(x^\ast):= [  f( x^\ast ; 0, t_p )   ]_{p=0,1,\ldots,P}   \in \mathbb{R}^{ P+1 }$; where $  \mathcal{MN}( \cdot, \cdot, \cdot ) $ represents mixture of independent normal distributions with the three arguments corresponding to the vector of the mixture probabilities, the means, and the standard deviations, respectively.

\subsection*{Three S framework: complete algorithm}\label{complete_algorithm}
The suggested complete workflow for implementing our framework (the three \enquote{S} framework) in nonlinear effect analysis, addressing Objectives 1-4 in Section \ref{Scalar-on-function regression} using all techniques introduced above, is outlined in Appendix G of the Supplementary Materials.

Overall, one typically starts with individual data \( ( Z_i , X_i , Y_i )_{i=1,\ldots, n} \) and obtains the stratum-specific scalar and functional variables after applying a stratification approach. 
For the number of strata, one may choose the strata number to avoid weak stratum-specific instrument strength, or else consider multiple strata numbers as a sensitivity analysis.
When measurable covariates \(\boldsymbol{C}\) are available, they can either be incorporated into the stratification (e.g., for improved counterfactual exposure estimation or matching) or included in the stratum-specific IV fitting, provided that \(\boldsymbol{C}\) is independent of the instrument \(Z\).  
When constructing the stratum-specific functional variable (weight function), it can be treated as a functional object using the R package \texttt{fda} or stored as a vector of weight function values over various exposure grids, given an appropriate grid precision.  

The inference strategy can then be selected based on the research goal. If the objective is to assess whether the effect is linear, one can conduct a linearity test (which remains robust under certain invalid instrument scenarios; see Appendix F of the Supplementary Materials). Alternatively, if the aim is to study the effect shape, both scalar-on-function and scalar-on-scalar regression models can be considered.  

When the focus is specifically on change-point detection, scalar-on-function regression can be employed and subsequently transformed into a change-point model, which can then be fitted using SuSiE. In a more general case, one may first adopt the complete method (stratification, scalar-on-function regression, and SuSiE) and then determine the fitting strategy (nonparametric or parametric) based on the complete method results. For example, if too many change points are detected, a parametric approach using basis functions, such as polynomials, may be preferable.  

When implementing the complete SSS analysis, three hyperparameters need to be specified: the number of strata, the set of change-point candidates, and the iteration number in SuSiE. For the number of strata, one may choose the strata number to avoid weak stratum-specific instrument strength, or else consider multiple strata numbers as a sensitivity analysis. For change-point candidates, we recommend including as many as feasible so that the true change-points can be well covered; the analysis remains stable as long as the true change-point is included. For the iteration number in SuSiE, we suggest using the default value of 10 (i.e., allowing at most 10 true change-points) or more, since SuSiE’s results are known to be robust when the iteration number is overestimated.

% Remark 1: when covariate exits, can be added into fitting for all steps given that the covariate are independent of the instrument
% Remark: also robust to multiple invalid IV scenarios
% Remark 2: the functional objective could be stored by FDA package or simply vector over multiple grid points

\section{Simulation}\label{simulation}
We conduct various simulations to evaluate the performance of our proposed framework for nonlinear analysis. The simulation consists of three parts, each with specific objectives:
\begin{itemize}
    \item In Part I, we demonstrate that the stratification-based strategy outperforms common nonlinear methods, including IV-regression and control-function methods, in predicting effect shapes.

    \item In Part II, we show that, when applied to stratification results, scalar-on-function fitting provides better estimation performance than scalar-on-scalar fitting in certain scenarios.

    \item In Part III, we illustrate that the complete SSS framework (stratification + scalar-on-function model + SuSiE fitting) offers a reliable approach for investigating effect change-points and provides a general, nonparametric method for effect shape analysis.  
\end{itemize}

\subsection*{Part I}
Denote the instrument, exposure, and outcome by $Z$, $X$, and $Y$, respectively.
Let the confounder $U \sim \mathcal{N}(0,1)$ and the error terms $\epsilon_X \sim \mathcal{N}(0,1)$ and $ \epsilon_Y \sim \mathcal{N}(0,1) $.
We consider the following basic structural model $   X= 0.15Z + U  + \epsilon_X $ with the following
scenarios for the instrument type and outcome structural models with linear effect: 
\begin{align}
    \begin{array}{l} 
      \text{Scenario 1: } \\
       \text{Scenario 2: }  \\
       \text{Scenario 3: } \\
       \text{Scenario 4: } 
    \end{array} 
    &\quad \begin{array}{l} 
       Z \sim \text{Bernoulli}(0.5)-0.5    \\
        Z \sim \text{Bernoulli}(0.5)-0.5 \\
        Z \sim \mathcal{N}(0,1)    \\
        Z \sim \mathcal{N}(0,1)    
    \end{array} 
    &\quad \begin{array}{l} 
        Y=X+ U + \epsilon_Y   \\
       Y=X+ |U| + \epsilon_X^2 + 2 |U| |\epsilon_X|  +  \epsilon_Y   \\
      Y=X+ U + \epsilon_Y     \\
      Y=X+ |U| + \epsilon_X^2 + 2 |U| |\epsilon_X|  +  \epsilon_Y   
    \end{array}
\end{align}
where Scenarios 1 and 2 contain a binary instrument which with at most two parameters relating to the effect shape can be identified in theory and cannot be supported by many nonlinear IV methods. The instrumental effect is set to be small to mimic the weak instrument scenario (in all scenarios, $R^2 \lesssim 0.01$). Scenarios 2 and 4 correspond to the complicated confounding effect scenarios with the confounding effect form is arbitrarily selected so that the confounding effect on the outcome is not well expressed by the confounding effect on the exposure, which poses challenges to control-function-based methods.

Without loss of generality, we can set $h(0) = 0$, as we are interested in the differences in the function h(x) with respect to a fixed reference point. To study the effect function $h(x)$ with $h(0)=0$ by definition, we consider multiple oracle methods and representative ML methods as well as our SSS method. For oracle methods, we take the true information that the effect form is polynomial, and it is up to second power, so two effect-shape-related parameters to be identified; that is, $ h(x) = \beta_1 x + \beta_2 x^2 $. The considered methods, denoted by M1-M7, are 
\begin{itemize}
    \item M1 (oracle control function method). This procedure involves two-step fitting. In the first step, the residual $r$ is obtained by regressing $X$ on $Z$. In the second step, a residual-inclusion regression with the known effect terms (a polynomial up to the second order) is fitted: $Y \sim X + X^2 + r$, and the fitted parameters corresponding to $X$ and $X^2$, $\{ \hat{\beta}_1, \hat{\beta}_2 \}$, are recorded. The estimated effect function is given by $ \hat{h}(x) = \hat{\beta}_1 x + \hat{\beta}_2 x^2 $. Note that the control function estimators here are not unbiased.

    \item M2 (oracle IV-regression method). Given the oracle effect form $ h(x) = \beta_1 x + \beta_2 x^2 $, the IV regression \eqref{IV_regression} can be expressed as
    \begin{equation}
    \mathbb{E}( Y |  z^\ast ) = \mathbb{E}(   \beta_1 X + \beta_2 X^2  |  z^\ast  ) = \beta_1  \mathbb{E}( X |   z^\ast  )  + \beta_2 \mathbb{E}( X^2 |  z^\ast  ) 
    \end{equation}
    for $z^\ast \in \{ 0:= \{Z=0\} , 1:= \{Z=1\} \}$ in the case of a binary instrument, or $ z^\ast \in \{ 0:= \{Z \leq F^{-1}_Z(0.5)\} , 1:= \{Z > F^{-1}_Z(0.5)\} \}$ in the case of a continuous instrument. 
    The two-parameter IV regression has the exact-identification solution:
    \begin{equation}
    \begin{pmatrix} \hat{\beta}_1 \\ \hat{\beta}_2 \end{pmatrix} =
    \begin{pmatrix} \hat{\mathbb{E}}( X |z^\ast =0 ) & \hat{\mathbb{E}}( X^2 |z^\ast =0 )  \\ \hat{\mathbb{E}}( X |z^\ast =1 )  & \hat{\mathbb{E}}( X^2 |z^\ast =1 )  \end{pmatrix}^{-1}
    \begin{pmatrix} \hat{\mathbb{E}}( Y |z^\ast =0 )  \\ \hat{\mathbb{E}}( Y |z^\ast =1 )  \end{pmatrix}
    \end{equation}
    without requiring the estimation of the conditional exposure density $ F(x|z^\ast) $.

   \item M3 (stratification with oracle scalar-on-scalar regression). We first obtain the stratum-specific IV estimates $\{ \hat{\beta}^\ast_k\}$ and their standard errors using doubly-ranked stratification with $K=100$ strata. Then, we build the scalar-on-scalar regression, given that $h(x)$ is a polynomial of up to the second order. Thus, we model:
\begin{equation}
    \hat{\beta}^\ast_k  = b_1 + b_2 \bar{x}_k + \epsilon_k, \quad \text{where} \quad \epsilon_k \sim \mathcal{N}( 0,s.e.( \hat{\beta}^\ast_k  )^2  ), \quad k =1,\ldots,K.
\end{equation}
where $ \bar{x}_k $ represents the exposure mean of the $k$-th stratum.
The fitted parameters are then transformed into the effect shape parameters: $ \hat{\beta}_1 = \hat{b}_1 $ and $ \hat{\beta}_2 = \hat{b}_2/2 $ for effect function estimation. The stratification is implemented using the R package \texttt{DRMR}.

   \item M4 (control function with model selection). This method is based on M1 but without oracle information. When fitting the second-step regression, we consider potential polynomial terms of $X$ and $r$ and then apply stepwise model selection to determine the optimal residual-inclusion model. The fitted parameters corresponding to the polynomial terms of $X$ are then used for effect function estimation. This control function approach in MR is known as PolyMR \citep{sulc2022polynomial}. M4 is implemented using the R package \texttt{PolyMR}.

   \item M5 (SSS). This method extends M3 by incorporating scalar-on-function regression and SuSiE for general effect function estimation. We adopt the same stratification setting as in M1. In SuSiE, we construct a change-point model with the change-point candidates chosen as the middle $90\%$ quantiles, along an intercept term corresponding to the potential linear effect. We fit SuSiE using an iteration count of $L=10$ and a uniform prior for the inclusion probabilities, ensuring that $ \boldsymbol{\pi}_l $ remains a constant vector for $l=1,\ldots,L$. The SuSiE fitting was implemented using the R package \texttt{susieR}.

   \item M6 (DeepIV). DeepIV \citep{hartford2017deep} is a representative nonparametric IV regression method which expresses the functions $h(x) $ with a neural network mapping as $ h_{\boldsymbol{\theta}}(x) $ and trains the optimization problem for the neural network parameters $\boldsymbol{\theta}$:
    \begin{equation}  
    \hat{\boldsymbol \theta } = \arg\min _{\boldsymbol \theta } \left[  \frac{1}{n}\sum_{i=1}^n\left( Y_i -  \frac{1}{M} \sum_{m=1}^M h_{\boldsymbol{\theta}}( x^\ast_{i,m}) \right)^2 \right] 
    \end{equation}
    where ${M}^{-1} \sum_{m=1}^M h_{\boldsymbol{\theta}}( x^\ast_m )$ acts as a Monte Carlo estimate for $\mathbb{E}( h_{\boldsymbol{\theta}}(X)|Z_i )$ and $ \{  x^\ast_{i,m} \}$ are $M$ i.i.d samples from $ \hat{F}(X|Z_i) $, which is also trained by neural network. The trained function $ \hat{h}(x) = h_{\hat{\boldsymbol{\theta}}}(x) $ will be centered such that $ \hat{h}(0) = 0 $. This centering is necessary for method comparison in this article, as the IV-regression model examines the effect function under the assumption that the remaining term in the exposure-outcome model has a mean of zero given the instrument. Without centering, the interpretation of the effect function would differ from that of other methods.

    \item M7 (KernelIV). KernelIV \citep{singh2019kernel} is another representative nonparametric nonlinear IV method. It is designed to address the ill-posed problem in solving $ \mathbb{E}(Y|Z )= \mathbb{E}(h(X)|Z ) $ by imposing assumptions on the effect function $h$ within the reproducing kernel Hilbert space (RKHS) framework. This approach essentially represents the effect function as a linear additive combination of infinitely many basis functions using feature maps, without requiring explicit specification in the kernel calculation. The model is then fitted using a two-stage kernel ridge regression, yielding a closed-form solution with hyperparameters. The data is split into two equal subsets to determine the hyperparameters. Similar to DeepIV, the final effect prediction is centered to ensure that the functional objective maintains the same interpretation as in other methods for comparison.
\end{itemize}

We consider 5,000 samples and conduct effect function estimation over 1,000 simulations. In each simulation, we record the predicted effect function values across five theoretical exposure quantiles (10\%, 30\%, 50\%, 70\%, 90\%) and calculate the mean squared error (MSE). Since the 50\%-quantile exposure level is always 0 in our model scenarios, its MSE is expected to be 0. 
DeepIV and KernelIV require significantly more computational time for fitting compared to other methods. We implement these algorithms using code adapted from the most recent paper by \cite{fonseca2025nonparametric}. Since DeepIV and KernelIV cannot handle binary instruments (Scenarios 1 and 2), and in practice, their fitting procedures often fail to complete, we only report the MSE for these methods in continuous IV scenarios (Scenarios 3 and 4). 
Among all method candidates, M1–M3 are oracle methods, while M4–M7 are either flexible or nonparametric methods. As oracle methods, M1–M3 are expected to achieve lower MSE. We are also interested in the performance of the flexible methods in the presence of additional noise, particularly with weak IVs.

\subsection*{Part II}
In this part, we investigate the two different model choices after stratification; namely, the scalar-on-function (SoF) regression and the scalar-on-scalar (SoS) regression. We consider same structural exposure model that $ X = 0.15 Z  + U + \epsilon_X $ with $  Z \sim \text{Bernoulli}(0.5) - 0.5 $.  We consider outcome structural model $Y=h(X)+ U + \epsilon_Y$ with two different nonlinear effect scenarios for better presenting the difference between SoF regression and the SoS regression; they are 
\begin{equation}
   h(X)= X + 0.5 X^2     \, \text{, so $h'(x)=1 + 2 x$   }
\end{equation}
\begin{equation}
   h(X)= I\{  X \geq 0 \}   \, \text{, so $h'(x)= I\{x >0 \} $   }
\end{equation}
Since the stratification-based method directly estimates the effect intensity \( h'(x) \), we will evaluate the MSE of \( h'(x) \) over the 10\%, 30\%, 50\%, 70\%, and 90\%-quantiles using SoS and SoF regression. To compare the two types of regression fairly, we use oracle information for the effect intensity form, ensuring that the inference procedures for both regressions follow the same parametric approach. 
For this part, we will increase the sample size to 50,000 to better mimic real-world applications. We will also consider three commonly used strata number settings in practice: \( K = 10 \), \( K = 50 \), and \( K = 100 \). While there is no optimal choice for the number of strata in stratification, we will demonstrate that the number of strata may influence the selection between SoS and SoF regression.

\subsection*{Part III}
In this part, we further investigate the SSS framework and demonstrate how it can be used to study change-points (i.e., when there is at least one change-point in the effect function, can SSS detect it and predict its value?) and to predict the effect shape function in a general manner.

We follow the same structural models as before. In causal inference and nonlinear analysis, to preserve the original interpretation of the intervention, one may not want to transform or rescale the exposure. Therefore, we also consider other continuous exposure distributions beyond the Gaussian distribution, including scenarios with more skewed distributions such as lognormal. We examine the following four scenarios for the instrument-exposure models:
\begin{align}
    \begin{array}{l} 
      \text{Scenario 1: } \\
       \text{Scenario 2: }  \\
       \text{Scenario 3: } \\
       \text{Scenario 4: } 
    \end{array} 
    &\quad \begin{array}{l} 
       X = 0.15 Z + U + \epsilon_X     \\
        X = 0.15 Z + U + \epsilon_X    \\
        X = \exp\{0.3 Z + U + \epsilon_X\}    \\
         X = \exp\{0.3 Z + U + \epsilon_X\}    
    \end{array} 
    &\quad \begin{array}{l} 
        \text{with } Z \sim \text{Bernoulli}(0.5) - 0.5 \\
       \text{with } Z \sim \mathcal{N}(0,1)  \\
      \text{with } Z \sim \text{Bernoulli}(0.5)  - 0.5 \\
      \text{with } Z \sim \mathcal{N}(0,1) 
    \end{array}
\end{align}
Scenarios 1 and 2 correspond to the (mixture of) normal distribution for the exposure, while Scenarios 3 and 4 correspond to the (mixture of) lognormal distribution for the exposure. In all scenarios, the instrumental effect is selected so that the proportion of exposure variance explained by the instrument is approximately $0.01$ (when the IV is normal) or $0.003$ (when the IV is binary). Such levels of coefficients of determination are common in Mendelian randomization. 
As compensation for the low coefficients of determination, modern genetic data usually have a large sample size. We set the sample size to $50\,000$, and the number of simulations is $1\,000$.

We express the outcome model as 
$
    Y = h(X) + U + \epsilon_Y
$
where $\epsilon_Y \sim \mathcal{N}( \mathbb{E}( Y(0) ),1) $ so that $h(0)=0$. In our simulation, we set $x=0$ as the baseline exposure level so that $h(x)$ represents the causal effect of the intervention changing the exposure level from $0$ to $x$ on the outcome. Note that for notation simplification, we let the confounding effect in the outcome model be similar to the confounding effect in the exposure model, but this is not required in our methods as previously discussed.

We consider the following cases of the effect function $h(x)$ for either the (mixture of) normal exposure  scenarios $1  \& 2$ or the (mixture of) lognormal exposure scenarios $3  \& 4$
\begin{align}
      \, & &  \,  &  \text{Scenario $1  \& 2$ (normal exposure)} &   \,  &\text{Scenario $3  \& 4$ (lognormal exposure)}    \nonumber \\
     \, & \text{Case 1:} &  \, & h(x) = x       & \, &   h(x) = x     \\
     \, & \text{Case 2:} &  \, & h(x) = (x-0)_+ & \, & h(x) = (x-2.5)_+  \\ 
   \, & \text{Case 3:}   &  \, & \begin{aligned}[t] % 强制顶端对齐
      h(x) = & \, 0.5x +0.5 (x+0.5)_+ - 0.25 + \\
               & \, 0.5 (x-0.5)_+
    \end{aligned} 
    &  \, & 
    \begin{aligned}[t] % 强制顶端对齐
        \! h(x) = & \, 0.5x + 0.5 (x-0.5)_+ + \\
               & \, 0.5 (x-2.5)_+
    \end{aligned}  \\
    \, & \text{Case 4:} & \, &  h(x) = -x + 0.5x^2 &  \, &  h(x) = -2x + 0.5x^2  
\end{align}
Case 1 represents the linearity case; Case 2 corresponds to the nonlinear case with one change-point; Case 3 corresponds to the two change-points case, containing 3 parameters in total, jointly determining the effect shape. The effect function of Case 3 cannot be identified by binary instruments using ordinary IV regression. Case 4 represents the case with a continuous model rather than the change-point model for $h'(x)$; where $h'(x)$ is continuous and varies over $x$. Therefore, the change-point model should have infinitely many non-zero parameters, which would induce high uncertainty in $\hat{h}(x)$. For Case 4, which is better fitted by a parametric model with a known polynomial basis, we are interested in seeing whether the change-point model can still provide useful fitting results.

We conduct the doubly-ranked stratification to construct $100$ strata. Our simulation results are not sensitive to the strata number.
We build the scalar-on-function regression according to the formula \eqref{SoF} using the first-order error. When building the change-point model, we consider the change-point candidates as the $90\%$ middle quantiles for the normal exposure (Scenario 1 and 2), along an intercept term (i.e. $0\%$ quantile), and the $95\%$ left quantiles (including $0$) for the lognormal exposure (Scenario 3 and 4). We fit the model via SuSiE using the iteration time $L=10$ and a uniform prior for the inclusion probabilities so that $ \boldsymbol{\pi}_l $ is a constant vector for $l=1,\ldots,L$.

We first study the change-point inference for all four instrument-exposure scenarios, using the one change-point model (Case 2), so the change-point is known as $x^\ast=0$ for Scenario 1 and 2, and $x^\ast = 2.5$ for Scenario 3 and 4. We consider two common predicted change-point values: the posterior mean and posterior mode. We record the value for the first detected parameter indicated by the SuSiE result. We then examine the performance of our approaches in inferring the effect shape function $h(x)$ using SuSiE fitting for all effect cases. We record the pointwise predicted values of $h(x)$ and obtain their mean and $0.025$ and $0.975$ quantiles (i.e., corresponding to the $95\%$ predicted values) for result presentation.

\subsection{Results}

\subsection*{Part I}
The MSE results of the methods M1-M7 for effect function estimation are given in Table \ref{methods_comparsion}. As expected, the oracle methods (M1-M3) outperform the other methods in all scenarios. Among these, the stratification-based method (M3) exhibits the minimal MSE in most scenarios, indicating that stratification makes better use of data information, even compared to the oracle IV-regression and control function methods in weak instrument scenarios.  

The weak instrument typically induces singular matrix issues in all fitting procedures of IV methods, including both IV-regression fitting and the second-step residual-inclusion fitting in control-function methods. However, such singular matrix issues are less prominent in the downstream regression after stratification. For instance, when fitting a scalar-on-scalar regression based on stratum-specific variables, the stratum-specific mean is maximally distinguished across strata due to the stratification property, which makes the model more robust to weak instruments. A similar rationale applies to the scalar-on-function regression.

Among the flexible methods (M4-M7), our SSS method (M5) consistently achieves a lower MSE than the other methods across more model and quantile scenarios. In all scenarios, the stratification method maintains good and stable performance, regardless of the instrument type or the confounding pattern. This is due to the stratification that ensures that the stratum-specific IV estimates are effectively utilized for nonlinear analysis.

Results were similar when estimating other nonlinear exposure--outcome models (Appendix H of the Supplementary Materials) with the method M5 (SSS) sometimes having better performances.

\begin{table}[htb]
    \centering
    \resizebox{\linewidth}{!}{ % 让表格等宽
    \begin{tabular}{c ccccc ccccc cc}
        \toprule
        & \multicolumn{5}{c}{Scenario 1} & \multicolumn{7}{c}{Scenario 3} \\
        & \multicolumn{5}{c}{(binary IV, simple confounding)} & \multicolumn{7}{c}{(continuous IV, simple confounding)} \\
         \cmidrule(lr){2-6} \cmidrule(lr){7-13}
          &   M1 & M2 & M3 & M4 & M5 & M1 & M2 & M3 & M4 & M5 & M6  & M7 \\
        \midrule
10\% & 0.418 & 0.421 & \textbf{0.409} & 0.775 & 0.821 & \textbf{0.065} & 0.111 & 0.086 & 0.076 & 0.136 & 3.363  & 1.635 \\ 
30\% & 0.070 & 0.070 & \textbf{0.054} & 0.139 & 0.110 & \textbf{0.011} & 0.018 & 0.011 & 0.014 & 0.015 & 0.564  & 0.170  \\ 
50\% & 0  & 0 & 0  & 0  & 0  & 0  & 0 & 0 & 0 & 0   & 0  & 0 \\ 
70\% & 0.070 & 0.070 & \textbf{0.052} & 0.135 & 0.087 & \textbf{0.011} & 0.018 & 0.012 & 0.014 & 0.013 & 0.563  & 0.167 \\ 
90\% & 0.417 & 0.418 & \textbf{0.384} & 0.734 & 0.495 & \textbf{0.065} & 0.110 & 0.094 & 0.076 & 0.076 & 3.349  &  1.587 \\ 
        & \multicolumn{10}{c}{} \\ % 空行
         & \multicolumn{5}{c}{Scenario 2} & \multicolumn{7}{c}{Scenario 4} \\
        & \multicolumn{5}{c}{(binary IV, complex confounding)} & \multicolumn{7}{c}{(continuous IV, complex confounding)} \\
         \cmidrule(lr){2-6} \cmidrule(lr){7-13}
        & M1 & M2 & M3 & M4 & M5 & M1 & M2 & M3 & M4 & M5   & M6 & M7 \\
        \midrule
10\% & 5.588 & 26.987 & \textbf{1.550} & 2.713 & 1.749 & 4.922 & 25.918 & 0.495 & 0.537 & 0.698 & 25.616  & \textbf{0.284} \\ 
30\% & 0.279 & 1.020 & \textbf{0.184} & 0.446 & 0.262 & 0.165 & 0.803 & \textbf{0.052} & 0.108 & 0.087 & 3.400  & 0.217 \\ 
50\% & 0 & 0 & 0 & 0 & 0 & 0 & 0 & 0 & 0 & 0 & 0 & 0 \\ 
70\% & 0.293 & 1.101 & \textbf{0.201} & 0.473 & 0.238 & 0.160 & 0.838 & \textbf{0.046} & 0.110 & 0.066 & 0.215 &  0.991\\ 
90\% & 5.791 & 28.163 & 1.806 & 3.145 & \textbf{1.547} & 4.847 & 26.427 & \textbf{0.405} & 0.578 & 0.406 & 2.399  &  9.406 \\ 
        \bottomrule
    \end{tabular}
    }
    \vspace{1em}
    \caption{\normalfont The MSE results of the effect function over several exposure quantiles across multiple nonlinear IV methods (denoted by M1-M7) under four different model scenarios in 1000 simulations.
    The objective of the effect function is defined to be zero when the exposure level is $0$, corresponding to the 50\% quantile in each scenario so the MSE is $0$. The M6 (DeepIV) and M7 (KernelIV) cannot work with binary instruments, hence were deprecated in scenario 1 and 2.
    For each scenario and quantile value, the minimal MSE value is highlighted. M1: Oracle control function. M2: Oracle IV-regression. M3: Stratification with oracle SoS regression. M4: PolyMR. M5: SSS. M6: DeepIV. M7: KernelIV.}
    \label{methods_comparsion}
\end{table}

\subsection*{Part II}
The MSE of predicting the effect intensity using two types of regression (SoS and SoF) is given in Table \ref{sos_vs_sof}. When the effect intensity is linear and the approximation equation \eqref{approx} hence holds well, SoS regression performs slightly better in terms of MSE across all strata number scenarios. However, when the effect intensity exhibits a jump, SoS regression results in a larger MSE than SoF regression due to the poor approximation in some strata. As the strata number increases, this difference in MSE diminishes. This is because a larger strata number corresponds to a narrower stratum-specific exposure distribution, making it more likely for the approximation equation \eqref{approx} to hold in most strata.

Since scalar-on-scalar regression serves as an algebraic approximation to the original scalar-on-function regression based on the approximation equation \eqref{approx}, it may not perform well when the approximation is poor. However, when the approximation is reasonable, scalar-on-scalar regression avoids the additional estimation of the functional variable (the weight function) required in scalar-on-function regression. This provides the motivation to use scalar-on-scalar regression as the primary method, particularly where the number of strata is large, as the nonparametric estimation of the stratum-specific weight function may be unreliable in scalar-on-function regression with small samples.

In conclusion, when the number of strata is large, one may consider using a simpler scalar-on-scalar regression for effect shape estimation. However, in practice, a smaller number of strata is often used due to several considerations: (i) ensuring each stratum has a sufficient sample size, (ii) minimizing selection bias for coarsened exposure \citep{tian2023relaxing}, and (iii) reducing storage costs for stratum-specific variables. When fewer strata are used, scalar-on-function regression is more suitable if the effect intensity is expected to exhibit a jump. Moreover, if the primary goal is to infer the change-point, scalar-on-function regression should be utilized.

\begin{table}[ht]
\centering
\begin{tabular}{lccccccc}  % 新增一列 'l'
   \toprule
   &  & \multicolumn{2}{c}{K=10} & \multicolumn{2}{c}{K=50}  & \multicolumn{2}{c}{K=100} \\
   \cmidrule(lr){3-4} \cmidrule(lr){5-6} \cmidrule(lr){7-8} 
   &  & SoS & SoF & SoS & SoF & SoS & SoF \\ 
   \hline
   & 10\% & 3.324 & 3.438 & 3.310 & 3.541 & 3.308 & 3.886 \\ 
   & 30\% & 0.560 & 0.595 & 0.558 & 0.598 & 0.558 & 0.654 \\ 
$ h'(x)=1+2x $  & 50\% & 0.009 & 0.008 & 0.008 & 0.008 & 0.008 & 0.008 \\ 
   & 70\% & 0.574 & 0.553 & 0.568 & 0.603 & 0.566 & 0.663 \\ 
   & 90\% & 3.360 & 3.334 & 3.335 & 3.552 & 3.329 & 3.907 \\ 
   \hline 
   & 10\% & 0.028 & 0.018 & 0.017 & 0.015 & 0.015 & 0.015 \\ 
   & 30\% & 0.028 & 0.018 & 0.017 & 0.015 & 0.015 & 0.015 \\ 
$h'(x) = I\{x>0 \}$  & 50\% & 0.028 & 0.018 & 0.017 & 0.015 & 0.015 & 0.015 \\ 
   & 70\% & 0.036 & 0.017 & 0.018 & 0.015 & 0.016 & 0.014 \\ 
   & 90\% & 0.036 & 0.017 & 0.018 & 0.015 & 0.016 & 0.014 \\ 
   \bottomrule
\end{tabular}
\vspace{1em}
\caption{\normalfont The MSE of the predicted effect intensity, $h'(x)$, over multiple exposure quantiles using two regression types with three numbers of strata ($K=10,50,100$) in $1000$ simulations. SoS: scalar-on-scalar; SoF: scalar-on-function.}
 \vspace{-2em}
\label{sos_vs_sof}
\end{table}

\subsection*{Part III}
\begin{figure}[tbp]
   % \begin{minipage}[t]{0.5\textwidth} % 左图占页面宽度的45%
   %     \centering
   %     \includegraphics[width=\textwidth]{sim_Fig2_left.eps} % 替换为你的图片路径
   % \end{minipage}
   % \hfill % 添加水平间距
   % \begin{minipage}[t]{0.5\textwidth} % 右图占页面宽度的45%
   %     \centering
   %     \includegraphics[width=\textwidth]{sim_Fig2_right.eps} % 替换为你的图片路径
   % \end{minipage}
     \centering  % 图片居中
    \includegraphics[width=1.0\textwidth]{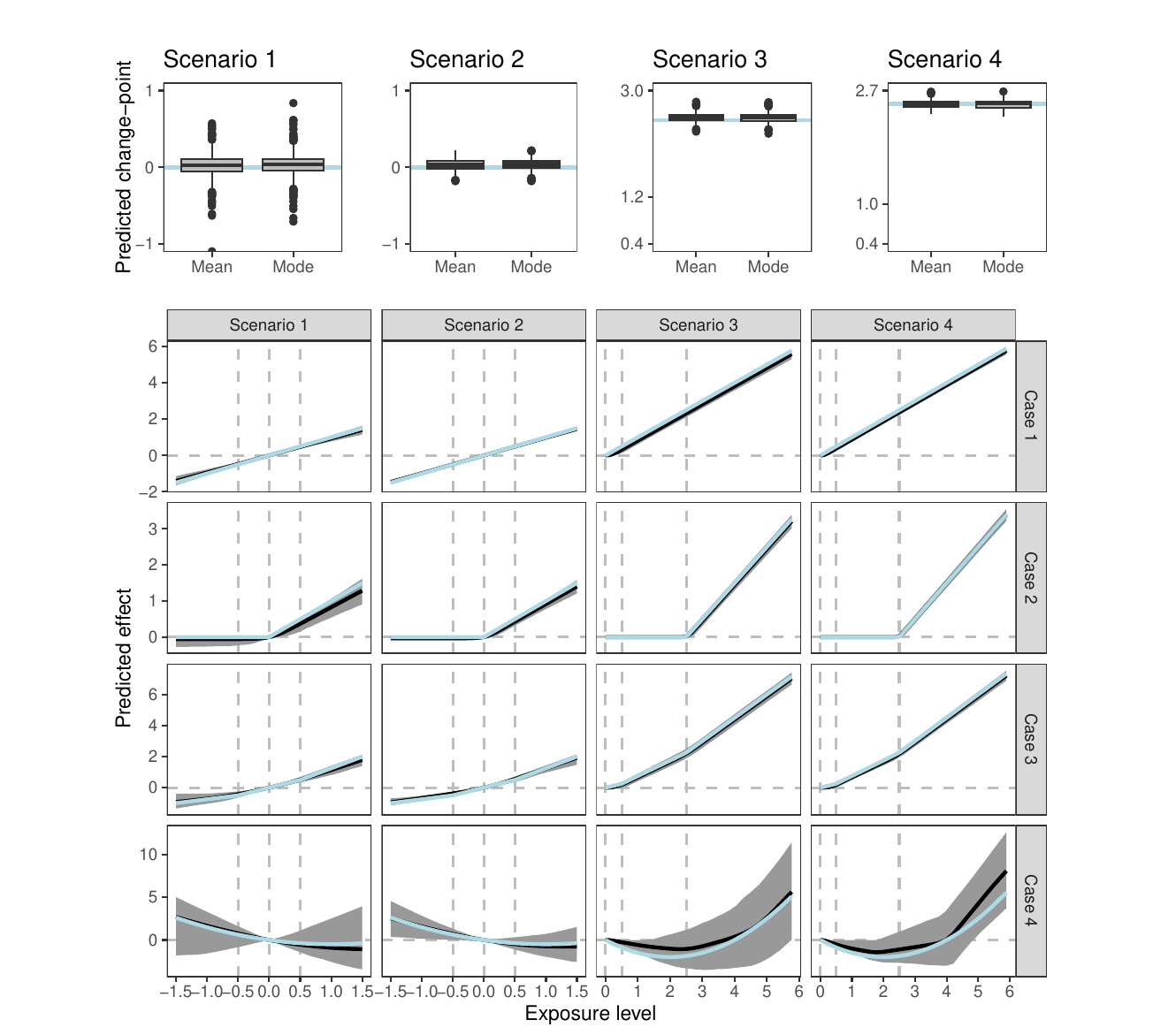}  % 图片路径和大小
    \vspace{-2em}
    \caption{\normalfont The results of the simulation. (Upper) The boxplot of the posterior predicted change-point of the first detected change-point for different instrument-exposure model scenarios (Scenario 1-4) under the one-change-point underlying model (Case 2) using SuSiE fitting with $1\,000$ simulations and the number of strata $K=100$. For each scenario, we provide the posterior mean and posterior mode of the posterior distribution for the change-point. The blue line represents the true change-point, and the values marked in the y-axis correspond the $0.25$, $0.5$, and $0.75$ quantiles of the exposure in each scenario. (Lower) The pointwise posterior mean and distribution of the effect function using SuSiE fitting with $1\,000$ simulations and the number of strata $K=10$ for different instrument-exposure models (Scenario 1-4) and effect cases (Case 1-4). The black line represents the mean of the posterior predicted effect, and the grey region represents the $95\%$ predicted effect values in our simulation. The blue curve is the true effect shape and the dashed vertical lines represent the exposure positions at which the change-point for some effect cases may be located. Scenario 1: binary IV and normal exposure. Scenario 2: continuous IV and normal exposure. Scenario 3: binary IV and lognormal exposure. Scenario 4: continuous IV and lognormal exposure.}
     \vspace{-2em}
        \label{sim_comb_fig}
\end{figure}

The change-point prediction results for effect case 2 are given in the upper panel of Figure \ref{sim_comb_fig}. In all simulations, at least one change-point is detected, and in most simulations, only one detected change-point is returned even when we set the iteration number $L=10$, which allows at most $10$ change-points to be detected. The distribution of the predicted change-point values among the simulations is close to the underlying true change-point. The bandwidth of the posterior predicted values is quite narrow, and most predicted values are close to the true change-point. Our result indicates that at least for the one change-point model, our approaches can correctly identify the number of change-points as well as predict the specific value of the change-point with low variance. One advantage of our approach for investigating the change-point is that we do not need to pre-specify the number of change-points in the SuSiE fitting, and the predicted results are insensitive to the choice of $L$, given that $L$ is larger than the actual number of change-points in the exposure-outcome model. We suggest using $L=10$ for the fitting as it is also the default setting in the SuSiE package. When the number of detected change-points is low (e.g., $1$), the fitting results can directly give a highly credible conclusion to the true exposure-outcome model and the underlying effect function. When the number of detected change-points is large (e.g., $L$), this indicates that the underlying effect function is generally continuous (like the effect case 4), hence not appropriately expressed by the change-point or piecewise constant model. In this scenario, one may instead consider the parametric fitting using continuous basis functions like polynomials.

The results of the pointwise predicted effect function using the change-point model are provided in the upper panel of Figure \ref{sim_comb_fig}. For the effect cases 1-3, which represent the change-point models, the effect shape is well captured by our approaches, with the pointwise predicted effect distribution highly close to the true effect. For effect case 4, where the differentiated effect shape is continuous, the SuSiE fitting based on the change-point model has high uncertainty (i.e., high variance of the predicted effect); this is as expected because the continuous effect model here can be understood as the extreme change-point model with infinitely many change-points. Therefore, our fitting with the maximal detected number $L=10$ does not produce posterior shrinkage. In this case, as discussed above, parametric fitting using a polynomial basis should be preferred over SuSiE fitting using the change-point model. Even though, the change-point model fitting still returns a nearly unbiased predicted effect over many exposure regions.

\vspace{-0em}
\section{Illustrative application}\label{application}
We implement the three `S' framework (Stratification, SoF regression, Sum of Single Effect) to investigate the causal effect of alcohol intake on SBP with a genetic instrument using observational data from UK Biobank. UK Biobank is a prospective cohort study of around half a million UK residents aged 40 to 69 years at baseline, recruited in 2006-2010 from across the United Kingdom \citep{sudlow2015}, and has been widely used for genetic causal inference. A recent Mendelian randomization study has implemented the stratification approach to investigate the stratum-specific effects of alcohol intake \citep{tian2023relaxing}, though the complete effect shape and the potential change-point are not well investigated. We consider 143,963 unrelated male individuals of European ancestries, who passed various quality control filters as described previously \citep{astle2016allelic}. We also conducted the analysis in female individuals, with the results provided in Appendix I of the Supplementary Materials. We construct a weighted genetic score for each individual from 93 genetic variants which have previously been shown to be associated with alcohol intake in 941,280 individuals at a genome-wide level of statistical significance \citep{liu2019alc}. Only around 124,590 individuals in this analysis were UK Biobank participants, minimizing bias due to sample overlap \citep{taylor2014mendelian}. The single weighted genetic score is used as the instrument in our nonlinear effect analysis. We are interested in the effect shape of alcohol intake on the SBP, particularly the potential change-point that divides the alcohol intake into the `safe' region where no causal effect is detected and the `danger' region within which causal effect exists.

In stratification, we adopt the doubly-ranked stratification to construct 10 strata. We take 10 strata for presenting stratum-specific results for the interpretability of findings. We perform the doubly-ranked stratification, as it is a more relaxed approach in terms of assumptions than the residual stratification. Previous studies using doubly ranked stratification have shown that the genetic association with alcohol intake increases across the strata, implying that the instrument’s effect on alcohol intake is not constant and thus violates the assumption underlying residual stratification \citep{tian2023relaxing}.
Additionally, alcohol intake is a coarsened exposure, with which doubly-ranked stratification deals much better than residual stratification \citep{tian2023relaxing}. Besides the SBP as the outcome of interest, we also consider age as the outcome. It is believed that the valid genetic variants are (nearly) independent of the confounders like age, and in our example, age is not plausibly affected by alcohol intake; hence, age serves as a negative control outcome. Any significant instrumental association with age beyond those expected by chance should indicate that the instrument is invalid. The negative control outcome can thus be used for evaluating the reliability of stratification for a given instrument-exposure combination, using the stratum-specific instrumental association with the negative control outcome. This is because poor stratification will induce the violation of within-stratum exchangeability.

In scalar-on-function regression, we only consider the first-order errors for the stratum-specific IV estimator, as in our example the genetic association with alcohol intake is sufficiently large to avoid weak instrument issues. We obtained the weight function in a nonparametric way as in the simulation with the following alcohol levels: 
\[
 \{ t_p := \hat{F}^{-1}_X\left(   \frac{  p }{100}   \right) \}_{p=0,1,\ldots,99}.
\]
We do not center or standardize the alcohol values, so the results are directly interpretable in the original scale (with unit/day and non-negative values).

In the sum of single effect analysis, we consider the change-point candidates as the left 95\% alcohol intake values, 
\[
 \{  F^{-1}_X\left(  \frac{p}{100} \right) \}_{p=0,1,\ldots,95}.
\]
Here we avoid the extreme candidate values for presentation purposes, though including them in SuSiE fitting does not affect the results. All the remaining setups are consistent with those in the simulation; that is, the largest number of possibly non-zero parameters \( L=10 \), the uniform inclusion priors in each \( l \) for \( \boldsymbol{\pi}_l \), and the variance of the parameter prior \( \sigma_0 \) in each \( l \) is estimated via the IBSS algorithm. Based on the number of detected change-points \( L^\ast \), for each detected change-point we construct the credible interval using the two-sided interval with \( 0.95 \) cumulative posterior density. For the effect shape estimation, we take two strategies: in the first strategy, we presume the change points are `confirmed' to be \( \{x^\ast_m \} \), taking the value of their posterior mode or mean; therefore, parametric fitting with known basis function \( \{ b_m := I\{ x \geq x^\ast_m \} \} \) can be drawn according to frequentist inference in Section \ref{Parametric_fitting}. There is no regularization term required, so the tuning parameter \( \lambda = 0 \). In the second strategy, we take into account the uncertainty of the change-point and predict the effect shape using the point-wise posterior mean formula \eqref{poterior_mean} and its point-wise 95\% credible interval based on formula \eqref{credible_interval1} and \eqref{credible_interval2}. When building the credible intervals of \( h(x) \), we simulated 10,000 i.i.d. samples according to formula \eqref{credible_interval2} to mimic the posterior distribution.

\subsection*{Application results}
\begin{figure}[tbp]  % 'h' 表示图片位置，h 表示尽可能放在当前位置
    \centering  % 图片居中
    \includegraphics[width=1.0\textwidth]{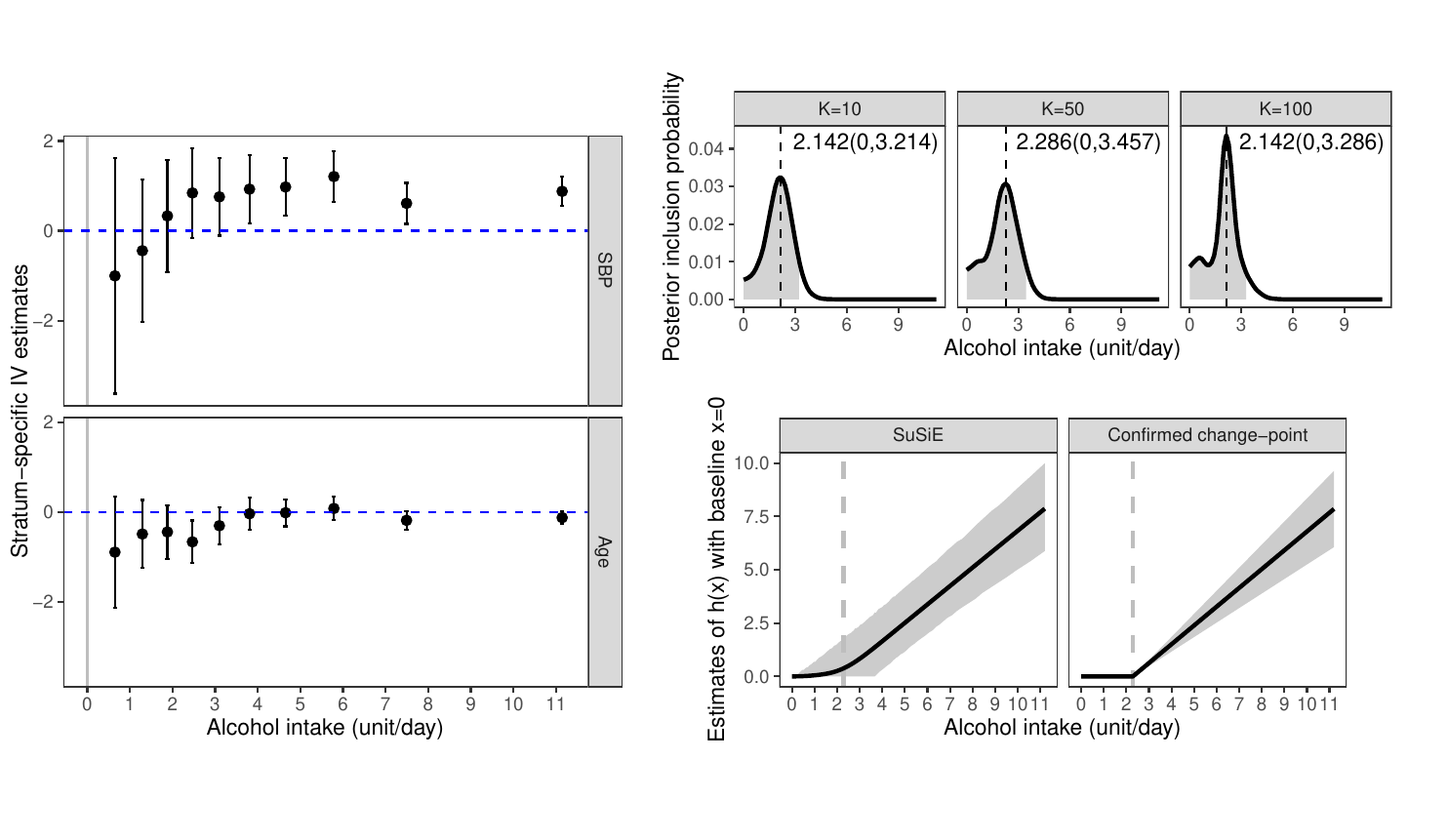}  % 图片路径和大小
    \vspace{-4em}
    \caption{\normalfont The results of the real application. Left: Stratum-specific IV estimates (point ratio estimates and the corresponding 95\% confidence interval) using the doubly-ranked stratification approach for the effect of the alcohol intake on the SBP (the outcome of interest) and age (the negative control outcome) with the genetic score as the instrument. The number of strata is $10$ and the stratum-specific mean of alcohol intake is used as the x-axis value for stratum-specific IV estimates. All the confidence intervals are not adjusted for multiple tests.
    Upper right: The posterior inclusion probability $ \boldsymbol{\pi}^\ast $ of the SuSiE fitting using three different number of strata ($K=10,50,100$). For all strata cases, only one change-point is detected hence we only need to present the posterior inclusion probability for the first parameter. The grey region corresponds to the one-sided $95\%$ credible interval for the change-point, and the numeric values on the upper right panel correspond to the specific value of the posterior mode and the credible interval.
    Lower right: The estimate of the effect shape $h(x)$ using two fitting strategies with the number of strata $K=100$: (left) SuSiE fitting where the point estimate $\hat{h}(x)$ is the posterior mean and the interval estimate is the $95\%$ credible interval; and (right) the parametric fitting with the confirmed change-point ($x=2.142$, the dashed vertical line, using the posterior mode) for basis functions, and the interval estimate is the $95\%$ confidence interval. The effect shape $h(x)$ is defined with the baseline value $x=0$ so that $h(0)=0$.}  % 图片标题
    \label{real_comb_fig}  % 图片标签，方便引用
\end{figure}

We first implement the stratification to obtain 10 strata to do the initial analysis. The stratum-specific results are given in the left panel of Figure \ref{real_comb_fig}. The stratum-specific IV estimates over different alcohol intake levels provide local information, and hence can be useful for nonlinear analysis without fitting the complete effect shape function. It is found that for the low alcohol intake region (\(\leq 2\) unit/day) the stratum-specific IV estimates on SBP are non-significant, while for the larger alcohol levels there exist stable positive effects. Though the confidence intervals are not adjusted by multiple tests, the continuous positive stratum-specific results give more evidence to the positive causal effect of alcohol intake on SBP, particularly over the regions with greater alcohol consumption, hence forming a nonlinear effect shape. As a contrast, most stratum-specific IV results are non-significant for the effect on age (the negative control outcome), and there is also no evidence to support the effect of alcohol intake on age by Cochran's Q test. The significant test results for SBP versus the non-significant test results for age over regions with greater alcohol consumption indicate that the instrument is likely to be valid after stratification.

We then proceed to investigate the potential change-point of the effect shape function \( h(x) \) using the SuSiE model. The results of the posterior density of the change-points are provided in the upper right panel of Figure \ref{real_comb_fig}, where we also consider other scenarios with strata number \( K=50 \) and \( 100 \) for sensitivity analysis. All scenarios support the highly similar results, where only one non-zero parameter is detected and the posterior distribution of the parameter is similar in terms of the posterior mode, mean, median, and credible intervals. Taking \( K=100 \) as the main illustration, the one-value predicted change-point for the effect of alcohol intake on the SBP is \( 2.142 \) (posterior mode) with the \( 95\% \) one-sided credible interval \( (0, 3.286) \). This means that although there could be a 'safe' region where alcohol intake does not form an evidenced effect on the SBP under a threshold (i.e. change-point in our model), the threshold is still quite low and less than \( 3.286 \) unit/day under a high credibility level. For most general alcohol drinking, this low threshold value typically indicates that long-term regular alcohol drinking can increase SBP.

The complete effect shape results are provided in the lower right panel of Figure \ref{real_comb_fig}, containing the results of SuSiE (Bayesian perspective considering the uncertainty of change-point) and parametric fitting (Frequentist perspective) using the confirmed change-point value \( 2.142 \) and strata number \( K=100 \). As the effect shape is defined with the baseline alcohol intake level \( x=0 \) (i.e. no alcohol intake) so that \( h(0)=0 \), the effect shape function \( h(x) \) is interpreted as the long-term causal effect of increasing the alcohol intake from \( 0 \) unit/day to \( x \) unit/day on the SBP.
Both results are similar and support that the alcohol intake effect is positive after a relatively low threshold. The SuSiE fitting results have more uncertainty due to its more flexibility in terms of model and also considering the uncertainty of the change-point. The estimated effect shape results are also insensitive to the number of strata (see Appendix J of the Supplementary Materials using other numbers of strata).

We note that the results from Mendelian randomization are usually more overestimated than the actual intervention effect. The reason is due to the property of using genetic variants as the instrument; because the genetic effect on the exposure is long-term while the intervention effect is typically short-term, the results of Mendelian randomization are more close to the long-term or cumulative effect of the exposure on the outcome, which is usually higher than the changed outcome level by the short-term intervention. Our results are consistent with published medical guidance on safe alcohol consumption, which recommends not regularly drinking more than 14 units per week ({\url{www.nhs.uk/live-well/alcohol-advice/calculating-alcohol-units/}).

\section{Discussion}\label{discussion}
In this article, we present a comprehensive framework, the SSS framework, for investigating nonlinear effects in instrumental variable analysis. Our framework consists of three sequentially key “S” layers: stratification, scalar-on-function or scalar-on-scalar regression, and sum of single-effect fitting. Through simulations, we demonstrate that the SSS framework outperforms other representative nonlinear IV methods in predicting effect shapes, particularly when the instrument is binary or weak or when the confounding pattern is complex. Beyond accurately capturing effect shapes, our framework can also identify potential change points by detecting their presence and estimating their precise locations.

Our framework is primarily built on the concept of stratification, which partitions the sample into multiple strata that satisfy the valid IV condition while capturing local exposure information. Once stratification is established, downstream analyses can be applied directly to the stratum-specific estimates using the methods introduced in this article to investigate the effect shape.
Importantly, this post-stratification approach eliminates the need to impose assumptions on the confounding pattern and remains effective in scenarios involving binary or weak IVs—situations that pose challenges for existing IV methods, including those based on IV regression and control functions. This provides an alternative strategy for conducting nonlinear analyses using our framework, not only in Mendelian randomization but also across a broad range of IV applications in various fields. Essentially, our framework replaces the stringent requirements on IV type and confounding structure imposed by most nonlinear IV methods with a stratification requirement. This is particularly attractive in practice, as reliable stratification can be achieved by selecting on measurable covariates that affect the exposure and are not colliders of the instrument and confounding factors.
When stratification is performed directly on counterfactual exposures using the approaches introduced (e.g. the doubly-ranked stratification approach), assumptions such as the rank-preserving assumption become necessary. However, in certain experimental or observational settings, ensuring the plausibility of this stratification assumption can be more feasible than satisfying the assumptions required by other nonlinear IV methods. Given that weak and binary instruments are common in practice, a stratification-based approach provides a straightforward and practical solution for analyzing potentially complex effect shapes. When high-dimensional instruments (e.g., multiple genetic variants) are available, one can first merge them into a single instrument, such as a polygenic score, and then conduct stratification based on this one-dimensional instrument, hence making the analysis easily scalable for high-dimensional instruments. Once strata have been created, IV estimation within the strata can be performed either using the one-dimensional score or using the high-dimensional instruments.
Nevertheless, caution is needed regarding the reliability of stratification. As with most assumptions involving counterfactual variables in causal inference, our stratification assumption is inherently unverifiable. It can only be partially assessed through methods such as triangulation frameworks, including the negative control design we implemented in our real-world application. With advancements in counterfactual-variable-based stratification methodologies and the integration of additional data into the stratification process, achieving reliable stratification will become increasingly feasible. This reinforces the potential of our framework as a promising direction for nonlinear effect analysis.

We provide two regression models after stratification: scalar-on-function regression and its approximated version, scalar-on-scalar regression. We do not recommend one over the other in general for studying effect shapes. In most scenarios, the approximation condition underlying scalar-on-scalar regression is well satisfied, making it a more practical choice due to its simplicity. This scalar-on-scalar regression is most commonly used in the current nonlinear MR literature. However, when the primary objective is to identify change points, such as investigating potential thresholds in the effect of alcohol intake, scalar-on-function regression should be considered because this method enables the application of change-point models and further sparse regression fitting techniques.
When constructing the weight function in scalar-on-function regression, one can estimate it nonparametrically, as we did, or use a parametric assumption based on the weight function properties introduced in this article (see Remark 4 and 5).
For fitting sparse regression models in change-point-based nonlinear effect analysis, we introduce the Bayesian nonparametric method, SuSiE. Compared to LASSO-based sparse regression methods, SuSiE is computationally efficient and has already gained significant popularity and success in fine-mapping applications. We demonstrate that SuSiE is well suited for change-point-based nonlinear effect estimation, as it not only provides credible regions for change-point locations but also offers a general approach for effect shape prediction.
If the effect shape is not pre-specified, our SSS framework can serve as a general solution. When the effect shape can be reasonably approximated using basis functions, parametric inference is recommended. Based on SSS results, if the detected number of change points is low, a change-point model should be considered. Conversely, if a large number of change points are detected—typically indicating that the underlying effect intensity $h'(x)$ is smoother and more continuous—a parametric basis function approach is suggested instead.

% cautions on stratification? the rank preserving assumption - really depends on instrument -expoxure relationship;

There are some important cautionary notes when applying our SSS framework. First, the framework places most of the assumptions for IV nonlinear effect analysis on the stratification process itself. Therefore, careful attention should be given to the reliability of the stratification, as discussed in \cite{davey2025erroneous} and \cite{hamilton2025non}.
Second, although we introduced our methods within the three `S' framework, the second `S' layer (scalar-on-function regression) and the third `S' layer (SuSiE fitting) are optional. If the objective of the nonlinear effect analysis does not involve tasks related to change points, one can opt for scalar-on-scalar regression to fit the stratum-specific IV estimates. 
Third, the stratification-based framework is particularly suited for pure nonlinear effect models, where the exposure effect is primarily driven by the exposure level itself, rather than being heavily influenced by unmeasured effect modifiers. This is because the stratification process also differentiates the distribution of covariates across the strata. In scenarios where the exposure effect is modified by covariates, the observed heterogeneity in stratum-specific IV estimates may be attributable to differences in the distribution of these covariates rather than the exposure effect itself. Further discussion of effect modification in nonlinear IV analysis can be found in \cite{tian2024data} and \cite{small2014commentary}.

In summary, the proposed framework, which has emerged in genetic epidemiology, has the potential to be widely implemented for various objectives in nonlinear effect analysis across many other fields.

\section*{Supplementary Materials}
Supplementary Materials is available at the end of the manuscript.

\section*{Software and reproducibility}
The R package “SSS” and the R code for simulations and data analyses are available at
\url{https://github.com/HDTian/SSS}.

\section*{Data availability}
UK Biobank individual participant data are available upon appropriate application to the UK Biobank study: \url{https://www.ukbiobank.ac.uk/}

\section*{Acknowledgments}
We thank the editors and the three anonymous reviewers for their insightful comments. H.T. thanks Pradeep Natarajan for his support of this research. S.B. is supported by the Wellcome Trust (225790/Z/22/Z) and the United Kingdom Research and Innovation Medical Research Council (MC\_UU\_00040/01). This research has been conducted using the UK Biobank resource under Application Number 98032.

\newpage
\mysection{Appendix A: Algorithm of the stratification methods}

\begin{algorithm}[H]
\DontPrintSemicolon  % 去掉每行后的分号

\textbf{Assumption:} $X= t [f(Z) + g_X(U, \epsilon_X)]$\;

\textbf{Input:} The individual-level data $(Z_i, X_i, Y_i)_{i=1,\ldots,n}$  \; 

$K \leftarrow$ the number of strata \;

$t[\cdot] \leftarrow$ the monotone bijective function for $Z$-$X$ relationship, or $t[x] \equiv x$\;

\If{$f(\cdot)$ is known}{
    $e_i \leftarrow t^{-1}[  X_i  ]  - f( Z_i ) $; obtain individual residuals \;
}\Else{
    Let $f(z) = \theta_1 f_1(z) + \theta_2 f_2(z) + \cdots + \theta_L f_L(z)  $\;

    Model selection based on $   \{   t^{-1}[X]  , Z ,  \{  \hat{\theta}_l , f_l(Z) \}_{l=1,\ldots,L}     \}  $ via, e.g., BIC or cross validation   \;

    $  \tilde{f}(z)  \leftarrow $ the best model \;

    $e_i \leftarrow t^{-1}[  X_i  ]  - \tilde{f}( Z_i ) $ ; obtain individual residuals \;
}

$\{e^\ast_i\} \leftarrow$ rank $\{e_i\}$\;

\For{$k \leftarrow 1$ \KwTo $K$}{
    $\mathcal{P}_k \leftarrow$ the individual index of the $[(k-1) \lfloor\frac{n}{K}\rfloor+(1:\lfloor\frac{n}{K}\rfloor)]$-th ranked individual in $\{e^\ast_i\}$\\
    $\mathcal{S}_k \leftarrow $ the individual data set $(Z_i, X_i, Y_i)$ with index  $ \mathcal{P}_k $\;
}
\KwOut{$\mathcal{S}_k, k=1,\ldots,K$}

%\KwOut{The test results (e.g., p-value) of the $Z$-$X$ homogeneity}\;

\caption{Prediction-based stratification (residual stratification)}
\label{R_algorithm}
\end{algorithm}

\vspace{5em}

\begin{algorithm}[H]
\DontPrintSemicolon  % 去掉每行后的分号
\textbf{Assumption:} rank-preserving assumption between the instrument and exposure

\textbf{Input:} the individual-level data $(Z_i, X_i, Y_i)_{i=1,\ldots,n}$ 

$K \leftarrow$ the number of strata

$S \leftarrow$ the size of pre-strata, an integer multiple of $K$. By default, $S=K$

$\{Z^\ast_i\} \leftarrow$  rank $\{Z_i\}$

\For{$s \leftarrow 1$ \KwTo $\lfloor  \frac{n}{S}  \rfloor$}{
$\mathcal{P}_s \leftarrow  $ the individual index of the $[(s-1)S+(1:S)]$-th ranked individual in $\{Z^\ast_i\} $

 $\{X^\ast_i;i\in \mathcal{P}_s\} \leftarrow$ rank $\{X_i;i\in \mathcal{P}_s\}$ 
}

\For{$k \leftarrow 1$ \KwTo $K$}{
    $ \mathcal{S}_k \leftarrow  $  the set of $(Z_i, X_i, Y_i)$ with individual corresponding to the $[ (k-1)\frac{S}{K} + (1:\frac{S}{K}) ] $-th individual in $\{  X_i^\ast; i \in \mathcal{P}_s \}$ for all $s$ }

\KwOut{$\mathcal{S}_k, k=1,\ldots,K$}

 \caption{Matching-based stratification (doubly-ranked stratification)}
 \label{DR_algorithm}
\end{algorithm}

\newpage
\mysection{Appendix B: Potential use cases under the rank-preserving assumption}
We list some potential use cases and discuss the plausibility of the rank-preserving assumption in some examples.

\medskip
\noindent
{\bf Example 1} {\it Inspired by \cite{angrist1991does}, consider a country where children born in different months of the year start school at the same calendar time but with different ages. Additionally, compulsory schooling laws mandate that students must remain in school until a specific birthday (e.g., 16th or 17th). In this case, birth time or season could serve as an instrument for schooling length, impacting outcomes such as earnings. Particularly, the rank-preserving assumption between birth time and schooling length is reasonable, as for any two individuals with the same birth time, the person with a longer schooling length should still maintain a longer schooling length, even if their birth times were counterfactually swapped with another identical birth time. We provided more explanation with the data of \cite{angrist1991does} in the next Appendix.}

%\medskip
%\noindent
%{\bf Example 2} {\it Inspired by \cite{bronars1994economic}, for individuals who prefer to have more children and are willing to give birth, whether the first birth was a twin or not is virtually randomly assigned. This random assignment makes it a reasonable instrumental variable (the "twin instrument") for the number of children, which can then be used to investigate the effect of the number of children on outcomes such as labor-force participation or poverty. The rank-preserving assumption between the twin instrument and the number of children is reasonable, as the effect of the twin instrument on the number of children can be modeled as a linear additive model with an effect size of one. A similar concept applies to the sex-mix instrument discussed by \cite{angrist1996children}.}

\medskip
\noindent
{\bf Example 2} {\it Drawing from the Judge IV idea \citep{kling2006incarceration}, strict judges and lenient judges may impose more or less prison time for the same criminal case. In many instances, the allocation of judges is random, making judge type a valid instrumental variable for analyzing the causal effect of prison time on various outcomes, such as income. Therefore, restricting the analysis to certain criminal cases, the rank-preserving assumption between judge type and prison time can be reasonable, as for specific case types, under any given judge, the person who serves a longer sentence in a pair will still serve longer.}

\medskip
\noindent
{\bf Example 3} {\it In Mendelian randomization, assume there exists a genetic variant or gene score that affects one phenotype of interest, such as weight or alcohol intake, for a group. Since the genetic variant can be reasonably considered to be randomly assigned, it can serve as an instrument for investigating the causal effect of the phenotype on an outcome of interest, such as cardiovascular disease. Given that the genetic effect follows a similar form (either constantly additive or multiplicative) within this group (e.g. male), the rank-preserving assumption between the gene and the phenotype may be plausible. This provides an opportunity to use the stratification approach to investigate the nonlinear causal effect.}

\newpage 
\mysection{Appendix C: Rank preserving assumption in \textit{Angrist\&Krueger (1991)} }\label{AK}
We provide a more detailed introduction to the rank-preserving assumption using real data from \cite{angrist1991does}, where birth quarter is considered an instrument for years of education. The dataset, derived from the 1970 Census, consists of men born between 1920 and 1929. It includes 222,389 samples, each with recorded birth quarter (Q1–Q4) and years of education, which are rounded to the nearest integer (ranging from a minimum of 0 to a maximum of 18). The data is readily accessible via the R package \texttt{sketching}. We construct four groups, each corresponding to a birth quarter, and examine the distribution of years of education.  

The distribution of years of education across birth quarters is presented in Figure \ref{fig:AK}, showing the proportion of individuals with each education level within each birth quarter group. The plot reveals a clear and consistent pattern: individuals born in later quarters tend to have fewer years of education. Specifically, for a given education level, the proportion of individuals with fewer years of education than that level generally decreases over the birth quarters Q1-Q4.  
This trend is expected due to the effect of birth quarter: among individuals who left school at a specific age, those born in later quarters typically have longer schooling durations than those born earlier in the same calender year. As a result, later birth quarters should have a lower proportion of individuals with fewer years of education and a higher proportion with more years of education. In this scenario, the effect of birth quarter on years of education appears relatively constant, supporting the rank-preserving assumption—namely, that the relative ordering of education lengths among individuals born in the same time period remains unchanged when their birth times shift.  
Although in this example, the instrument (birth quarter) is rounded to discrete values, its impact on doubly-ranked stratification should be minimal. This is because doubly-ranked stratification first aggregates similar instrument values, and the instrument's effect size is relatively small.

\begin{figure}[H]
    \centering
    \centering  % 图片居中
    \includegraphics[width=0.5\textwidth]{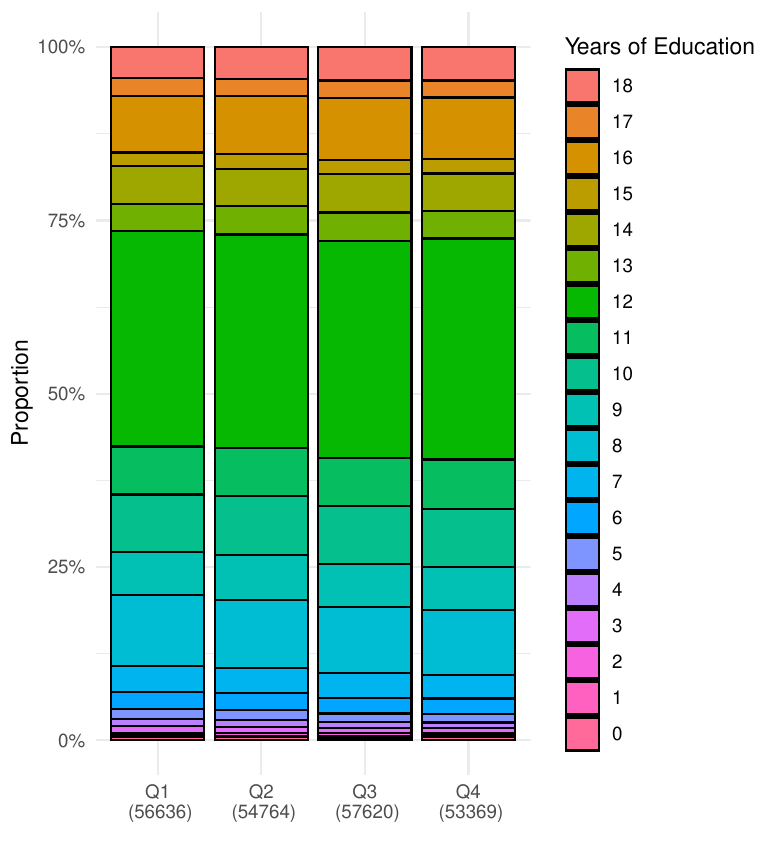}  % 图片路径和大小
    \vspace{-2em}
    \caption{\normalfont The education length distribution by quarter of birth (Q1-Q4) in the data of \cite{angrist1991does}. The value in the brackets indicate the sample size in each quarter group.} 
    \label{fig:AK}
\end{figure}

The plausibly valid rank-preserving condition is the only additional requirement beyond the core assumptions of instrumental variables (IV) for nonlinear effect analysis in our SSS method. We proceed to analyze the data from \cite{angrist1991does} to investigate the effect of education length on earnings (log weekly wages). Since the exposure (education length) is coarsened, we adopt the doubly-ranked stratification due to its proven ability to produce reliable results under coarsened exposure. For coarsened exposures, the number of strata should be carefully determined to avoid inducing excessive selection bias from measurement errors in the exposure \citep{tian2023relaxing}. The optimal number of strata can be determined using the Gelman-Rubin statistics from the \texttt{DRMR} package, and in our example, we chose to set the number of strata to 10. We conducted the downstream analysis using our developed software, \texttt{SSS}.

There is one change-point detected by our methods, and the posterior inclusion probability (PIP) of this change-point is shown in the left panel of Figure \ref{fig:AK_SSS}. The predicted change-point value is 5.3 (posterior mean) and 7.2 (posterior mode). This change-point corresponds to the primary education level, typically 6 years of education, above which the effect of education length on earnings becomes evident.
The estimated effect function, $h(x)$, with the baseline level at $x = 0$, is presented in the right panel of Figure \ref{fig:AK_SSS}. As shown, the effect of education length on earnings becomes significant after 10 years of education.

\begin{figure}[H]
    \centering
    \begin{minipage}{0.4\textwidth}
        \centering
        \includegraphics[width=1.0\linewidth]{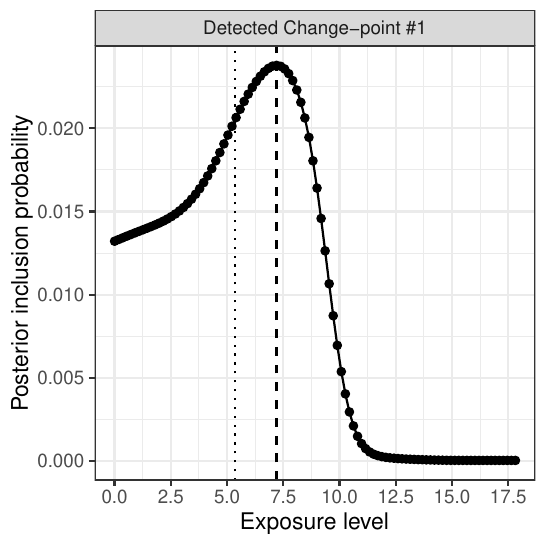}
        \label{fig:left}
    \end{minipage}
    \hfill
    \begin{minipage}{0.4\textwidth}
        \centering
        \includegraphics[width=1.0\linewidth]{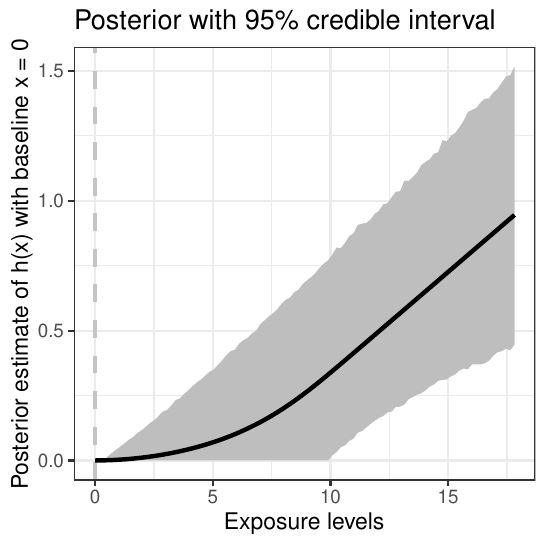}
        \label{fig:right}
    \end{minipage}
    %\vspace{-2em}
    \caption{\normalfont Nonlinear effect analysis results using the SSS method applied to the data from \cite{angrist1991does}. The exposure variable is years of education, and the outcome is log weekly wages from the 1970 Census. Left panel: The posterior inclusion probability of the only detected change-point. The dashed and dotted lines represent the posterior mode and mean values, respectively. Right panel: The estimated effect function. The gray region represents the $95\%$ credibility interval.}
    \label{fig:AK_SSS}
\end{figure}

\newpage
\mysection{Appendix D: Scalar-on-function model properties}\label{app_A}

% Note: in this sample, the section number is hard-coded in. Following
% proper LaTeX conventions, it should properly be coded as a reference:

%In this appendix we prove the following theorem from
%Section~\ref{sec:textree-generalization}:

In this appendix we prove the scalar-on-function (SoF) model for the IV framework with stratification, and show the properties of the weight function.

Assume $ Y= h(X) + U $ and the valid instrument $Z$. Assume that there exits the lower bound $l$ for the exposure domain $\mathcal{X}$ such that $\mathbb{P}( l < X )=1 $ almost surely. We let the effect function is continuous.
We have 
\begin{equation} \label{app_e1}
    \begin{split}
        \frac{Cov( Z ,Y )}{ Cov( Z,X )} &= \frac{Cov( Z , h(X) + U )}{ Cov( Z,X )} \\
               &= \frac{Cov( Z , \int _l^X h'(x) \text{d}x + h(l) + U )}{ Cov( Z,X )} \\& =  \frac{Cov( Z , \int _l^\infty h'(x) I\{X\geq x \} \text{d}x + h(l) + U )}{ Cov( Z,X )} \\
               &= \frac{Cov( Z , \int _l^\infty h'(x) I\{X\geq x \} \text{d}x )}{ Cov( Z,X )} \\
               &=    \int_\mathcal{X} h'(x)  \underbrace{\frac{ Cov(  Z, I\{X\geq x \} ) }{ Cov(Z,X)}}_{=: W(x)} \text{d}x
    \end{split}
\end{equation}
Note that this conclusion is easily extended to the complex case $h(X,\epsilon)$ with exogenous term $\epsilon$, given that $h(X,\epsilon)=\int _l^X h'(x,\epsilon) \text{d}x + h(l,\epsilon)$ where $h'(x,\epsilon) = \delta h(x,\epsilon)/ \delta x $ exists. In this case, we have $h'(x) = \mathbb{E}_\epsilon(  h'(x,\epsilon) ) $ in Equation \eqref{app_e1}.

In terms of the weight function $W(x)$, we can show
\begin{equation}
    \begin{split}
          \int_\mathcal{X}  W(x) \text{d}x  &=  \frac{  \int_\mathcal{X} Cov(Z, I\{ 
X\geq x \})   }{  Cov(Z,X)  } \\
     &= \frac{  Cov(   Z , \int_\mathcal{X}  I\{ X\geq x\} \text{d}x   ) }{ Cov(Z,X) } \\ 
     &= \frac{Cov(Z,X)}{ Cov(Z,X)} =1
    \end{split}
\end{equation}

Given the instrument-exposure model with linearity and homogeneity so that $X=\alpha Z + g_X(U,\epsilon_X)$ with confounder $U$ and exogenous variable $\epsilon_X$ and w.o.l.g. the instrumental effect $\alpha   >0$. Denote $ g_X(U,\epsilon_X) $ by $E$, we have 
\begin{equation}
    \begin{split}
        W(x)&=  \frac{   [  \mathbb{E}( Z|X \geq x) - \mathbb{E}( Z ) ] \mathbb{P}(X\geq x)   }{Cov(Z,X)} \\
            &= \frac{   [  \mathbb{E}( Z|X \geq x) - \mathbb{E}( Z | X < x) ]  \mathbb{P}(X< x)\mathbb{P}(X\geq x)   }{Cov(Z,X)} \\ 
            &=  \frac{ \mathbb{P}(X< x)\mathbb{P}(X\geq x) }{  \alpha^2 Var(Z) } 
            \left[   \int_\mathcal{Z} z  [    f_Z(z|X \geq x ) -f_Z(z|X \leq x )     ]\text{d}z          \right]
    \end{split}
\end{equation}
Since $  f_Z(  z | X \geq x ) =  \frac{   f_Z(z) f_E( x-\alpha z )   }{  \mathbb{P}(  X \leq x  )  } $, while $ \mathbb{P}( E \geq x-\alpha z )  $ is a monotone non-decreasing function over $z$; similar to $f_Z(  z | X \leq x )  $ with the monotone non-increasing function $ \mathbb{P}( E \leq x-z )  $; hence there exists an constant $z^\ast$ such that $  f_Z(  z | X \geq x ) \geq f_Z(  z | X \leq x )  $ when $ z \geq z^\ast $ and $  f_Z(  z | X \geq x ) \leq f_Z(  z | X \leq x )  $ otherwise. Therefore, 
\begin{equation}
\begin{split}
     \int_\mathcal{Z} z  [    f_Z(z|X \geq x ) -f_Z(z|X \leq x )     ]\text{d}z = &     \int_{\mathcal{Z}<z^\ast} z  [    f_Z(z|X \geq x ) -f_Z(z|X \leq x )     ]\text{d}z +\\
     & \int_{\mathcal{Z}\geq z^\ast} z  [    f_Z(z|X \geq x ) -f_Z(z|X \leq x )     ]\text{d}z  \\
      \geq & \, z^\ast  \int_{\mathcal{Z}<z^\ast}  [    f_Z(z|X \geq x ) -f_Z(z|X \leq x )     ]\text{d}z + \\
      & \, z^\ast  \int_{\mathcal{Z}\geq z^\ast}  [    f_Z(z|X \geq x ) -f_Z(z|X \leq x )     ]\text{d}z  \\
      =& z^\ast \int_{\mathcal{Z}}  [    f_Z(z|X \geq x ) -f_Z(z|X \leq x )     ]\text{d}z =0 
\end{split}    
\end{equation}
which means that $W(x) \geq 0$. This conclusion is easy to extend to the more general case that $  X=m(Z) + g_X(U,\epsilon_X) $ with a monotone instrumental effect function $m(\cdot)$.

In the case that $\{ Z,X  \}$ follows a joint normal distribution that 
\begin{equation}  
\left( \begin{array}{c} 
Z \\ 
X 
\end{array} \right)     \sim \mathcal{N}\left(
\left( \begin{array}{c} 
\mu_Z \\ 
\mu_X
\end{array} \right)  ,
\left( \begin{array}{cc} 
\sigma_Z^2 &  \sigma_{Z,X} \\ 
\sigma_{Z,X} & \sigma_X^2 
\end{array} \right)		\right)   
\end{equation}
which means $$   \mathbb{E}(  Z|X=x ) = \mu_Z+ \frac{\sigma_{Z,X}}{\sigma_X^2 }( x- \mu_X ) $$
We have the weight function

\begin{equation}
\begin{split}
W(s)=&   \frac{1}{\sigma_{Z,X}} \left[ \mathbb{E}( I\{X\geqslant s\}Z ) -\mathbb{E}( I\{X\geqslant s\})\mathbb{E}(Z)  \right]   \\
=&  \frac{1}{\sigma_{Z,X}} \left[  \int_{ [s,\infty )} f_X(x)   \int z  f_{Z|X}(z|x) dz dx  -  \int_{ [s,\infty )} f_X(x)   \int z  f_{Z}(z) dz dx\right]  \\
=&  \frac{1}{\sigma_{Z,X}} \left[ \int_{ [s,\infty )}
f_X(x)   \left[     \mathbb{E}(Z|X=x) - \mathbb{E}(Z)  \right]dx
\right] \\
=&  \frac{1}{\sigma_{Z,X}} \left[ \int_{ [s,\infty )}
f_X(x)   \left[ \frac{\sigma_{Z,X}}{\sigma_X^2 }( x- \mu_X )     \right]dx\right] \\
=&   \frac{1}{\sigma_X^2 } \int_{ [s,\infty )}
f_X(x)    ( x- \mu_X )     dx \\
=& \frac{1}{\sqrt{2 \pi} \sigma_X}   \int_{[s,\infty]} \frac{1}{\sigma_X^2} \exp\left\{  - \frac{  (x-\mu_X)^2 }{  2 \sigma_X^2  }  \right\} (x-\mu_X)  dx\\
=& \frac{1}{\sqrt{2 \pi} \sigma_X} \left. \left\{   -  \exp\left(  - \frac{  (x-\mu_X)^2 }{  2 \sigma_X^2  }  \right) \right \}  \right|_s^\infty  = 
\frac{1}{\sqrt{2 \pi} \sigma_X} \exp\left\{  - \frac{  (s-\mu_X)^2 }{  2 \sigma_X^2  }  \right\} = f_X(s)
\end{split}
\end{equation}
which means $W(s)$ is the marginal normal density of $X$. \hfill\BlackBox

\newpage
\mysection{Appendix E: Linearity testing}\label{linearity_test}
The null hypothesis of effect linearity implies that $ h'(x) $ remains constant for all $x$, equaling the average treatment effect $\beta$. Denoting the stratum-specific genetic association estimators for the exposure and the outcome as $ \hat{\alpha}_k$ and $  \hat{\theta}_k $, respectively, we obtain the asymptotic result for each stratum:
\begin{equation} \label{linearity_con}
    \frac{   \hat{\theta}_k }{  \hat{\alpha}_k }  \overset{ p }{\to}  \beta \qquad \text{for }\, s = 1,\ldots, K.
\end{equation}
Since stratification results in multiple strata ($K > 1$), we can leverage over-identification or coherence approaches to test for effect linearity based on Equation \eqref{linearity_con} \citep{hansen1982large}. A common approach is to use Cochran’s $Q$-based statistic:
\begin{equation}\label{Q_statistic}
    Q(  \beta ) := \sum_{k=1}^K    \left\lbrace  
\frac{      \left(   \hat{\theta}_k   -   \beta \hat{{\alpha}}_k   \right) ^2    }{
	s.e.(  \hat{\theta}_k  )^2    +  \beta^2  s.e.(  \hat{\alpha}_k  )^2  }
\right\rbrace,  
\end{equation}
which is a simplified version (e.g., ignoring the correlation between $\hat{\theta}_k$ and $\hat{\alpha}_k$) commonly used in meta-analysis and Mendelian randomization \citep{hardy1998detecting,greco2015detecting}. 

In practice, we compare the observed $Q$ statistic, $Q(  \hat{\beta} )$, where $ \hat{\beta}  = \argmin_{\beta} Q(\beta)$ or any fundamental IV estimator under the linearity assumption, to its approximated $\chi^2$ distribution with degrees of freedom $K-1$ under the null hypothesis. The $Q$ statistic can also be interpreted as the maximal profile likelihood in an error-in-variable model that accounts for measurement error in $\alpha_k$, treating $\alpha_k$ as a nuisance parameter \citep{zhao2020statistical}, or as the residual sum of squares from an inverse-variance weighted (IVW) regression of $\hat{\theta}_k$ on $\hat{\alpha}_k$ using second-order weights \citep{bowden2019improving}. 
Rejection of the $Q$ test provides evidence for effect nonlinearity.

\mysection{Appendix F: Complicated model scenario and their model properties}\label{invalid_scenarios}
In this section, we extend our analysis to a more complex model 
\begin{equation}
    Y = h(X, Z, \epsilon) + g(U, Z, \epsilon).
\end{equation}
We assume that the instrument $Z$ can be used for stratification, ensuring that Features 1 and 2 hold. Under this assumption, we construct the SoF model introduced before within each stratum. 
For reference, we define the simplest model, $Y = h(X) + U$, as \textbf{Scenario 0}. Below, we outline several complex model scenarios, each designed for specific objectives, along with illustrative examples. 

\medskip
\noindent 
{\bf Complex scenario 1.} {\it When there exists exogenous effect modifiers such that 
\begin{equation}
    Y=h(X,\epsilon) + g(U,\epsilon) 
\end{equation}
with regular conditions, all the ideas introduced above can be applicable to the objectives described by the mean effect function, $h(x):= \mathbb{E}_{\epsilon}( h(x,\epsilon) ) $, or the modifier-controlled effect function, $h(x, \epsilon^\ast)$, for a modifier layer $\epsilon^\ast$ when the modifier is measurable. For example, sex can be the measured modifier of the effect function, hence sex-strata may be first constructed for any downstream sex-specific nonlinear effect analysis. In addition, sex generally cannot be the downstream effect of the instrument and the confounder, hence selection on sex should not induce collider bias.}

\medskip
\noindent
{\bf Complex scenario 2.} {\it In the case that there exists the additive instrumental direct effect on the outcome (one typical case that violation of the exclusion restriction), in the sense that
\begin{equation}
    Y=h(X,\epsilon) + g(Z,U,\epsilon) 
\end{equation}
if we have the decomposition that $g(Z,U,\epsilon) =g_1(Z,\epsilon)   + g_2(U,\epsilon)$ so that the instrumental direct effect is not modified by confounders, one can still utilize the stratification and SoF model to test the effect linearity via testing the constant value among the stratum-specific IV estimates. The decomposition property is a quite common condition used in invalid IV models, like \cite{bowden2015mendelian,tchetgen2021genius}. 
The decomposition property allows that the linearity test is now robust to certain invalid IV cases, and also indicates that certain confounders in some applications can serve as the instrument for linearity testing. For example, in the exposure scenario where one believes the rank-preserving assumption between age and exposure is plausible and the age also has a direct effect on the outcome, one may still able to treat age as the 'instrument' to test the linear effect of the exposure on the outcome. Similarly in Mendelian randomization, some genes may have the direct effect on the outcome rather than through the exposure, which in genetics is called horizontal pleiotropy \citep{paaby2013many}; if one believes the genetic direct effect is not modified by the confounder, they are still valid for effect linearity testing even though they are invalid in IV context.}

\medskip
\noindent
{\bf Complex scenario 3.} {\it In the case that the instrument modifies the effect of the exposure on the outcome  (i.e. another typical case that violation of the exclusion restriction), in the sense that
\begin{equation}
     Y=h(X,Z,\epsilon) + g(U,\epsilon) 
\end{equation}
if we have the factorization that $ h(X,Z,\epsilon)=h_0(X,\epsilon) f(Z,\epsilon) $, one can still utilize the stratification and SoF model to test the effect linearity regarding $h_0(\cdot)$ via testing the constant value among the stratum-specific IV estimates (details see the previous appendix). Note that this conclusion can be combined with the further decomposition scenario that $g(Z,U,\epsilon) =g_1(Z,\epsilon)   + g_2(U,\epsilon)$. For example, in some applications where the age (treated as an invalid instrument) has both the additive direct effect and the modified effect on the outcome, the linearity of the exposure effect can still be tested, given that the stratification assumption and the decomposition/factorization model assumption are plausible.    }

% constant additive direct IV effect on the outcome
% confounder as the IV  (eg U = age)
% factorized case: h(X,Z)=h(X)g(Z)
% all for nonlinear effect testing
% original change-point in h(x)

% but cannot deal with the non-exchangeability case that Z |\| U  as in DR stratification, Sum_s  cov_s(Z,U) = cov(Z,U) so cannot ensure that the stratification cannot make constant    

We prove the properties with a more complex model. Since the variable $\epsilon$ is exogenous and any stratification based on $(Z,X)$ will not change the distribution of $\epsilon$, according to the previous appendix we will drop it for notation simplification. 

Given the model with decomposition
\begin{equation}
    Y = h(X) + g_1(Z) + g_2(U)
\end{equation}
Assume that stratification can be implemented to $(Z,X)$ to construct $K$ strata, and the stratum-specific exchangeability $ Z \indep U | S $ hold. The stratum-specific IV estimator has the asymptotic value
\begin{equation}
    \begin{split}
        \frac{  Cov_k( Z, Y ) }{  Cov_k(  Z, X )  }  =&   \frac{  Cov_k( Z, \int_\mathcal{X}h'(x)I\{X \geq x\}  \text{d}x + g_1(Z)  ) }{  Cov_k(  Z, X )  } \\
        =& \int_\mathcal{X} h'(x) \frac{  Cov_k(Z, I\{X\geq x\}) }{ Cov_k(Z,X)} \text{d}x + \frac{ Cov_k( Z, g_1(Z) ) }{ Cov_k(Z,X) }   \\
        =&  \int_\mathcal{X} h'(x)  W_k(x) \text{d}x +  \frac{c_0}{ Cov_k( Z,X ) }
    \end{split}
\end{equation}
where $Cov_k( Z, g_1(Z) )  =c_0 $ is a constant regardless of strata index due to the stratification property that the distribution of the instrument is constant across strata (this also indicates that $ Var_k(Z) $ is constant over $s$). Therefore, under the null hypothesis of effect linearity that $h'(x)= c_1$, we have the condition, similar to meta-analysis
\begin{equation}
    \hat{\theta}_k = c_0 + c_1 \alpha_k +  \epsilon_k \qquad \text{with} \, \,  \epsilon_k \sim \mathcal{N}(  0, s.e.(  \hat{\theta}_k )^2 )
\end{equation}
With the over-identification idea, we can test the model via either the simplified form treating $ \hat{\alpha}_k = \alpha_k $ (i.e. no measured error) 
or consider the uncertainty of $ \{ \hat{\alpha}_k \} $ under errors-in-variable model framework with the Cochran's Q test statistic
\begin{equation}
    Q(  c_0, c_1 )  = \sum_{s=1}^K    \left\lbrace  
\frac{      \left(   \hat{\theta}_k   -   c_0 - c_1 \hat{{\alpha}}_k   \right) ^2    }{
	s.e.(  \hat{\theta}_k  )^2    +  c_1^2  s.e.(  \hat{\alpha}_k  )^2  }
\right\rbrace,  
\end{equation} 
the Q test statistic $  Q(  \hat{c}_0, \hat{c}_1 )  $, where $  ( \hat{c}_0, \hat{c}_1 )  = \argmin_{( c_0,c_1 )}  Q(  c_0, c_1 ) $ under null has the approximated Chi-squared distribution with degree-of-freedom $ K - 2$.

Given the model with the factorization
\begin{equation}
    Y =h(X) f(Z)  + g(U)
\end{equation}
We have
\begin{equation}
    \begin{split}
       \theta_k =  \frac{Cov_k(Z, Y )}{ Var_k(Z) } =&   \frac{   \int_\mathcal{X}Cov_k(  Z, f(Z) h'(x) I\{ X \geq x \} )  \text{d}x  }{ Var_k(Z)  }    \\
     =&    \beta  \frac{   \int_\mathcal{X}Cov_k(  Z, f(Z)  I\{ X \geq x \} )  \text{d}x  }{ Var_k(Z)  }   \quad \text{under the null that $h'(x)\equiv \beta$}  \\
     =&  \beta  \frac{   Cov_k(  Z, f(Z)(X-l) )  }{ Var_k(Z)  }  \quad \text{where $X\geq l$ surely} \\
     =&   \beta  \frac{   Cov_k(  Z, f(Z)X ) }{ Var_k(Z)  }   - l\, \beta \frac{Cov_k(   Z, f(Z))}{Var_k(Z)}   \\
     =& \beta  \frac{   Cov_k(  Z, f(Z)X ) }{ Var_k(Z)  }   - l\, \beta \,c
    \end{split}
\end{equation}
If one wish to make assumption on $f(Z)$ or for binary instrument $Z$ such that $ f(Z)=1+ \tau Z $ nonparametrically, the equation can be further expressed as
\begin{equation}
    \begin{split}
         \theta_k = \frac{Cov_k(Z, Y )}{ Var_k(Z) } = &  \beta  \frac{   Cov_k(  Z, X + \tau Z X ) }{ Var_k(Z)  }   - l\, \beta \,c  \\
         =& \beta   \frac{   Cov_k(  Z, X ) }{ Var_k(Z)  }   + \beta \, \tau \frac{   Cov_k(  Z, ZX ) }{ Var_k(Z)  }  - l\, \beta \,c   \\
         =& \beta   \alpha_k + c_1 \gamma_k + c_0
    \end{split}
\end{equation}
where $ c_0 = l\, \beta \,c    $, $ c_1 = \beta \, \tau $ and $  \gamma_k  $ is the $s$-th stratum-specific instrumental association with $ZX$. Hence, the effect linearity test using Cocharan's Q is based on the Q statistic
\begin{equation}
    Q(c_0,c_1,\beta) = \sum_{s=1}^K    \left\lbrace  
\frac{      \left(   \hat{\theta}_k   -   c_0 - c_1 \hat{{\gamma}}_k  - \beta \hat{\alpha}_k   \right) ^2    }{
	s.e.(  \hat{\theta}_k  )^2    +  c_1^2  s.e.(  \hat{\gamma}_k  )^2 + \beta^2 s.e.(  \hat{\alpha}_k  )^2  }
\right\rbrace,  
\end{equation}
the Q test statistic $  Q(  \hat{c}_0, \hat{c}_1, \hat{\beta} )  $, where $  ( \hat{c}_0, \hat{c}_1,\hat{\beta} )  = \argmin_{( c_0,c_1,\beta )}  Q(  c_0, c_1, \beta ) $ under null has the approximated Chi-squared distribution with degree-of-freedom $ K - 3$.

\mysection{Appendix G: Complete algorithm of the SSS framework}

\begin{figure}[H]  % 用 figure 包裹，使算法和 Remark 视为整体
\begin{minipage}{\textwidth}  % 让算法和 Remark 一起移动
\begin{algorithm}[H]
\DontPrintSemicolon  % 去掉每行后的分号

\textbf{Input:} individual-level data $(Z_i, X_i, Y_i)_{i=1,\ldots,n}$

\textbf{Output:} results for Objective 1-4 (e.g. estimated effect function, the predicted change-point)

\vspace{0.5em}

$ K \leftarrow  $ the strata number

$ P \leftarrow  $ the basis-functions number in parametric fitting or quantile number in the change-point model

$ L \leftarrow $ the iteration number in SuSiE fitting

\vspace{0.5em}

\textbf{First `S': Stratification}

$  \mathcal{S} \leftarrow  $ strata set (size of $K$) by the doubly-ranked or residual stratification by Algorithm \ref{R_algorithm} or \ref{DR_algorithm}

\For{each stratum $k \leftarrow 1$ \KwTo K}{
 Obtain the point summary statistics from the $k$-th strata: 
   \begin{itemize}[noitemsep, nosep]
       \item estimated instrumental association with $X$ and its s.e. : $ \hat{\alpha}_k $, $ s.e.(  \hat{\alpha}_k )  $

       \item estimated instrumental association with $Y$ and its s.e. : $ \hat{\theta}_k $, $ s.e.(  \hat{\theta}_k )  $
   \end{itemize}
 
 Obtain the weight function from the $k$-th strata: 
 $$     \hat{W}_k(x)   = \frac{  \widehat{Cov}_k( Z, I\{ X \geq x  \} )  }{    \widehat{Cov}_k( Z,X ) }     $$
 
}

(Optional) linearity testing using $ \{ \hat{\alpha}_k $, $ s.e.(  \hat{\alpha}_k ), \hat{\theta}_k $, $ s.e.(  \hat{\theta}_k )   \}$, e.g., via $Q$ test in Eq \eqref{Q_statistic}

\vspace{0.5em}
\textbf{Second `S': Scalar-on-function or scalar-on-scalar regression}

\If{basis function assumption $ \{  \phi_l(x) \} $ made}{

construct the covariates: $   \langle   \phi_p , \hat{W}_k    \rangle   $ for all $p$ and $k$

construct the scalar-on-function or scalar-on-scalar regression using the covariates above and  $ \{ \hat{\alpha}_k $, $ s.e.(  \hat{\alpha}_k ), \hat{\theta}_k $, $ s.e.(  \hat{\theta}_k )   \}_{k=1,\ldots,K} $

\If{regularization required}{
   Determine the tuning parameter $\lambda$, e.g., using GCV
}

   Perform parametric fitting

\textbf{Return:} the fitted effect function $ \hat{h}( x ) $ and its pointwise confidence interval
   
}

\vspace{0.5em}
\textbf{Third `S': SuSiE (only with scalar-on-function regression)}

Construct the covariates: $   \langle   I\{ x\geq t_p \} , \hat{W}_k    \rangle  $ for all $p$ and $k$

Build the change-point model (piecewise constant effect model) 

Conduct sum of single effect fitting, and obtain
\begin{itemize}[noitemsep,nosep]
    \item $ \boldsymbol{\pi}^\ast $, the matrix posterior inclusion probabilities

    \item $   \boldsymbol{\mu}^\ast  $, the matrix of posterior mean

    \item $  \boldsymbol{\sigma}^\ast $, the matrix of the posterior standard deviation
\end{itemize}

\textbf{Return:} the predicted value or/and credible set of the changepoint based on $\boldsymbol{\pi}^\ast$

%\If{basis function assumption $ \{  \phi_l(x) \} $ made}{aaa}

\textbf{Return:} the predicted effect function $ \hat{h}(x) $ and its pointwise credibility set based on $  \boldsymbol{\pi}^\ast , \boldsymbol{\mu}^\ast , \boldsymbol{\sigma}^\ast  $

 \caption{The three `S' framework (SSS) for nonlinear effect analysis}
 \label{alg_complete}

\end{algorithm}

%\parbox{\textwidth}{%
%\vspace{-0.3cm}

%\noindent\textsuperscript{4} the predicted change-point values may be further used to construct the known basis function for parametric fitting
%}
\end{minipage}% 让算法和 Remark 一起移动
\end{figure}

\newpage
\mysection{Appendix H: Supplementary simulation}\label{app_sim}
We repeat the simulation from Part I using a nonlinear effect function. The setup remains similar, and we consider the following scenarios for the instrument type and outcome structural models with a nonlinear effect:
\begin{align}
    \begin{array}{l} 
      \text{Scenario 1:} \\
       \text{Scenario 2:}  \\
       \text{Scenario 3:} \\
       \text{Scenario 4:} 
    \end{array} 
    &\quad \begin{array}{l} 
       Z \sim \text{Bernoulli}(0.5)-0.5    \\
        Z \sim \text{Bernoulli}(0.5)-0.5 \\
        Z \sim \mathcal{N}(0,1)    \\
        Z \sim \mathcal{N}(0,1)    
    \end{array} 
    &\quad \begin{array}{l} 
        Y=I\{ X >0 \}  + U + \epsilon_Y   \\
       Y=I\{ X >0 \} + |U| + \epsilon_X^2 + 2 |U| |\epsilon_X|  +  \epsilon_Y   \\
      Y=I\{ X >0 \} + U + \epsilon_Y     \\
      Y=I\{ X >0 \} + |U| + \epsilon_X^2 + 2 |U| |\epsilon_X|  +  \epsilon_Y   
    \end{array}
\end{align}
where the effect shape function is defined as \( h(x) = I\{x>0\} \), which is nonlinear with a change-point at \( x = 0 \). We consider the same methods (M1–M5) as in the main text but exclude M6 (DeepIV) and M7 (KernelIV) in this supplementary simulation due to their significantly long runtime. For the originally oracle methods (M1–M3), we continue using a polynomial basis function of up to the second order because such a polynomial is a common choice in practice. While it may not perfectly capture the nonlinear effect \( h(x) = I\{x>0\} \), its second-order property allows it to approximate the shape reasonably well across different exposure levels.

The MSE results are provided in Table \ref{methods_comparsion_app}. The 'oracle' methods (M1–M3) no longer achieve the best MSE performance in many quantile cases, particularly for quantiles below \(50\%\), where the effect is zero. In contrast, the SSS method (M5) demonstrates the best MSE performance in more scenarios, despite having greater estimation uncertainty due to its model flexibility. This can be attributed to the SSS method’s ability to accommodate change-point effect patterns, whereas other methods rely solely on polynomial approximations of the underlying effect shape. When the polynomial order is low, as in M1–M3, it often fails to accurately approximate the effect shape in lower exposure quantiles. This may explain why many nonlinear methods tend to estimate a J-shaped effect function, even in cases where the true effect remains zero until a certain change-point.

We also consider another nonlinear effect shape case, so that 
\begin{align}
    \begin{array}{l} 
      \text{Scenario 1:} \\
       \text{Scenario 2:}  \\
       \text{Scenario 3:} \\
       \text{Scenario 4:} 
    \end{array} 
    &\quad \begin{array}{l} 
       Z \sim \text{Bernoulli}(0.5)-0.5    \\
        Z \sim \text{Bernoulli}(0.5)-0.5 \\
        Z \sim \mathcal{N}(0,1)    \\
        Z \sim \mathcal{N}(0,1)    
    \end{array} 
    &\quad \begin{array}{l} 
        Y=\exp\{ 0.5 X \}  + U + \epsilon_Y   \\
       Y=\exp\{ 0.5 X \}+ |U| + \epsilon_X^2 + 2 |U| |\epsilon_X|  +  \epsilon_Y   \\
      Y=\exp\{ 0.5 X \} + U + \epsilon_Y     \\
      Y=\exp\{ 0.5 X \} + |U| + \epsilon_X^2 + 2 |U| |\epsilon_X|  +  \epsilon_Y   
    \end{array}
\end{align}
The MSE results are provided in Table \ref{methods_comparsion_app2}. Due to the good approximation ability of the second-order polynomial to the underlying exponential effect, the MSE results for M1–M4 are similar to those in the main text using a linear effect. The main difference is that the SSS method (M5) exhibits better MSE performance in many cases.  

\newpage
\begin{table}[H]
    \centering
    %\resizebox{\linewidth}{!}{ % 让表格等宽
    \begin{tabular}{c ccccc ccccc}
        \toprule
        & \multicolumn{5}{c}{Scenario 1} & \multicolumn{5}{c}{Scenario 3} \\
        & \multicolumn{5}{c}{(binary IV, simple confounding)} & \multicolumn{5}{c}{(continuous IV, simple confounding)} \\
         \cmidrule(lr){2-6} \cmidrule(lr){7-11}
          &   M1 & M2 & M3 & M4 & M5 & M1 & M2 & M3 & M4 & M5  \\
        \midrule
  10\% & 0.519 & 0.460 & 0.599 & 0.720 & \textbf{0.125} & 0.250 & 0.123 & 0.281 & 0.292 & \textbf{0.042} \\ 
  30\% & 0.130 & 0.105 & 0.138 & 0.112 & \textbf{0.055} & 0.094 & 0.063 & 0.098 & 0.082 & \textbf{0.025} \\ 
  50\% & 0 & 0 & 0 & 0 & 0 & 0 & 0 & 0 & 0 & 0 \\ 
  70\% & 0.182 & \textbf{0.140} & 0.144 & 0.308 & 0.182 & 0.101 & 0.071 & 0.100 & 0.135 & \textbf{0.069} \\ 
  90\% & 0.707 & \textbf{0.449} & 0.631 & 1.584 & 0.892 & 0.276 & \textbf{0.118} & 0.299 & 0.458 & 0.227 \\ 
        & \multicolumn{10}{c}{} \\ % 空行
         & \multicolumn{5}{c}{Scenario 2} & \multicolumn{5}{c}{Scenario 4} \\
        & \multicolumn{5}{c}{(binary IV, complex confounding)} & \multicolumn{5}{c}{(continuous IV, complex confounding)} \\
         \cmidrule(lr){2-6} \cmidrule(lr){7-11}
        & M1 & M2 & M3 & M4 & M5 & M1 & M2 & M3 & M4 & M5    \\
        \midrule
  10\% & 3.856 & 27.509 & 1.387 & 1.163 & \textbf{0.189} & 3.180 & 26.206 & 0.626 & 0.636 & \textbf{0.146} \\ 
  30\%& 0.154 & 0.771 & 0.231 & 0.125 & \textbf{0.073} & \textbf{0.036} & 0.504 & 0.134 & 0.100 & 0.058 \\ 
  50\% & 0 & 0 & 0 & 0 & 0 & 0 & 0 & 0 & 0 & 0 \\ 
  70\% & \textbf{0.157} & 0.838 & 0.252 & 0.498 & 0.305 & \textbf{0.035} & 0.535 & 0.127 & 0.325 & 0.182 \\ 
  90\%  & 4.024 & 28.823 & \textbf{1.658} & 3.046 & 1.765 & 3.125 & 26.805 & \textbf{0.549} & 1.625 & 0.815 \\ 
        \bottomrule
    \end{tabular}
    %}
    \vspace{1em}
    \caption{\normalfont The MSE results of the effect function over several exposure quantiles across multiple nonlinear IV methods (denoted by M1-M5) under four different model scenarios in 1000 simulations. The underlying effect is $ h(x) = I\{  x>0 \} $.
    The objective of the effect function is defined to be zero when the exposure level is $0$, corresponding to the 50\% quantile in each scenario so the MSE is $0$. 
    For each scenario and quantile value, the minimal MSE value is highlighted. M1: Oracle control function. M2: Oracle IV-regression. M3: Stratification with oracle SoS regression. M4: PolyMR. M5: SSS. The oracle methods use the basis function as the polynomial up to second order in fitting.}
    \label{methods_comparsion_app}
\end{table}

\begin{table}[H]
    \centering
    %\resizebox{\linewidth}{!}{ % 让表格等宽
    \begin{tabular}{c ccccc ccccc}
        \toprule
        & \multicolumn{5}{c}{Scenario 1} & \multicolumn{5}{c}{Scenario 3} \\
        & \multicolumn{5}{c}{(binary IV, simple confounding)} & \multicolumn{5}{c}{(continuous IV, simple confounding)} \\
         \cmidrule(lr){2-6} \cmidrule(lr){7-11}
          &   M1 & M2 & M3 & M4 & M5 & M1 & M2 & M3 & M4 & M5  \\
        \midrule
  10\%  & 0.421 & 0.417 & 0.428 & 0.659 & \textbf{0.223} & \textbf{0.067} & 0.119 & 0.099 & 0.209 & 0.151 \\ 
  30\% & 0.071 & 0.071 & 0.061 & 0.103 & \textbf{0.060} & \textbf{0.016} & 0.025 & 0.017 & 0.038 & 0.038 \\  
  50\% & 0 & 0 & 0 & 0 & 0 & 0 & 0 & 0 & 0 & 0 \\ 
  70\% & 0.075 & 0.073 & \textbf{0.065} & 0.122 & 0.084 & 0.024 & 0.028 & \textbf{0.021} & 0.039 & 0.038 \\ 
  90\% & 0.427 & 0.417 & \textbf{0.459} & 0.939 & 0.583 & \textbf{0.111} & 0.131 & 0.122 & 0.250 & 0.159 \\
        & \multicolumn{10}{c}{} \\ % 空行
         & \multicolumn{5}{c}{Scenario 2} & \multicolumn{5}{c}{Scenario 4} \\
        & \multicolumn{5}{c}{(binary IV, complex confounding)} & \multicolumn{5}{c}{(continuous IV, complex confounding)} \\
         \cmidrule(lr){2-6} \cmidrule(lr){7-11}
        & M1 & M2 & M3 & M4 & M5 & M1 & M2 & M3 & M4 & M5    \\
        \midrule
  10\% & 5.429 & 25.954 & 1.450 & 1.376 & \textbf{0.294} & 4.743 & 24.875 & 0.456 & 0.635 & \textbf{0.221} \\ 
  30\% & 0.232 & 0.877 & 0.175 & 0.195 & \textbf{0.084} & 0.113 & 0.659 & \textbf{0.055} & 0.107 & 0.057 \\ 
  50\% & 0 & 0 & 0 & 0 & 0 & 0 & 0 & 0 & 0 & 0 \\ 
  70\% & 0.396 & 1.289 & 0.222 & 0.257 & \textbf{0.159} & 0.259 & 1.033 & \textbf{0.059} & 0.124 & 0.082 \\ 
  90\% & 6.809 & 29.729 & 1.945 & 2.384 & \textbf{1.400} & 5.856 & 28.062 & \textbf{0.468} & 0.910 & 0.526 \\ 
        \bottomrule
    \end{tabular}
    %}
    \vspace{1em}
    \caption{\normalfont The MSE results of the effect function over several exposure quantiles across multiple nonlinear IV methods (denoted by M1-M5) under four different model scenarios in 1000 simulations. The underlying effect is $ h(x) = \exp\{ 0.5 x \} $.}
    %\vspace{-2em}
    \label{methods_comparsion_app2}
\end{table}

\newpage
\mysection{Appendix I: UK Biobank real application with females}%\label{app_real}
We replicated the real-data analysis from the main text in the female sample to investigate the nonlinear effect of alcohol intake on SBP. The genetic association with alcohol consumption is known to differ between males and females (e.g., cultural factors may reduce drinking among females), so the stratification assumptions may be violated in a mixed-sex sample. In addition, the causal effect of alcohol on many traits may be heterogeneous across sexes since women and men may metabolize alcohol differently \citep{graham1998should}. Therefore, this analysis should not be viewed as a strict replication of the male-only analysis presented in the main text.

We used the same setup as in the main text. Specifically, we analyzed 150{,}600 female individuals of European ancestry and for each participant we constructed a weighted genetic score from the same 93 variants used in the male text. We adopted doubly-ranked stratification, forming 10 strata for presenting stratum-specific results and 100 strata for estimation. We also treated age as a negative control outcome. For each stratum, we obtained point IV estimates and the corresponding functional weight functions. We then built a scalar-on-function regression and considered 95 change-point candidates corresponding to the lower 95\% quantiles of the original alcohol-intake distribution in females,
$
 \{\, t_p := \hat{F}^{-1}_X(p/100) \,\}_{p=0,1,\ldots,95}.
$. We fitted the regression using SuSiE with the iteration number $L=10$ and obtained posterior inclusion probabilities for the change-points. We further computed the pointwise estimate and the 95\% credible interval for the effect shape function
\(h(x):=\mathbb{E}\{Y(x)-Y(0)\}\).
Complete results are shown in Figure~\ref{female_results}. Overall, the findings in females mirror those in males, supporting a positive effect of alcohol intake on SBP, with a comparable credible-interval range for the effect change-point location: $(0, 3.271)$ in females versus $(0, 3.286)$ in males under the same setting.
\vspace{-2em}

\begin{figure}[H]  % 'h' 表示图片位置，h 表示尽可能放在当前位置
    \centering  % 图片居中
    \includegraphics[width=1.0\textwidth]{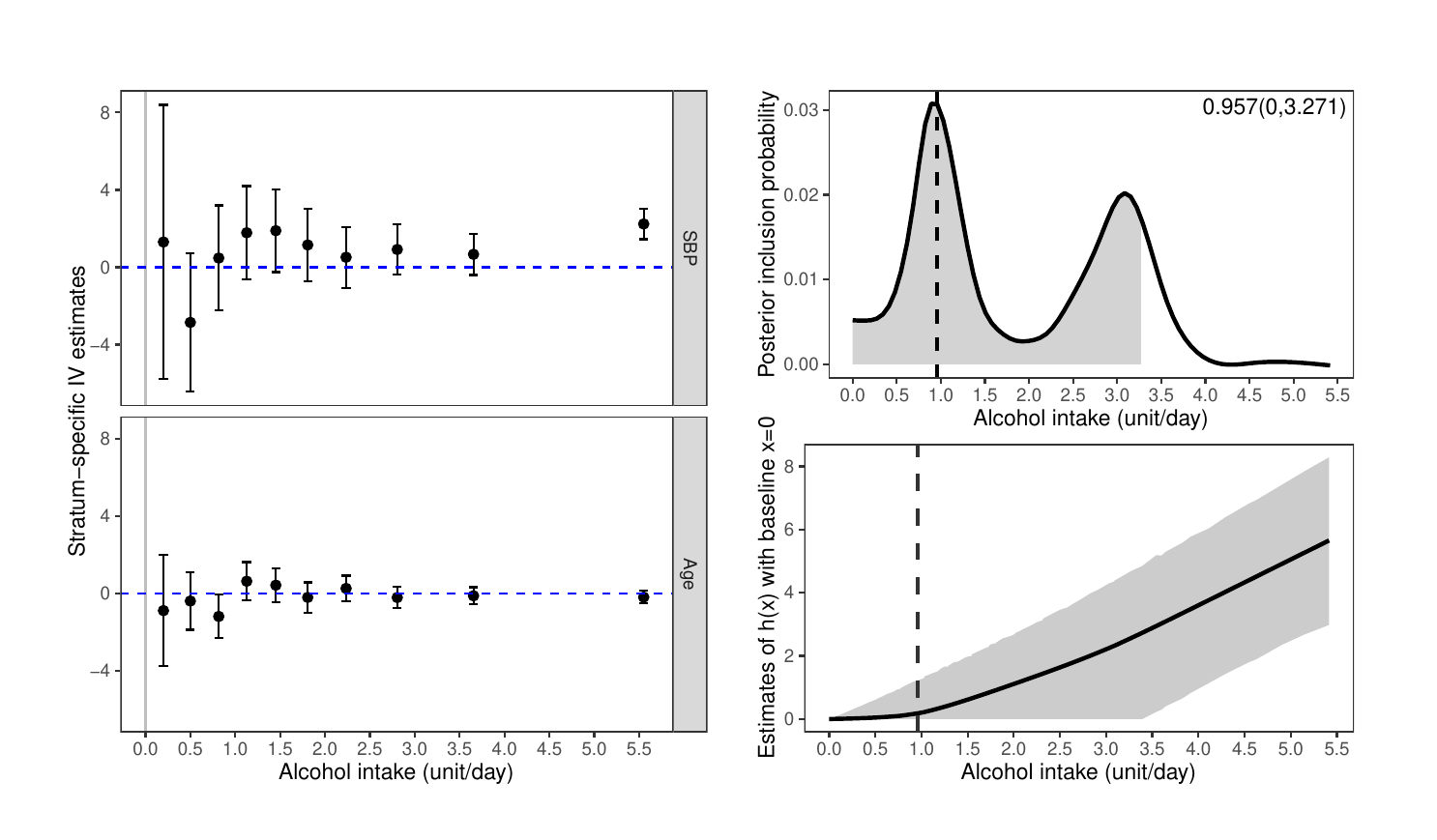}  % 图片路径和大小
    \vspace{-2em}
    \caption{\normalfont Left: Stratum-specific IV estimates for SBP and Age, obtained using doubly-ranked stratification with 10 strata (shown here for clarity of presentation). \\Right: Change-point and effect-function inference results from the “SSS” framework, using doubly-ranked stratification with 100 strata and Sum-of-Single-Effect (SuSiE) estimation. Upper panel: Posterior inclusion probabilities of change-points and the $95\%$ credible interval for the change-point location. Lower panel: Estimated effect function with its $95\%$ credible interval. The vertical dashed line marks the posterior mode of the change-point. Only one change-point was detected.}   % 图片标题
    \vspace{-2em}
    \label{female_results}  % 图片标签，方便引用
\end{figure}

\newpage
\mysection{Appendix J: Supplementary results of the real application}\label{app_real}
We conducted effect shape estimation for the real application using other common choices for the number of strata, namely $K=10$ and $K=50$. The estimation results are presented in Figures \ref{real_Fig3_app2} and \ref{real_Fig3_app3}. The results are highly similar for both choices of $K$, indicating that our findings are insensitive to the number of strata.

\begin{figure}[H]  % 'h' 表示图片位置，h 表示尽可能放在当前位置
    \centering  % 图片居中
    \includegraphics[width=0.8\textwidth]{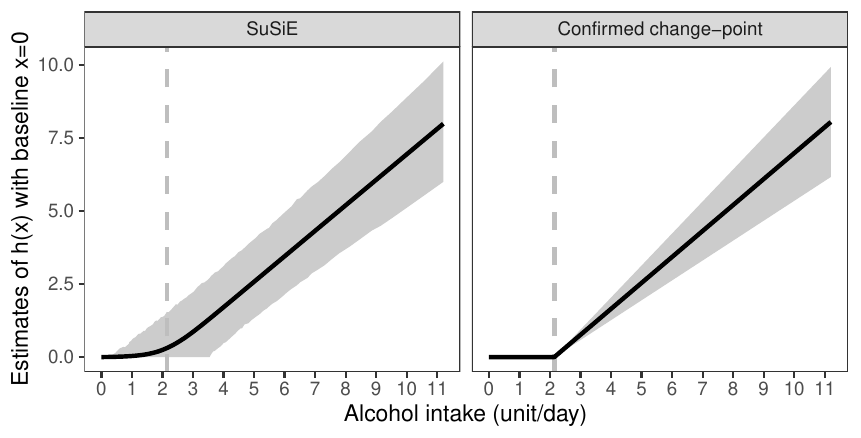}  % 图片路径和大小
    \vspace{-1em}
    \caption{\normalfont The estimate of the effect shape $h(x)$ using two fitting strategies with the number of strata $\boldsymbol{K=10}$: (left panel) SuSiE fitting where the point estimate $\hat{h}(x)$ is the posterior mean and the interval estimate is the $95\%$ credible interval; and (right panel) the parametric fitting with the confirmed change-point ($x=2.142$, the dashed vertical line, using the posterior mode) for basis functions, and the interval estimate is the $95\%$ confidence interval. The effect shape $h(x)$ is defined with the baseline value $x=0$ so that $h(0)=0$.}   % 图片标题
    \label{real_Fig3_app2}  % 图片标签，方便引用
\end{figure}

\begin{figure}[H]  % 'h' 表示图片位置，h 表示尽可能放在当前位置
    \centering  % 图片居中
    \vspace{-2em}
    \includegraphics[width=0.8\textwidth]{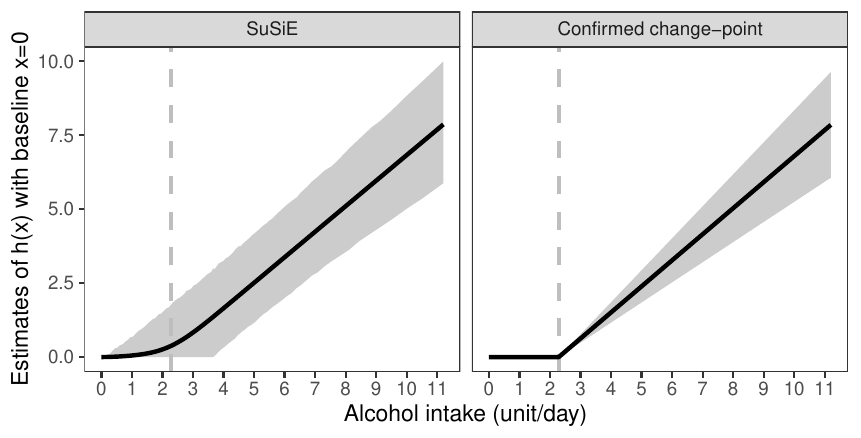}  % 图片路径和大小
    \vspace{-1em}
    \caption{\normalfont The estimate of the effect shape $h(x)$ using two fitting strategies with the number of strata $\boldsymbol{K=50}$: (left panel) SuSiE fitting where the point estimate $\hat{h}(x)$ is the posterior mean and the interval estimate is the $95\%$ credible interval; and (right panel) the parametric fitting with the confirmed change-point ($x=2.286$, the dashed vertical line, using the posterior mode) for basis functions, and the interval estimate is the $95\%$ confidence interval. The effect shape $h(x)$ is defined with the baseline value $x=0$ so that $h(0)=0$.}   % 图片标题
    %\vspace{-1em}
    \label{real_Fig3_app3}  % 图片标签，方便引用
\end{figure}

%\newpage
%\begingroup
%\small
%\setlength{\bibsep}{0pt} % 如果你用了 natbib，有效
%\linespread{1.0}\selectfont
%\bibliographystyle{plainnat}
%\bibliography{sample}
%\endgroup
\newpage
\begingroup
\setlength{\bibsep}{1pt}  % Set the space between entries to 0pt
\bibliography{sample}

\begin{thebibliography}{63}
\providecommand{\natexlab}[1]{#1}
\providecommand{\url}[1]{\texttt{#1}}
\expandafter\ifx\csname urlstyle\endcsname\relax
  \providecommand{\doi}[1]{doi: #1}\else
  \providecommand{\doi}{doi: \begingroup \urlstyle{rm}\Url}\fi

\bibitem[Andrews et~al.(2019)Andrews, Stock, and Sun]{andrews2019weak}
Isaiah Andrews, James~H Stock, and Liyang Sun.
\newblock Weak instruments in instrumental variables regression: Theory and practice.
\newblock \emph{Annual Review of Economics}, 11:\penalty0 727--753, 2019.

\bibitem[Angrist and Krueger(1991)]{angrist1991does}
Joshua~D Angrist and Alan~B Krueger.
\newblock Does compulsory school attendance affect schooling and earnings?
\newblock \emph{The Quarterly Journal of Economics}, 106\penalty0 (4):\penalty0 979--1014, 1991.

\bibitem[Astle et~al.(2016)Astle, Elding, Jiang, Allen, Ruklisa, Mann, Mead, Bouman, Riveros-Mckay, Kostadima, et~al.]{astle2016allelic}
William~J Astle, Heather Elding, Tao Jiang, Dave Allen, Dace Ruklisa, Alice~L Mann, Daniel Mead, Heleen Bouman, Fernando Riveros-Mckay, Myrto~A Kostadima, et~al.
\newblock The allelic landscape of human blood cell trait variation and links to common complex disease.
\newblock \emph{Cell}, 167\penalty0 (5):\penalty0 1415--1429, 2016.

\bibitem[Bowden et~al.(2015)Bowden, Davey~Smith, and Burgess]{bowden2015mendelian}
Jack Bowden, George Davey~Smith, and Stephen Burgess.
\newblock Mendelian randomization with invalid instruments: effect estimation and bias detection through egger regression.
\newblock \emph{International journal of epidemiology}, 44\penalty0 (2):\penalty0 512--525, 2015.

\bibitem[Bowden et~al.(2019)Bowden, Del Greco~M, Minelli, Zhao, Lawlor, Sheehan, Thompson, and Davey~Smith]{bowden2019improving}
Jack Bowden, Fabiola Del Greco~M, Cosetta Minelli, Qingyuan Zhao, Debbie~A Lawlor, Nuala~A Sheehan, John Thompson, and George Davey~Smith.
\newblock Improving the accuracy of two-sample summary-data mendelian randomization: moving beyond the nome assumption.
\newblock \emph{International journal of epidemiology}, 48\penalty0 (3):\penalty0 728--742, 2019.

\bibitem[Brinch et~al.(2017)Brinch, Mogstad, and Wiswall]{brinch2017beyond}
Christian~N Brinch, Magne Mogstad, and Matthew Wiswall.
\newblock Beyond late with a discrete instrument.
\newblock \emph{Journal of Political Economy}, 125\penalty0 (4):\penalty0 985--1039, 2017.

\bibitem[Burgess(2023)]{burgess2023violation}
Stephen Burgess.
\newblock Violation of the constant genetic effect assumption can result in biased estimates for non-linear mendelian randomization.
\newblock \emph{Human heredity}, 88\penalty0 (1):\penalty0 79--90, 2023.

\bibitem[Burgess et~al.(2014)Burgess, Davies, and Thompson]{burgess2014instrumental}
Stephen Burgess, Neil~M Davies, and Simon~G Thompson.
\newblock Instrumental variable analysis with a nonlinear exposure--outcome relationship.
\newblock \emph{Epidemiology (Cambridge, Mass.)}, 25\penalty0 (6):\penalty0 877, 2014.

\bibitem[Chernozhukov and Hansen(2006)]{chernozhukov2006instrumental}
Victor Chernozhukov and Christian Hansen.
\newblock Instrumental quantile regression inference for structural and treatment effect models.
\newblock \emph{Journal of Econometrics}, 132\penalty0 (2):\penalty0 491--525, 2006.

\bibitem[Cole et~al.(2010)Cole, Platt, Schisterman, Chu, Westreich, Richardson, and Poole]{cole2010}
Stephen~R Cole, Robert~W Platt, Enrique~F Schisterman, Haitao Chu, Daniel Westreich, David Richardson, and Charles Poole.
\newblock {Illustrating bias due to conditioning on a collider}.
\newblock \emph{International Journal of Epidemiology}, 39\penalty0 (2):\penalty0 417--420, 2010.
\newblock \doi{10.1093/ije/dyp334}.

\bibitem[Craven and Wahba(1978)]{craven1978smoothing}
Peter Craven and Grace Wahba.
\newblock Smoothing noisy data with spline functions.
\newblock \emph{Numerische mathematik}, 31\penalty0 (4):\penalty0 377--403, 1978.

\bibitem[Davey~Smith and Ebrahim(2003)]{davey2003mendelian}
George Davey~Smith and Shah Ebrahim.
\newblock ‘mendelian randomization’: can genetic epidemiology contribute to understanding environmental determinants of disease?
\newblock \emph{International journal of epidemiology}, 32\penalty0 (1):\penalty0 1--22, 2003.

\bibitem[Davey~Smith and Ebrahim(2025)]{davey2025erroneous}
George Davey~Smith and Shah Ebrahim.
\newblock Erroneous epidemiological findings on vitamins: coming full circle after two decades of mendelian randomization?, 2025.

\bibitem[Dawid(1979)]{dawid1979conditional}
A~Philip Dawid.
\newblock Conditional independence in statistical theory.
\newblock \emph{Journal of the Royal Statistical Society Series B: Statistical Methodology}, 41\penalty0 (1):\penalty0 1--15, 1979.

\bibitem[Didelez and Sheehan(2007)]{didelez2007mendelian}
Vanessa Didelez and Nuala Sheehan.
\newblock Mendelian randomization as an instrumental variable approach to causal inference.
\newblock \emph{Statistical methods in medical research}, 16\penalty0 (4):\penalty0 309--330, 2007.

\bibitem[Florens et~al.(2008)Florens, Heckman, Meghir, and Vytlacil]{florens2008identification}
Jean-Pierre Florens, James~J Heckman, Costas Meghir, and Edward Vytlacil.
\newblock Identification of treatment effects using control functions in models with continuous, endogenous treatment and heterogeneous effects.
\newblock \emph{Econometrica}, 76\penalty0 (5):\penalty0 1191--1206, 2008.

\bibitem[Fonseca et~al.(2025)Fonseca, Peixoto, and Saporito]{fonseca2025nonparametric}
Yuri Fonseca, Caio Peixoto, and Yuri Saporito.
\newblock Nonparametric instrumental variable regression through stochastic approximate gradients.
\newblock \emph{Advances in Neural Information Processing Systems}, 37:\penalty0 131756--131785, 2025.

\bibitem[Frangakis and Rubin(2002)]{frangakis2002principal}
Constantine~E Frangakis and Donald~B Rubin.
\newblock Principal stratification in causal inference.
\newblock \emph{Biometrics}, 58\penalty0 (1):\penalty0 21--29, 2002.

\bibitem[Graham et~al.(1998)Graham, Wilsnack, Dawson, and Vogeltanz]{graham1998should}
Kathryn Graham, Richard Wilsnack, Deborah Dawson, and Nancy Vogeltanz.
\newblock Should alcohol consumption measures be adjusted for gender differences?
\newblock \emph{Addiction}, 93\penalty0 (8):\penalty0 1137--1147, 1998.

\bibitem[Greco~M et~al.(2015)Greco~M, Minelli, Sheehan, and Thompson]{greco2015detecting}
Fabiola~Del Greco~M, Cosetta Minelli, Nuala~A Sheehan, and John~R Thompson.
\newblock Detecting pleiotropy in mendelian randomisation studies with summary data and a continuous outcome.
\newblock \emph{Statistics in medicine}, 34\penalty0 (21):\penalty0 2926--2940, 2015.

\bibitem[Greenland(2000)]{greenland2000}
S~Greenland.
\newblock {An introduction to instrumental variables for epidemiologists}.
\newblock \emph{International Journal of Epidemiology}, 29\penalty0 (4):\penalty0 722--729, 2000.
\newblock \doi{10.1093/ije/29.4.722}.

\bibitem[Guo and Small(2016)]{guo2016control}
Zijian Guo and Dylan~S Small.
\newblock Control function instrumental variable estimation of nonlinear causal effect models.
\newblock \emph{The Journal of Machine Learning Research}, 17\penalty0 (1):\penalty0 3448--3482, 2016.

\bibitem[Hamilton et~al.(2024)Hamilton, Hughes, Spiller, Tilling, and Smith]{hamilton2024non}
Fergus~W Hamilton, David~A Hughes, Wes Spiller, Kate~M Tilling, and George~Davey Smith.
\newblock Non-linear mendelian randomization: detection of biases using negative controls with a focus on bmi, vitamin d and ldl cholesterol.
\newblock \emph{European Journal of Epidemiology}, 2024.

\bibitem[Hamilton et~al.(2025)Hamilton, Hughes, Lu, Kutalik, Gkatzionis, Tilling, Hartwig, and Smith]{hamilton2025non}
Fergus~W Hamilton, David~A Hughes, Tianyuan Lu, Zolt{\'a}n Kutalik, Apostolos Gkatzionis, Kate~M Tilling, Fernando~P Hartwig, and George~Davey Smith.
\newblock Non-linear mendelian randomization: evaluation of effect modification in the residual and doubly-ranked methods with simulated and empirical examples.
\newblock \emph{European Journal of Epidemiology}, 2025.

\bibitem[Hansen(1982)]{hansen1982large}
Lars~Peter Hansen.
\newblock Large sample properties of generalized method of moments estimators.
\newblock \emph{Econometrica: Journal of the econometric society}, pages 1029--1054, 1982.

\bibitem[Hardy and Thompson(1998)]{hardy1998detecting}
Rebecca~J Hardy and Simon~G Thompson.
\newblock Detecting and describing heterogeneity in meta-analysis.
\newblock \emph{Statistics in medicine}, 17\penalty0 (8):\penalty0 841--856, 1998.

\bibitem[Hartford et~al.(2017)Hartford, Lewis, Leyton-Brown, and Taddy]{hartford2017deep}
Jason Hartford, Greg Lewis, Kevin Leyton-Brown, and Matt Taddy.
\newblock Deep iv: A flexible approach for counterfactual prediction.
\newblock In \emph{International Conference on Machine Learning}, pages 1414--1423. PMLR, 2017.

\bibitem[Hartung et~al.(2011)Hartung, Knapp, and Sinha]{hartung2011statistical}
Joachim Hartung, Guido Knapp, and Bimal~K Sinha.
\newblock \emph{Statistical meta-analysis with applications}.
\newblock John Wiley \& Sons, 2011.

\bibitem[He et~al.(2023)He, Liu, Lin, Zhuang, Shen, and Pan]{he2023delivr}
Ruoyu He, Mingyang Liu, Zhaotong Lin, Zhong Zhuang, Xiaotong Shen, and Wei Pan.
\newblock Delivr: a deep learning approach to iv regression for testing nonlinear causal effects in transcriptome-wide association studies.
\newblock \emph{Biostatistics}, 2023.

\bibitem[Heckman and Robb~Jr(1985)]{heckman1985alternative}
James~J Heckman and Richard Robb~Jr.
\newblock Alternative methods for evaluating the impact of interventions: An overview.
\newblock \emph{Journal of econometrics}, 30\penalty0 (1-2):\penalty0 239--267, 1985.

\bibitem[Hern{\'a}n et~al.(2004)Hern{\'a}n, Hern{\'a}ndez-D{\'\i}az, and Robins]{hernan2004structural}
Miguel~A Hern{\'a}n, Sonia Hern{\'a}ndez-D{\'\i}az, and James~M Robins.
\newblock A structural approach to selection bias.
\newblock \emph{Epidemiology}, pages 615--625, 2004.

\bibitem[Holland(1986)]{holland1986statistics}
Paul~W Holland.
\newblock Statistics and causal inference.
\newblock \emph{Journal of the American statistical Association}, 81\penalty0 (396):\penalty0 945--960, 1986.

\bibitem[Horowitz(2014)]{horowitz2014ill}
Joel~L Horowitz.
\newblock Ill-posed inverse problems in economics.
\newblock \emph{Annu. Rev. Econ.}, 6\penalty0 (1):\penalty0 21--51, 2014.

\bibitem[Imbens and Angrist(1994)]{imbens1994identification}
Guido~W Imbens and Joshua~D Angrist.
\newblock Identification and estimation of local average treatment effects.
\newblock \emph{Econometrica: journal of the Econometric Society}, pages 467--475, 1994.

\bibitem[Kassaw et~al.(2024)Kassaw, Zhou, Mulugeta, Lee, Burgess, and Hypp{\"o}nen]{kassaw2024alcohol}
Nigussie~Assefa Kassaw, Ang Zhou, Anwar Mulugeta, Sang~Hong Lee, Stephen Burgess, and Elina Hypp{\"o}nen.
\newblock Alcohol consumption and the risk of all-cause and cause-specific mortality—a linear and nonlinear mendelian randomization study.
\newblock \emph{International journal of epidemiology}, 53\penalty0 (2):\penalty0 dyae046, 2024.

\bibitem[Kling(2006)]{kling2006incarceration}
Jeffrey~R Kling.
\newblock Incarceration length, employment, and earnings.
\newblock \emph{American Economic Review}, 96\penalty0 (3):\penalty0 863--876, 2006.

\bibitem[Lawlor et~al.(2008)Lawlor, Harbord, Sterne, Timpson, and Davey~Smith]{lawlor2008mendelian}
Debbie~A Lawlor, Roger~M Harbord, Jonathan~AC Sterne, Nic Timpson, and George Davey~Smith.
\newblock Mendelian randomization: using genes as instruments for making causal inferences in epidemiology.
\newblock \emph{Statistics in medicine}, 27\penalty0 (8):\penalty0 1133--1163, 2008.

\bibitem[Liu et~al.(2019)Liu, Jiang, Wedow, Li, Brazel, Chen, Datta, Davila-Velderrain, McGuire, Tian, et~al.]{liu2019alc}
Mengzhen Liu, Yu~Jiang, Robbee Wedow, Yue Li, David~M Brazel, Fang Chen, Gargi Datta, Jose Davila-Velderrain, Daniel McGuire, Chao Tian, et~al.
\newblock Association studies of up to 1.2 million individuals yield new insights into the genetic etiology of tobacco and alcohol use.
\newblock \emph{Nature genetics}, 51\penalty0 (2):\penalty0 237--244, 2019.

\bibitem[Martens et~al.(2006)Martens, Pestman, de~Boer, Belitser, and Klungel]{martens2006}
E.P. Martens, W.R. Pestman, A.~de~Boer, S.V. Belitser, and O.H. Klungel.
\newblock {Instrumental variables: application and limitations}.
\newblock \emph{Epidemiology}, 17\penalty0 (3):\penalty0 260--267, 2006.
\newblock \doi{10.1097/01.ede.0000215160.88317.cb}.

\bibitem[Nelson and Startz(1988{\natexlab{a}})]{nelson1988distribution}
Charles Nelson and Richard Startz.
\newblock The distribution of the instrumental variables estimator and its t-ratiowhen the instrument is a poor one, 1988{\natexlab{a}}.

\bibitem[Nelson and Startz(1988{\natexlab{b}})]{nelson1988some}
Charles Nelson and Richard Startz.
\newblock Some further results on the exact small sample properties of the instrumental variable estimator, 1988{\natexlab{b}}.

\bibitem[Newey and Powell(2003)]{newey2003}
Whitney~K Newey and James~L Powell.
\newblock {Instrumental variable estimation of nonparametric models}.
\newblock \emph{Econometrica}, 71\penalty0 (5):\penalty0 1565--1578, 2003.
\newblock \doi{10.1111/1468-0262.00459}.

\bibitem[Paaby and Rockman(2013)]{paaby2013many}
Annalise~B Paaby and Matthew~V Rockman.
\newblock The many faces of pleiotropy.
\newblock \emph{Trends in genetics}, 29\penalty0 (2):\penalty0 66--73, 2013.

\bibitem[Pearl(2009)]{pearl2009causality}
Judea Pearl.
\newblock \emph{Causality}.
\newblock Cambridge university press, 2009.

\bibitem[Ramsay and Silverman(2005)]{ramsay2005functional}
James~O Ramsay and Bernard~W Silverman.
\newblock \emph{Functional data analysis}.
\newblock Springer, 2005.

\bibitem[Richardson and Robins(2013)]{richardson2013single}
Thomas~S Richardson and James~M Robins.
\newblock Single world intervention graphs (swigs): A unification of the counterfactual and graphical approaches to causality.
\newblock \emph{Center for the Statistics and the Social Sciences, University of Washington Series. Working Paper}, 128\penalty0 (30):\penalty0 2013, 2013.

\bibitem[Rubin(1974)]{rubin1974estimating}
Donald~B Rubin.
\newblock Estimating causal effects of treatments in randomized and nonrandomized studies.
\newblock \emph{Journal of educational Psychology}, 66\penalty0 (5):\penalty0 688, 1974.

\bibitem[Rubin(1980)]{rubin1980randomization}
Donald~B Rubin.
\newblock Randomization analysis of experimental data: The fisher randomization test comment.
\newblock \emph{Journal of the American statistical association}, 75\penalty0 (371):\penalty0 591--593, 1980.

\bibitem[Singh et~al.(2019)Singh, Sahani, and Gretton]{singh2019kernel}
Rahul Singh, Maneesh Sahani, and Arthur Gretton.
\newblock Kernel instrumental variable regression.
\newblock \emph{Advances in Neural Information Processing Systems}, 32, 2019.

\bibitem[Small(2014)]{small2014commentary}
Dylan~S Small.
\newblock Commentary: Interpretation and sensitivity analysis for the localized average causal effect curve.
\newblock \emph{Epidemiology}, 25\penalty0 (6):\penalty0 886--888, 2014.

\bibitem[Sofianopoulou et~al.(2024)Sofianopoulou, Kaptoge, Afzal, Jiang, Gill, Gundersen, Bolton, Allara, Arnold, Mason, et~al.]{sofianopoulou2024estimating}
Eleni Sofianopoulou, Stephen~K Kaptoge, Shoaib Afzal, Tao Jiang, Dipender Gill, Thomas~E Gundersen, Thomas~R Bolton, Elias Allara, Matthew~G Arnold, Amy~M Mason, et~al.
\newblock Estimating dose-response relationships for vitamin d with coronary heart disease, stroke, and all-cause mortality: observational and mendelian randomisation analyses.
\newblock \emph{The Lancet Diabetes \& Endocrinology}, 12\penalty0 (1):\penalty0 e2--e11, 2024.

\bibitem[Staley and Burgess(2017)]{staley2017semiparametric}
James~R Staley and Stephen Burgess.
\newblock Semiparametric methods for estimation of a nonlinear exposure-outcome relationship using instrumental variables with application to mendelian randomization.
\newblock \emph{Genetic epidemiology}, 41\penalty0 (4):\penalty0 341--352, 2017.

\bibitem[Sudlow et~al.(2015)Sudlow, Gallacher, Allen, Beral, Burton, Danesh, Downey, Elliott, Green, Landray, et~al.]{sudlow2015}
Cathie Sudlow, John Gallacher, Naomi Allen, Valerie Beral, Paul Burton, John Danesh, Paul Downey, Paul Elliott, Jane Green, Martin Landray, et~al.
\newblock {UK Biobank: an open access resource for identifying the causes of a wide range of complex diseases of middle and old age}.
\newblock \emph{PLOS Medicine}, 12\penalty0 (3):\penalty0 e1001779, 2015.

\bibitem[Sulc et~al.(2022)Sulc, Sjaarda, and Kutalik]{sulc2022polynomial}
Jonathan Sulc, Jennifer Sjaarda, and Zolt{\'a}n Kutalik.
\newblock Polynomial mendelian randomization reveals non-linear causal effects for obesity-related traits.
\newblock \emph{Human Genetics and Genomics Advances}, 3\penalty0 (3):\penalty0 100124, 2022.

\bibitem[Tamer(2010)]{tamer2010partial}
Elie Tamer.
\newblock Partial identification in econometrics.
\newblock \emph{Annu. Rev. Econ.}, 2\penalty0 (1):\penalty0 167--195, 2010.

\bibitem[Taylor et~al.(2014)Taylor, Davies, Ware, VanderWeele, Smith, and Munaf{\`o}]{taylor2014mendelian}
Amy~E Taylor, Neil~M Davies, Jennifer~J Ware, Tyler VanderWeele, George~Davey Smith, and Marcus~R Munaf{\`o}.
\newblock Mendelian randomization in health research: using appropriate genetic variants and avoiding biased estimates.
\newblock \emph{Economics \& Human Biology}, 13:\penalty0 99--106, 2014.

\bibitem[Tchetgen~Tchetgen et~al.(2021)Tchetgen~Tchetgen, Sun, and Walter]{tchetgen2021genius}
Eric Tchetgen~Tchetgen, BaoLuo Sun, and Stefan Walter.
\newblock The genius approach to robust mendelian randomization inference.
\newblock \emph{Statistical Science}, 36\penalty0 (3):\penalty0 443--464, 2021.

\bibitem[Terza et~al.(2008)Terza, Basu, and Rathouz]{terza2008two}
Joseph~V Terza, Anirban Basu, and Paul~J Rathouz.
\newblock Two-stage residual inclusion estimation: addressing endogeneity in health econometric modeling.
\newblock \emph{Journal of health economics}, 27\penalty0 (3):\penalty0 531--543, 2008.

\bibitem[Tian et~al.(2023)Tian, Mason, Liu, and Burgess]{tian2023relaxing}
Haodong Tian, Amy~M Mason, Cunhao Liu, and Stephen Burgess.
\newblock Relaxing parametric assumptions for non-linear mendelian randomization using a doubly-ranked stratification method.
\newblock \emph{PLoS genetics}, 19\penalty0 (6):\penalty0 e1010823, 2023.

\bibitem[Tian et~al.(2024)Tian, Tom, and Burgess]{tian2024data}
Haodong Tian, Brian~DM Tom, and Stephen Burgess.
\newblock A data-adaptive method for investigating effect heterogeneity with high-dimensional covariates in mendelian randomization.
\newblock \emph{BMC Medical Research Methodology}, 24\penalty0 (1):\penalty0 34, 2024.

\bibitem[Wang et~al.(2020)Wang, Sarkar, Carbonetto, and Stephens]{wang2020simple}
Gao Wang, Abhishek Sarkar, Peter Carbonetto, and Matthew Stephens.
\newblock A simple new approach to variable selection in regression, with application to genetic fine mapping.
\newblock \emph{Journal of the Royal Statistical Society Series B: Statistical Methodology}, 82\penalty0 (5):\penalty0 1273--1300, 2020.

\bibitem[Yang et~al.(2024)Yang, Mason, Wood, Schooling, and Burgess]{yang2024dose}
Guoyi Yang, Amy~M Mason, Angela~M Wood, C~Mary Schooling, and Stephen Burgess.
\newblock Dose-response associations of lipid traits with coronary artery disease and mortality.
\newblock \emph{JAMA Network Open}, 7\penalty0 (1):\penalty0 e2352572--e2352572, 2024.

\bibitem[Zhao et~al.(2020)Zhao, Wang, Hemani, Bowden, and Small]{zhao2020statistical}
Qingyuan Zhao, Jingshu Wang, Gibran Hemani, Jack Bowden, and Dylan~S Small.
\newblock Statistical inference in two-sample summary-data mendelian randomization using robust adjusted profile score.
\newblock \emph{The Annals of Statistics}, 48\penalty0 (3):\penalty0 1742--1769, 2020.

\end{thebibliography}
\endgroup

\end{document}